\documentclass[12pt]{elsart}
\usepackage{graphicx}
\def\be{\begin{eqnarray}}
\def\ee{\end{eqnarray}}
\def\bc{\begin{center}}
\def\ec{\end{center}}

\def\lsim{\stackrel{\scriptstyle <}{\phantom{}_{\sim}}}
\def\gsim{\stackrel{\scriptstyle >}{\phantom{}_{\sim}}}

\begin{document}
\begin{frontmatter}
\title{Hydrodynamical description of   first-order  phase
transitions: analytical  treatment and numerical modeling\\ }
\author[GSI,FIAS,JINR]{V.V. Skokov} and \author[GSI,MEPhI]{D.N. Voskresensky}
\address[GSI]{GSI, Planckstra\ss{}e 1, D-64291 Darmstadt, Germany}
\address[FIAS]{Frankfurt Institute for
Advanced Studies, Universit\"at Frankfurt, D-60438 Frankfurt am
Main, Germany}
\address[JINR]{Joint Institute for Nuclear Research, 141980 Dubna, Moscow Region, Russia}
\address[MEPhI]{Moscow Engineering Physical Institute,\\ Kashirskoe
  Avenue 31, RU-115409 Moscow, Russia}
\begin{abstract}
Solutions of hydrodynamical equations are presented  for an
equation of state allowing for a  first-order phase transition.
The numerical analysis is supplemented by analytical treatment
provided the system is close to the critical point. The processes
of growth and dissolution of  seeds of various sizes and shapes in
meta-stable phases (like super-cooled vapor and super-heated
liquid) are studied, as well as the dynamics of unstable modes in
the spinodal region. We show that initially nonspherical seeds
acquire spherical shape with passage of time. Applications to the
description of the first-order phase transitions in nuclear
systems, such as the nuclear gas-liquid transition occurring in
low energy heavy-ion collisions and the hadron-quark transition in
the high energy heavy-ion collisions are discussed. In both cases
we point out the important role played by effects of  viscosity
and surface tension. It is shown that fluctuations dissolve and
grow as if the fluid were effectively very viscous. Even in the
spinodal region seeds may grow  slowly due to viscosity and
critical slowing down. This prevents the enhancement of
fluctuations in the near-critical region, which is frequently
considered as a signal of the critical point in heavy-ion
collisions.
\end{abstract}\end{frontmatter}

\section{Introduction}
The description of first-order phase transitions is usually based
on  phenomenological approaches. One constructs a thermodynamical
 potential depending on an order parameter, similar to that in the
Landau theory of  phase transitions. Exploiting the fact that the
time evolution of the collective mode is usually slow compared to
the dynamics of microscopic processes one introduces a
time-dependent equation for the order parameter \cite{LP}
 \be\label{GL}
 \partial_{t}\phi =-\Gamma(\Delta)({\delta F}/{\delta \phi}), \quad
 \Delta =\partial_{x_1}^2 +....+\partial_{x_d}^2 .
  \ee
Here $d$ is the dimensionality of the space, $t$ is the time, $F$
is a  thermodynamic functional  expressed in proper variables
(e.g., the $\phi$- dependent correction to the Helmholtz  free
energy in $T,V$ variables), $T$ is the temperature, $V$ is the
volume and $\phi$ is an order parameter. For a non-conserving
order parameter $\Gamma(\Delta)=a_0$ and for a conserving one
$\Gamma(\Delta)=a_1 \Delta $. Appropriate quantities should be
chosen as order parameters for different physical systems under
consideration. These could be a macroscopic wave function for
 metallic superconductors and non-relativistic superfluids, as
superfluid $\mbox{He}^4$ and $\mbox{He}^3$, dipole moment for
 ferroelectrics, magnetic moment for ferromagnetics,
$\sigma$, $\omega$, $\rho$, $\pi$, $K$-mean fields for  various
phenomena of nuclear physics, etc.

There are many phenomena, where   first-order phase transitions
occur between phases with different densities. The description of
such phenomena should be similar to that for a  gas-liquid phase
transition. Therefore it is worthwhile to find the solutions of
corresponding hydrodynamical equations. Although some simplified
analytical \cite{PS,MSTV90,V93} and fragmentary two-dimensional
numerical \cite{Onuki} solutions have been found, many problems
still remain unsolved. In the general case one should construct an
appropriate numerical scheme, which could describe the phenomenon.
The hydrodynamical approach is fairly efficient for the
description of heavy-ion collisions in a broad range of collision
energy  from SIS to RHIC energies (see e.g.
\cite{SG,IR,IdealRHIC,Teaney,Romatschke,Romatschke:2007mq,Shuryak:2008eq,Muronga,Lallouet,NA,Dumitru}).
As a simplification, most of three-dimensional hydrodynamical
models use ideal hydrodynamics. Effects of viscosity and thermal
conductivity are simulated then with the help of an artificially
introduced friction between different components of the liquid.
See, e.g. Ref. \cite{IR}
 for  a three-fluid hydrodynamical scheme
applicable in a broad energy range from SIS to SPS.

There are arguments \cite{PS,MSTV90,V93} that the dynamics of a
first-order phase transition is controlled by non-zero values of the
kinetic coefficients.
 For
nuclear systems   transport coefficients are poorly known.
Transport coefficients in nuclear matter have been evaluated in
\cite{GIK,Danielewicz} for temperatures and nucleon densities
relevant for the nuclear gas-liquid (NGL) phase transition.
Estimates \cite{GIK} have demonstrated that the bulk viscosity is
much smaller  than the shear viscosity. Equations of
non-relativistic non-ideal hydrodynamics have been  solved to
construct  a description of heavy-ion collisions at SIS energies
\cite{SG}. Some models, \cite{Romatschke:2007mq,Shuryak:2008eq},
describing the expansion of matter at RHIC energies   solve
equations of relativistic non-ideal hydrodynamics in two spatial
dimensions and indicate that effects of viscosity in the state of
strongly coupled quark-gluon plasma (sQGP)  are minor. One
concluded that the ratio of the shear viscosity to the entropy
density is $\eta/s<0.2$, see \cite{Shuryak:2008eq}. For certain
materials, such as  helium, nitrogen and water  the experimental
values for this ratio $\eta/s$  show a minimum at the phase
transition \cite{CKM}. The properties of the bulk viscosity near
the hadronic--to--sQGP phase transition critical point (endpoint)
are discussed in Refs. \cite{Kharzeev,SR,Gubster}. For different
models have been derived results which disagree with each other.
For example, Refs. \cite{Kharzeev} argue for an essential increase
of the bulk viscosity at the critical point, contrary to the
results of \cite{SR}. Lattice QCD calculations \cite{Sakai}
applicable to the case of the sQGP demonstrate that the bulk
viscosity is much smaller  than the shear viscosity.

It is expected that at  large baryon densities and  not too high
temperatures the hadron--sQGP phase transition is of first-order
\cite{Berges,Stephanov}. Lattice results \cite{Alton,FK} support
this conclusion. Signatures of such a transition might manifest
themselves  in heavy-ion collisions in a broad interval of
energies, like those available at SPS (CERN) and at future low
energy campaign  RHIC (Brookhaven), and FAIR (GSI) and
 NICA (JINR) facilities. One  expects  strangeness trapping and enhancement
of the kaon multiplicity fluctuations  \cite{KMR}, enhancement of
the soft pion yield \cite{Berdnikov} and  baryon number density
fluctuations \cite{FrimanRedlich}, as  signals of the first-order
phase transition. The dynamics of classical modes of the chiral
condensate in presence of an expanding fluid of quarks was studied
in \cite{Dumitru} in terms of the $\sigma$-model. Quarks were
treated in the framework of relativistic ideal hydrodynamics. It
was indicated that significant density inhomogeneities appear
around the critical point of the first-order phase transition.
Ref. \cite{Stephanov} argued that the fireball may linger longer
in the vicinity of the critical point due to a divergence of
susceptibilities, e.g., the specific heat.  Refs.
\cite{Berdnikov,NA} paid attention to the critical slowing down
that limits the growth  of the $\sigma$-field correlation length in
the vicinity of the critical point. Some models speculate about
explosive  freeze-out assuming an increase of the viscosity close to
the critical point, see \cite{Scavenius}.

Another relevant phenomenon is the NGL first-order phase
transition that manifests itself in heavy-ion collisions in the
expansion stage at low densities \cite{Bondorf}. Possible effects
of super-cooled vapor and super-heated liquid phases, as well as
those of the spinodal region, have been considered in
\cite{SVB,Siemens}. The occurrence of a negative specific heat
($C_P$) was reported, as the first experimental evidence of the
liquid-gas phase transition in heavy-ion collision reactions
\cite{Agostino}. The complete list of references is too long to be
included here. For  a review of this interesting topic, see
\cite{Randrup}.

Mixed (so-called pasta) phases may occur in  systems with two or
more conserved charges (e.g. electric and baryon charges)
undergoing first-order phase transitions, see \cite{Glendenning}.
The physical origin of this phenomenon is based on the fact that
the electric charge can be conserved globally rather than locally.
The pasta (gas-liquid) phases appear in the  inner crusts of
neutron star. Besides, pasta phase may arise  also  in the
interior regions of compact stars, provided the equation of state
(EoS) allows for first-order phase transitions to the pion or the
kaon condensate, or to  quark matter \cite{VYT}. In equilibrium,
pasta is constructed of Wigner-Seitz cells. The conservation of
the electric charge is assumed within each Wigner-Seitz cell. Each
Wigner-Seitz cell contains a seed (nucleus) of one phase
surrounded by  matter of the another phase. Seeds are droplets for
a small concentration of the new phase, and  rods and slabs for
higher concentrations. Only little is known how pasta phases are
cooked dynamically, see \cite{MSTV90,Watanabe}.

In this paper, we will describe the  dynamics of  first-order phase
transitions by means of the  standard system of equations of
non-ideal non-relativistic hydrodynamics: the  Navier-Stokes
equation,
 \be \label{Navier}
\nonumber &&\rho \partial_{t} {u}_i +\rho
(\vec{u}\nabla) {u}_i \\ &&= -\nabla_i P
+ \nabla_k \left\{  \eta ( \nabla_k u_i + \nabla_i u_k -\frac{2}{d} \delta_{ik}
\mbox{div} \vec{u}   )   +
\zeta \delta_{ik}
\mbox{div} \vec{u} \right\} ,
 \ee
the continuity equation,
 \be\label{cont} \partial_{t}\rho +\mbox{div} (\rho
\vec{u})=0,
 \ee
and the general equation for the heat transport,
 \be\label{en}
&& T\left[\frac{\partial s}{\partial t} +\mbox{div}(s\vec{u}
)\right]\nonumber\\ &&=\mbox{div}(\kappa \nabla T)+\eta
(\nabla_k u_i + \nabla_i u_k -\frac{2}{d} \delta_{ik}
\mbox{div} \vec{u} )^2 +\zeta (\mbox{div} \vec{u})^2 .
 \ee
Here $\rho =mn$, $m$ is the mass of the constituents, $n$ is
 the density of  the conserving charge (e.g, the baryon charge); $P$ is
the pressure; $\eta$ and $\zeta$ are the first (shear) and second
(bulk) viscosities; $\vec{u}$ is the velocity of the element of
the fluid; $S$ is the entropy, $s=dS/dV$; $\kappa$ is the thermal
conductivity; as before,  $d$  is the dimensionality of space.
We solve these equations numerically in two spatial dimensions,
$d=2$, and analytically for arbitrary $d$  in the vicinity of the
critical point. Then we perform estimations for the cases of the NGL
and the hadron-sQGP phase transitions.

The paper is organized as follows. In sect. \ref{Small} we
analytically treat the  dynamics of a system  in the vicinity of the
critical point of the first-order phase transition. First
we perform a   reduction of the general system of
equations of non-ideal hydrodynamics to  equations for the order
parameters.
 Then, in order
 to solve the problem analytically, we
 use
a  density  expansion of the Landau free energy functional in the
vicinity of  the critical point
 and  derive the equation of motion for the density variable in
 dimensionless units.
 We study
evolution of  density fluctuations in metastable regions:
the super-heated liquid and
 the super-cooled vapor.
 Then we consider the  evolution of the seed shape
and study the dynamics of  density
 fluctuations in the spinodal region.
In sect. \ref{Numerical} we
 find numerical solutions of the general system of hydrodynamical
 equations (\ref{Navier}) -- (\ref{en}) in two spatial dimensions
 and  compare  these numerical results with
those of the analytical treatment. We also pay attention  to the
specific patterns,
 which are not seen in a simplified analytical formulation, such as
 the dependence of the time evolution of the seed  on  its shape and on the power of the inflow of the surrounding matter.
 To be specific, throughout   numerical calculations and analytical estimations
 we use  phenomenological   modified  Van der Waals  (mVW) EoS. Its expansion in the vicinity of the critical
 point is done in the Appendix A and
implies validity of the mean field approximation. A modification
of the VW EoS which we do does not change the universality class,
being the same for many substances named the VW fluids.
 In Appendix B
assuming thermal equilibrium  we evaluate a fluctuation region.
Although our consideration is very general, allowing for
 further applications of the results to  the description of specific nuclear
 dynamics, in sect. \ref{Quark}
 we specify the parameters first  for a nuclear matter  system undergoing a  NGL phase
 transition and then  for a system undergoing a hadron-to-sQGP phase transition.
 Sect. \ref{Conclusion}  formulates  conclusions.

 Throughout the paper we use units $\hbar =c=1$. The
 results of this paper are briefly summarized in the letter \cite{SV08}.

\section{Small overcriticality}\label{Small}
\subsection{ Reduction of equations of nonideal hydrodynamics
 to  equations for the order parameter}\label{Reduction}

Assume that EoS allows for a  first-order  phase transition and
conditions are such that the system is  somewhere in the vicinity
of the critical point. In this case pressure isotherm as function
of the density has a convex-concave shape, as for the Van der
Waals EoS.

Further in order to construct hydrodynamical description we need
an expression for a thermodynamical potential depending on
appropriate thermodynamical variables in each space-time point.
Working in ($T,\rho$) variables one may use relation between the
pressure $P$ and the Helmholtz free energy $F$:
 \be P=
 \rho({\delta [F(T,\rho )]}/{\delta \rho})|_{T}.
 \ee
If one wanted to work in ($s$, $\rho$) variables, one could use
that
 \be P
 =\rho({\delta [E(s,\rho )]}/{\delta \rho})|_{S},
 \quad T=
{\rho}^{-1} ({\delta [E(s,\rho)]}/{\delta s})|_{\rho},
 \ee
where $E(S,V)$ is the  energy. These expressions generalize
standard thermodynamical relations $P=-({\partial
F(T,V)}/{\partial V})|_{T} =-({\partial E(S,V)}/{\partial
V})|_{S}$, $ T =({\partial E(S,V)}/{\partial S})|_{V} $ to the
slightly spatially inhomogeneous configurations.

To treat the problem analytically let us expand  quantities,
entering EoS and equations of hydrodynamics, near some reference
point ($T_{\rm r} ,\rho_{\rm r}$)  on the curve $P(T,\rho )$ (or
($s_{\rm r} ,\rho_{\rm r}$) on the curve $P(s,\rho )$,
respectively), where ($T_{\rm r} , \rho_{\rm r}$) are assumed to
be close to the values in the critical point, i.e.
$0<T_{cr}-T_{\rm r} \ll T_{cr}$, $ 0<(\rho_{liq}^{\rm MC}
-\rho_{gas}^{\rm MC})/\rho_{\rm r} \ll 1$.
 It is convenient to take $\rho_{\rm r}$ satisfying the condition  $(\partial^2
P(\rho ,T)/\partial \rho^2 )|_{\rm r}=0$, or,  as an alternative,
$\rho_{\rm r} =\frac{1}{2}(\rho_{liq}^{\rm MC}+\rho_{gas}^{\rm
MC})$ can be chosen, where $\rho_{liq}^{\rm MC}$ and
$\rho_{gas}^{\rm MC}$ are 
 densities at the liquid-gas equilibrium ($\mu_{liq}=\mu_{gas}=\mu_{\rm MC}$,
$\mu$ is the chemical potential) determined by the Maxwell
construction (equal squares on the plot $P(1/\rho)|_{T}$, cut off
by the horizontal line $P=const$). Further to be specific  we will
take $T_{\rm r} =T_{cr}, \rho_{\rm r} =\rho_{cr}$. Then the above
mentioned condition $(\partial^2 P(T,\rho )/\partial \rho^2
)|_{\rm r}=0$ is  fulfilled.  In case of a mVW EoS, which we
exploit in this paper, all necessary explicit expressions are
presented in Appendix A.

We will consider evolution of  fluctuations. These are, e.g.,
seeds of one phase in another phase and  fluctuations like waves.
Seeds can be of different shapes. Simplest forms of seeds in $d=3$
spatial dimensions are spherical droplets and bubbles, rods and
slabs. Also seeds can have more peculiar shapes.

Since the evolution of collective modes is slower than that for
microscopic modes, we further consider small velocities
$\vec{u}(t,\vec{r})$ of the growth/damping of the density and
temperature fluctuations. Thus we linearize hydrodynamical
equations in the velocity "$u$" and in the density $\delta\rho
=\rho -\rho_{\rm r}$ and  temperature $\delta T=T-T_{\rm r}$
variables. Introducing auxiliary variable $z=\mbox{div}\,\vec{u}$
and applying operator "$\mbox{div}$" to both sides of the
Navier-Stokes equation  we obtain \cite{PS,V93}:
 \be\label{au}
\rho_{\rm r}({\partial z}/{\partial t })=-\Delta \left[ \delta P
-\left(\widetilde{d} \eta_{\rm r} +\zeta_{\rm r} \right) z \right]
,\quad \widetilde{d}={2(d-1)}/{d}.
 \ee
Continuity equation becomes
 \be\label{divz}
  ({\partial \delta\rho}/{\partial
t})=-\rho_{\rm r} z.
 \ee
Replacement of $z$ into Eq. (\ref{au})  produces
\cite{V93}
 \be\label{v-t}
\frac{\partial^2 \delta\rho}{\partial t^2}=\Delta \left[ \rho_{\rm
r }\frac{\delta [F(T,\delta\rho )]}{\delta (\delta\rho)}|_{T}+C
+\rho_{\rm
r}^{-1}\left(\widetilde{d}\eta_{\rm r} + \zeta_{\rm r}
\right)\frac{\partial \delta\rho}{\partial t} \right] ,
 \ee
where we expressed the pressure  in terms of the free energy in
$T,\rho$ variables. We added an additional constant term $C$ in
square brackets which will be specified below. Obviously $\Delta C
=0$. Note that for $\delta \rho
>0$ ($\rho
>\rho_{cr}$) we deal with one phase and for  $\delta \rho <0$ ($\rho <\rho_{cr}$), with
another phase. Transport coefficients can  differ in those phases.
To simplify the problem
 we assume a smooth density and temperature dependence of
transport coefficients at the transition through the critical
point although in a narrow fluctuation region  some of these
quantities could have a singular behavior. We ignore this
complication since such a temperature region is narrow and  since
it may take a long time to develop these singularities.

 Equation for the
density should be  supplemented with the equation for the entropy,
which in case of slow evolution of the seed acquires the form
 \be\label{en1}
T_{\rm r}({\partial s}/{\partial t}) =c_{V,\rm r}({\partial \delta
T}/{\partial t})=\kappa_{\rm r} \Delta \delta T ,
 \ee
where $\eta_{\rm r}$, $\zeta_{\rm r}$ and $\kappa_{\rm r}$ are the
kinetic coefficients taken at $T=T_{\rm r}$ and $\rho=\rho_{\rm
r}$.

In terms of $s$, $\rho$ variables Eq. (\ref{v-t}) reads
 \be \label{noncon}\frac{\partial^2 \delta\rho}{\partial
t^2}=\Delta \left[ \rho_{\rm r}\frac{\delta [E(s,\delta\rho
)]}{\delta (\delta\rho)}|_{s} +\rho_{\rm
r}^{-1}\left(\widetilde{d}\eta_{\rm r} + \zeta_{\rm r}
\right)\frac{\partial \delta\rho}{\partial t} \right] ,
 \ee
and Eq. (\ref{en1}) acquires the form of the equation for the
conserving order parameter,  also known as Cahn-Hilliard equation
\cite{CH}:
 \be\label{en2}
  T_{\rm r}({\partial
s}/{\partial t}) =\kappa_{\rm r} \Delta  ({\delta
[E(s,\rho)]}/{\delta s})|_{\rho}.
 \ee
We should note that  Eq. (\ref{v-t}) differs from that is usually
exploited in the framework of the phenomenological Landau
approach, see (\ref{GL}), and from equations used  for the
description of the dynamics of first-order phase transitions in
heavy-ion collisions, e.g. see
\cite{Gavin,Kapusta,Koide,Berdnikov}, and in relativistic
astrophysical problems \cite{astro}. Difference with Eq.
(\ref{GL}) disappears, if one sets  zero the square bracketed term
in the r.h.s. of (\ref{v-t}).  Then Eq. (\ref{v-t}) becomes
 \be\label{rs} \rho_{\rm r}^{-1}\left(\widetilde{d}\eta_{\rm r}
+\zeta_{\rm r} \right)\frac{\partial \delta\rho}{\partial t}
=\rho_{\rm r}\frac{\delta [F(T,\delta\rho )]}{\delta
(\delta\rho)}|_{T} +C =\rho_{\rm r}\frac{\delta [F_L (T,\delta\rho
)]}{\delta (\delta\rho)}|_{T}.
 \ee
From the first glance, such a reduction procedure is legitimate,
if space-time gradients are small. However for
a seed,  being prepared in a fluctuation at $t=0$ with a
distribution $\delta\rho (t=0,\vec{r})=\delta\rho (0,\vec{r})$,
the condition $\frac{\partial \delta\rho (t, \vec{r}) }{\partial
t}|_{t=0}\simeq 0$ should also be fulfilled. Otherwise there
appears a positive kinetic energy contribution. Probability of
such fluctuations should be suppressed.  On the other hand, two
initial conditions
 cannot be simultaneously fulfilled, if
equation contains  time derivatives of the  first-order only. Thus
{\em{there exists an initial stage of the dynamics of phase
transitions ($t\lsim t_{\rm init})$, which  is not described by
the standard Landau equation, see Eq. (\ref{GL}) or (\ref{rs}),
being broadly exploited in condensed matter physics.}} R.h.s. of
(\ref{rs}) presents thermodynamical force driving the system to
the final (equilibrium) state. This force should become zero for
$t\rightarrow \infty$, i.e.  when the system reaches the final
($f$) equilibrium state. Therefore $C=-P_{f}$. Thus instead of the
Helmholtz free energy we may introduce the Landau free energy
requiring that
 \be\label{landP}\rho_{\rm r}\frac{\delta [F_L (T,\delta\rho
)]}{\delta (\delta\rho)}|_{T}=P-P_{f}=0
 \ee
  in the final equilibrium
state.

Let the time scale  for the relaxation of the density, following
Eq. (\ref{v-t}), is $t_{\rho}$ and the time scale  for the
relaxation of the entropy/temperature, following (\ref{en1}),
is $t_T$. The latter quantity is estimated  as
 \be\label{kT}
t_T = R^2 c_{V,\rm r}
/\kappa_{\rm r} .
 \ee
Thus $t_T$ grows $\propto R^2$ with increase of the size of the
seed $R$. On the other hand, following  Eq. (\ref{v-t})
$t_{\rho}\propto R$ (we show below that a seed of rather large
size grows with constant velocity). Evolution of the seed is
governed by  the slowest mode. Thus, dynamics of seeds with sizes
$R<R_{\rm fog}$ (for $t_T (R)<t_{\rho}(R)$) is controlled by Eq.
(\ref{v-t}) for the density. Here $R_{\rm fog}$ is  the typical
seed size at which $t_{\rho}=t_T$. For seeds with sizes $R>R_{\rm
fog}$, $t_T\propto R^2$ exceeds $t_{\rho}\propto R$ and growth of
seeds is slown down. Thereby, number of seeds with the size $R\sim
R_{\rm fog}$ grows with time.  If conditions are such that
$\kappa$ is sufficiently large, $t_T$ exceeds $t_{\rho}$ only for
seeds of rather large sizes. At terrestrial conditions, when
growing droplet in the cloud becomes sufficiently large and heavy,
it falls down under the action of the gravity. Falling down, the
seed causes an avalanche of secondary droplets, if there are
accumulated already many droplets of the size $R\sim R_{\rm fog}$.
 Thus {\em{solving
Eqs. (\ref{v-t}), (\ref{en1}) we are able to describe  such a
phenomenon as rain}}, provided gravitational forces are
 incorporated. Contrary, if $\kappa$ is sufficiently small, $t_T$
exceeds $t_{\rho}$ already for seeds of rather small sizes (at
terrestrial conditions, it may occur when gravity is not yet
efficient). Since for $R>R_{\rm fog}$  growth of  seeds  is
slowing down and thus number of seeds with the size $R\sim R_{\rm
fog}$ is increasing  with time, {\em{there appears the fog.}}

Note that seeds of the new phase are produced in the old phase
owing to   short-scale fluctuations. The latter fluctuations are
not incorporated in above hydrodynamical equations describing by
the mean field variables. Contributions of short-scale
fluctuations can be simulated  by a random force induced in Eqs.
(\ref{v-t}), (\ref{en1}) with the help of the $\delta$-correlated
source terms, cf. \cite{PS}. Being produced owing these source
terms, large scale fluctuations (seeds)  evolve in time following
hydrodynamical equations.

\subsection{EoS in the vicinity of the critical point
 and equation for the density in dimensionless units}\label{ES}

Consider evolution  in $d$-dimensional space of seeds in case when
the heat transport is not yet efficient and the dynamics is
controlled by Eq. (\ref{v-t}) for the density variable.
 For the sake of simplicity
 let us use a convenient parameterization of the Landau  free
 energy, i.e. the generating functional  in $\delta \rho$, $\delta T$ variables,
 which variation in  $\delta \rho$ produces  equation of motion
 \be\label{fren}
\delta F_L = \int \frac{d^3 x}{\rho_{\rm r}}\left[ \frac{c[\nabla
(\delta \rho)] ^2}{2}+\frac{\lambda (\delta
\rho)^4}{4}-\frac{\lambda v^2 (\delta \rho) ^2}{2}-\epsilon \delta
\rho \right],
 \ee
where $\delta F_L =F_L [T,\rho ] -F_L [T_{\rm r},\rho_{\rm r}]$,
that allows for the first-order phase transition.  Here
coefficients $c
>0$, $\lambda >0$ are some functions of $T_{\rm r}$
and $\rho_{\rm r}$, $v$ and $\epsilon $ are yet functions of
$\delta T$ and $|\epsilon | \ll \lambda v^3$.
 As we have
mentioned, for  convenience we choose $T_{\rm r}=T_{cr}$,
$\rho_{\rm r}=\rho_{cr}$. For the mVW EoS, that we further exploit
in our numerical calculations, see Appendix A, the value $v^2$
diminishes towards the critical point as $v^2 \propto (T_{cr}-T)$.

For slightly inhomogeneous configurations the pressure can be
expressed as
 \be\label{pre} \delta P&=&P- P(T_{\rm r},\rho_{\rm r})\nonumber\\
 &\simeq& \frac{\partial P}{\partial T}|_{T_{\rm r},\rho_{\rm
 r}}\delta T +\frac{1}{2}\frac{\partial^2 P}{\partial T^2}|_{T_{\rm r},\rho_{\rm
 r}}(\delta T)^2
 -\lambda
v^2 \delta \rho +\lambda  (\delta \rho)^3  -c\Delta \delta \rho .
 \ee
Here we used expansion of $\delta P$ near $T_{\rm r},\rho_{\rm r}$
and Eqs. (\ref{landP}), (\ref{fren}). Thus
 \be\epsilon
&=&P_{f}-P(T_{\rm r},\rho_{\rm r})-\frac{\partial P}{\partial
T}|_{T_{\rm r},\rho_{\rm
 r}}\delta T -\frac{1}{2}\frac{\partial^2 P}{\partial T^2}|_{T_{\rm r},\rho_{\rm
 r}}(\delta T)^2 -...\nonumber\\
 &\simeq& P_f-P_{\rm MC}\simeq n_{cr} (\mu_{in}-\mu_{f}),\nonumber
  \ee
   $\mu_{in}$ and $\mu_f$ are the chemical potentials of the initial and final
configurations (at fixed $P$ and $T$).
 For the mVW EoS the pressure expansion in $\delta T$, $\delta n$
near $T_{cr}$, $n_{cr}$ is performed in Appendix A. Then maximum
value $\epsilon^{max}=P^{max}-P_{\rm MC}\simeq
n_{cr}(\mu^{max}-\mu_{\rm MC}) \propto (T_{cr}-T)^{3/2}$, where
$P^{max}$ and the chemical potential $\mu^{max}$ correspond to the
state where $P(n)$ has local maximum $P^{max}$, and $P_{\rm MC}$,
$\mu_{\rm MC}$ are quantities on the Maxwell construction.

\begin{figure}
\centerline{%
\rotatebox{0}{\includegraphics[height=5.0truecm] {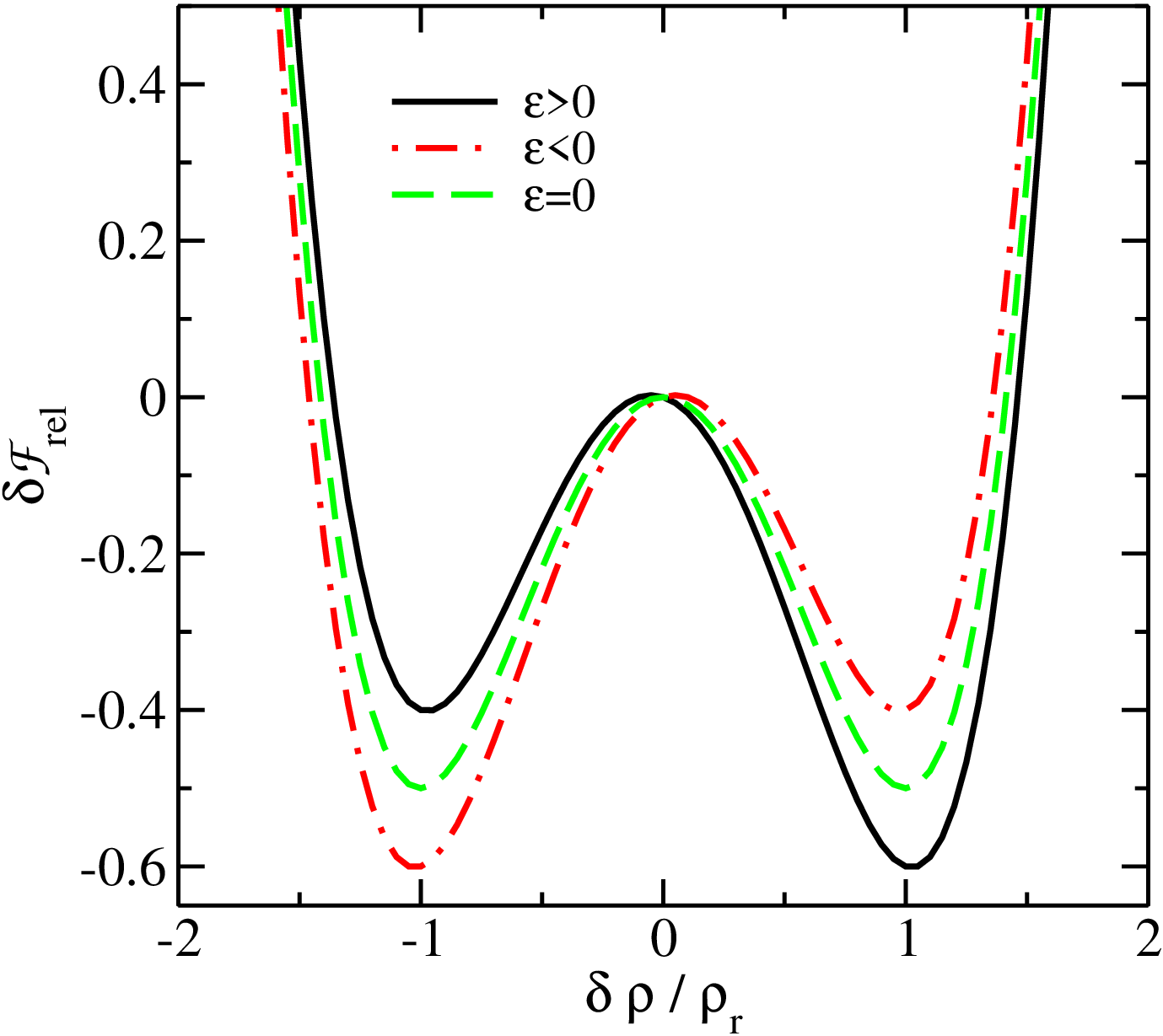}
} \rotatebox{0}{\includegraphics[height=5.0truecm]
{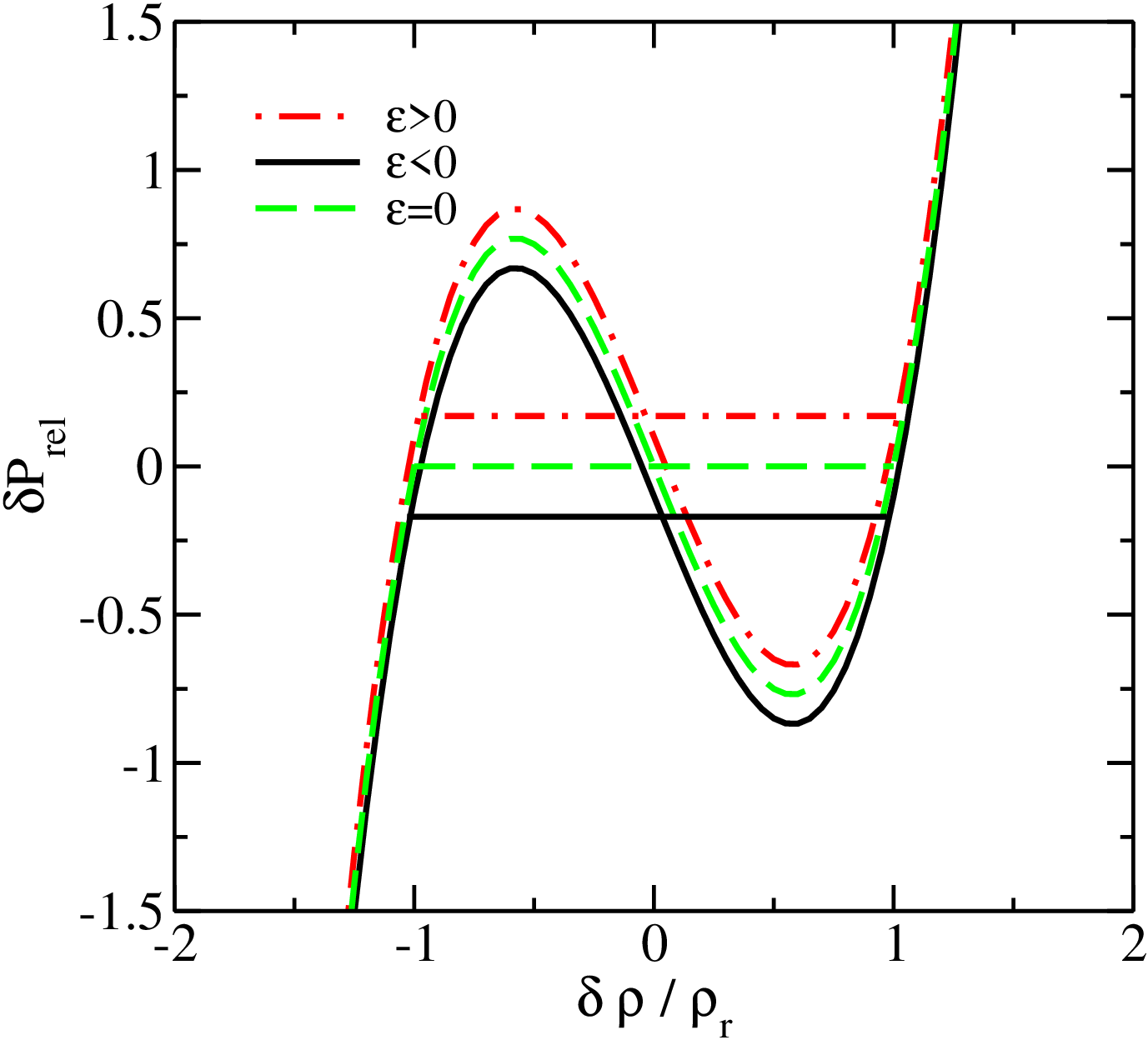}   }  } \caption{ The Landau free energy density
$\delta {\cal F}_{rel} =  \delta {\cal{F}}_L /{\cal{F}}_L (T_{cr},\rho_{cr})$ and the value
$ \delta { P}_{rel} = \rho_{\rm cr}\frac{\delta [F_L (T,\delta\rho )]}{\delta
(\delta\rho)}|_{T}/P (T_{cr},\rho_{cr})$
for uniform matter, as functions of the order parameter $\delta
\rho$ for the EoS determined  by Eq. (\ref{fren}).
Dash horizontal line ($\epsilon =0$) in the right panel shows
Maxwell construction. }\label{pres}
\end{figure}

The Landau free energy density $\delta {\cal{F}}_{rel} =  \delta {\cal{F}}_L /{\cal{F}}_L
(T_{cr},\rho_{cr})$ and the value  $\delta P_{rel} =  \rho_{\rm r}\frac{\delta [F_L
(T,\delta\rho )]}{\delta (\delta\rho)}|_{T}/P (T_{\rm r},\rho_{\rm
r}),$
 for $T_{\rm r}=T_{cr},\rho_{\rm
r}=\rho_{cr}$,
for spatially homogeneous configurations constructed following Eq.
(\ref{fren}),
as functions of the density $\delta
\rho$, are schematically shown in Fig. \ref{pres} left and right,
respectively. For $\epsilon
>0$ (solid lines) the liquid state is stable and the gas state is metastable,
and for $\epsilon <0$ (dash-dotted lines) the liquid state is
metastable, whereas the gas state is stable. For $\epsilon =0$ two
minima of the Landau free energy coincide
and correspond to the Maxwell construction on the curve $\delta P
(1/\rho)$ (shown by horizontal lines in the plot $\delta P
(\delta\rho)$ in the right panel).
Using Eq. (\ref{pre}), we rewrite Eq. (\ref{v-t}) as
 \be\label{v-t1}
-\frac{\partial^2 \delta\rho}{\partial t^2}=\Delta\left[c\Delta
\delta \rho +\lambda v^2 \delta \rho -\lambda (\delta \rho)^3
+\epsilon -\rho_{\rm r}^{-1}\left(\widetilde{d}{\eta_{\rm r}}
+{\zeta_{\rm r }} \right)\frac{\partial \delta\rho}{\partial
t}\right],
 \ee
In dimensionless variables $\delta \rho =v \psi$, $\xi_i =x_i /l$,
$i =1 ,\cdots , d$, ${\tau}=t/t_0$, Eq. (\ref{v-t1}) is simplified
as
 \be\label{dimens}
 &&- \beta \frac{\partial^2 \psi }{\partial
{\tau}^2} =\Delta_{\xi}\left(\Delta_{\xi}\psi +2\psi
 (1-\psi^2)+\widetilde{\epsilon}- \frac{\partial \psi}{\partial
 {\tau}}\right),\\
 &&l=\left(\frac{2c}{\lambda v^2}\right)^{1/2} ,\,\, t_0
 =\frac{2( \widetilde{d}\eta_{\rm r} +\zeta_{\rm r} )}{\lambda v^2
 \rho_{\rm r}},\,\,
 \widetilde{\epsilon}=\frac{2\epsilon}{\lambda v^3}, \,\, \beta
 =\frac{c\rho_{\rm r}^2 }{ ( \widetilde{d}\eta_{\rm r} +\zeta_{\rm
 r}
 )^2}.\nonumber
  \ee

Eq. (\ref{dimens}) is {\em{the key equation for our subsequent
analysis.}} For $\rho_{\rm r} =\rho_{cr}$, $T_{\rm r} =T_{cr}$
taken, as the reference point, and for the VW fluids (see Appendix
A) parameters demonstrate the following temperature dependence
near the critical point: $v\propto |\delta T|^{1/2}$, $\epsilon
\propto (\delta T)^{3/2}$ ($\epsilon^{max}=\sqrt{3}n_{cr}|\delta
  {{T}}|^{3/2}/T_{cr}^{1/2}$, $\widetilde{\epsilon}^{max}=4/(3\sqrt{3})$), and $l\propto
 |\delta T|^{-1/2}$ and $t_0\propto
 |\delta T|^{-1}$.  The latter value shows that  {\em{relaxation
 processes  in the vicinity of the phase transition
 critical point prove to be very slow.}}
This means that close to the critical point it is, indeed,
legitimate to describe the phase transition dynamics in the
framework of the non-relativistic approximation, as we do.

Note that values $c$, $\lambda$, $v$, $\epsilon$ in Eq.
(\ref{fren}) are purely phenomenological coefficients. Introducing
new variables $\lambda^{'}=\lambda m^2$, $v^{'}=v/m$, $\eta^{'}
=\eta /m$, $\zeta^{'}=\zeta/m$, $\epsilon^{'}=\epsilon/m$ one
could exclude, e.g., dependence on the mass of the constituent $m$
from all dynamical characteristics of the system. The value $m$ is
not well defined quantity in case of the quark system. In new
variables
 pressure and the Landau free energy are proportional to $m$, whereas
all  parameters entering Eq. (\ref{dimens}) do not already depend
on $m$.

As we have mentioned,
 Ref.  \cite{Stephanov} expressed an opinion
 that, if at some incident energy the trajectory passes  the vicinity of the critical point,
 the system may  linger longer in this region due to strong thermodynamical fluctuations resulting in the
 divergence of
 susceptibilities,
 that may reflect on observables.
 Contrary, we argue that {\em{fluctuation effects of the vicinity of
 the critical point in heavy-ion collisions
can  hardly be pronounced}},
 since all relevant  processes are proved to be frozen for $\delta T\rightarrow 0$, while the system
 passes this region during a finite time.

 As one can see from Eq. (\ref{dimens}), the dynamics of the phase
transition is governed by  single combination $(\widetilde{d}\eta
+\zeta )$ of  shear and bulk viscosities, cf. \cite{Kapusta}. This
observation might be interesting in connection with discussions of
the particular effects of   shear and bulk viscosities in sQGP,
see \cite{Shuryak:2008eq}.

The Landau free energy (\ref{fren}), being expressed in
dimensionless variables, becomes
 \be\label{dimF}
 \delta F_L =\frac{ l^3}{\rho_{\rm r}}\frac{\lambda
 v^4}{2}\int  d^3 \xi\left[\frac{(\nabla_{\xi}\psi)^2 +(\psi^2 -1)^2}{2}
 -\widetilde{\epsilon}\psi\right]-\frac{\lambda
 v^4}{4\rho_{\rm r}}\int d^3 x .
 \ee
 There exist homogeneous stationary solutions
of Eq. (\ref{dimens}):
 \be \psi \simeq\pm 1+\widetilde{\epsilon}/4,\quad |\widetilde{\epsilon}|/4 \ll
 1,
  \ee
corresponding to the Landau free energy density
 \be\label{de}
 \delta{\cal{F}}_L\simeq -\frac{\lambda
 v^4}{4\rho_{\rm r}}\mp
\frac{\epsilon v }{\rho_{\rm r}}.
 \ee
 Thus for $\widetilde{\epsilon}>0$ the
upper  sign solution describes stable liquid and the lower sign
solution  circumscribes super-cooled vapor, see Fig. \ref{pres}.
For $\widetilde{\epsilon}<0$ the lower sign solution describes
stable gas and the upper sign solution circumscribes  super-heated
liquid.

\subsection{Dynamics  of seeds in metastable and in stable
surroundings, for $t_{\rho}\gg t_T$}\label{metastable}

\subsubsection{Seed density profile}

In $d=3$ space one deals with configurations of different
symmetries, e.g., with spherically symmetric solutions ($d_{\rm
sol}=3$, droplets/bubbles), axially symmetric solutions ($d_{\rm
sol}=2$, liquid and gas rods) and one-dimensional solutions
($d_{\rm sol}=1$, liquid and gas slabs). To describe mentioned
configurations we search two-phase solution of Eq. (\ref{dimens})
in the form, see
 \cite{PS,MSTV90,V93},
 \be\label{sol}
 \psi \simeq \mp \tanh[\xi -\xi_0
 ({\tau})]+\widetilde{\epsilon}/4 ,
  \ee
$\xi =\xi_1$ for kinks (walls separating metastable and stable
phases), $\xi =\sqrt{\xi^2_1 +\xi_2^2}$, for rods, and $\xi
=\sqrt{\xi^2_1 +\xi_2^2 +\xi_3^2}$, for droplets/bubbles.
 For $\widetilde{\epsilon} >0$ upper sign
solution describes evolution of drop\-lets (or rods and kinks of
liquid phase) in a metastable super-cooled vapor medium.  The
lower sign solution circumscribes   evolution of bubbles (or rods
 and kinks of the gas phase) in a stable liquid medium.
For $\widetilde{\epsilon} <0$ the upper sign solution describes
evolution of droplets  in a stable gas medium, whereas the lower
sign solution circumscribes  then bubbles in a metastable
super-heated liquid medium. From the kink solution  for $\xi_0 \gg
1$ (thin wall)  one can easily construct the solution for the slab
($-\xi_0 <\xi<\xi_0$):
 \be\label{slab} \psi \simeq\mp \mbox{sgn}\xi
\tanh[\xi -\mbox{sgn}\xi\cdot \xi_0
 ({\tau})]+\widetilde{\epsilon}/4 .
  \ee
The boundary layer has the length $|\xi - \xi_0 (\tau)|\sim 1$.
Outside this layer corrections to homogeneous solutions are
exponentially small. Considering motion of the boundary for $\xi_0
(\tau)\gg 1$, we may put $\xi \simeq
 \xi_0 (\tau)$ in (\ref{sol}), (\ref{slab}).
Then keeping only linear terms in $\epsilon$ in Eq. (\ref{dimens})
we arrive at equation for $\xi_0 (\tau)$,
 \be\label{ks}
\frac{\beta}{2}\frac{d^2\xi_0}{d\tau^2} =
\pm\frac{3}{2}\widetilde{\epsilon}-\frac{d_{\rm sol}-1}{ \xi_0
(\tau)}-\frac{d\xi_0}{d\tau}.
 \ee

\subsubsection{Volume and surface contributions to the Landau free  energy}

Substituting (\ref{slab}) in (\ref{dimF}), subtracting infinite
constant term and supposing, as we have used above
$\xi_0(\tau)\gg1$,
we obtain
 \be\label{Fsol}
&&\delta F_L [\xi_{0}] =\frac{2\pi^{3/2}\Lambda^{3-d_{\rm
sol}}\lambda v^4 l^{d_{\rm sol}}}{\Gamma(d_{\rm sol}/2)\Gamma
(1+(3-d_{\rm sol})/2)\rho_{\rm r}}  \left[ \mp \frac{1}{d_{\rm
sol}}\widetilde{\epsilon} \xi_{0}^{d_{\rm sol}}
+\frac{2}{3}\xi_{0}^{d_{\rm sol}-1}\right],
  \ee
$2\Lambda$ is the diameter, height of cylinder and the length of
the squared plate
 for $d_{\rm sol}=3,2$ and $1$, respectively;  $\Gamma$ is the Euler $\Gamma$-function. The first term in
(\ref{Fsol}) is the volume term and the second one is the surface
contribution, $\delta  F_{L,\rm surf}$. At fixed volume in $d =3$
space, the surface contribution  for droplets/bubbles is
 smaller than for rods and slabs. Thereby, if  a seed prepared in a
 fluctuation  is initially non-spherical,  it   acquires spherical form
 with
passage of time (see  discussion in subsection \ref{Spherization}
below). The surface term is $\delta  F_{L,\rm surf} \equiv \sigma
S$, $S$ is the surface of the seed, $\sigma$ is the surface
tension.
 The gradient term
in (\ref{fren}) is as follows
 \be\label{frs_ad} \delta F_{L,\rm surf}^{\rm grad}
=\frac{1}{2\rho_{\rm r}} c \int (\nabla \rho )^2 dV =\frac{v^2
cl}{2\rho_{\rm r}}
 \int (\nabla_{\xi} \psi )^2 d^3\xi = \frac{2 v^2
c}{3l\rho_{\rm r}}S=\frac{1}{2}\delta  F_{L,\rm surf}.
 \ee
 Now we are
able to express the surface
 tension  through parameters of the EoS. For the mVW EoS, see Appendix A, we find
 \be \label{c}\sigma = \sigma_0
|\delta {\cal{T}}|^{3/2},\quad {\sigma_0^2}= {32mn_{cr}^2
T_{cr}}c;\quad
l=\frac{\sigma_0}{6T_{cr}n_{cr}|\delta{\cal{T}}|^{1/2}}, \ee
$\delta{\cal{T}}=(T-T_{cr})/T_{cr}$.

\subsubsection{Role of the viscosity and surface tension}
There are only two dimensionless parameters in Eqs.
(\ref{dimens}), (\ref{ks}), $\widetilde{\epsilon}$ and $\beta$.
 Parameter $\widetilde{\epsilon}$ shows the difference
 in  the Landau free energies  of metastable  and  stable
states, see Eqs. (\ref{de}) and (\ref{sol}).
 {\em{Dynamics is controlled by the parameter
$\beta$}}, which enters together with the second derivative in
time. This parameter can be expressed in terms of the surface
tension and the viscosity as
 \be
\beta
 = (32T_{cr})^{-1}[\widetilde{d}\eta_{\rm r} +\zeta_{\rm r} ]^{-2}\sigma_0^2 m.
  \ee
{\em{The larger viscosity and the smaller surface tension, the
effectively more viscous is the fluidity of seeds.}} For $\beta
\ll 1$ one deals with the regime of effectively viscous fluid and
at $\beta \gg 1$, with the regime of perfect fluid. Note that
$\beta$ does not  explicitly depend on the $n_{cr}$. As is shown
experimentally \cite{Beysens}, the ability of liquid domains to
coalesce is controlled by an interplay between the surface tension
and the viscosity. Parameter $\beta$, which we introduced, is
responsible for such an interplay.

\subsubsection{Critical radius}

 For the case of a metastable seed prepared
in a stable surrounding, both terms in the Landau free energy
(\ref{Fsol}) are positive (at $d_{\rm sol}\neq 1$). Thereby such
seeds should shrink. {\em{For stable seeds developing in
metastable surrounding}} the first (volume) term is negative.
Therefore in this case there exists a critical size of the seed,
which can be found by minimization of (\ref{Fsol}) in $\xi_0$:
$\xi^{cr}_0 = {2(d_{\rm sol}-1)}/({3|\widetilde{\epsilon}|}).$ For
the mVW EoS, see Appendix A, the condition $\xi_0^{cr}\gg 1$, that
we exploited  deriving Eq. (\ref{ks}), is fulfilled for
$|\widetilde{\epsilon}|\ll \widetilde{\epsilon}^{max}$. The closer
initial state to the Maxwell construction, the smaller is
$|\widetilde{\epsilon}|$.

The Landau free energy (\ref{Fsol}) decreases with decrease of the
droplet/bubble (or rod) size, provided $\xi_0(\tau)<\xi^{cr}_0$.
Such seeds of the new phase, being produced, are dissolved  with
passage of time. For $\xi_0(\tau)>\xi^{cr}_0$ the Landau free
energy decreases with increase of the seed size, that explains
growth of overcritical droplets/bubbles and rods of the stable
phase in the metastable medium.

 In dimensional units the critical size of the droplet/bubble
or rod becomes
 \be\label{crR}
R_{cr}&=&\frac{(d_{\rm sol}-1)v^2\sqrt{2c\lambda}}{3|\epsilon |}\\
R_{cr}(\gamma \epsilon^{max})&=&\frac{\sigma_0}{2\sqrt{3}\,\gamma
n_{cr}T_{cr}|\delta{\cal{T}}|^{1/2}},\nonumber
 \ee
  where the latter equality
  is valid for  the mVW EoS, see Appendix A, and we put $|\epsilon |  =
  \gamma\epsilon^{max}$, where constant $0\leq \gamma \leq 1$ characterizes deviation of $|\epsilon |$
from $\epsilon^{max}$. The  surface tension parameter $\sigma_0$
is given by Eq. (\ref{c}).
  The smaller  $|\epsilon |$, the larger is the critical radius of
the droplet/bubble (or the rod)   growing into the new phase. For
$\epsilon \rightarrow 0$ (Maxwell construction) $R_{cr}\rightarrow
\infty$.

Slabs of the stable phase, being placed in a metastable medium,
grow independently of what was the value of their initial size
($R\gsim l$).

\subsubsection{Probability of fluctuation}

   The probability of the initial
density fluctuation $\psi (\tau =0)$  of the size $\xi\simeq \xi_0
(\tau =0)$ (in dimensionless units), see (\ref{slab}), is as
follows
 \be\label{prob}
 W\sim e^{-\delta F_L [\xi_0 (\tau =0)]/T},
 \ee
 for $\xi_0 (\tau =0)<\xi_0^{cr}$.
For $\xi_0 (\tau =0)>\xi_0^{cr}$ (unstable region) $\delta F_L $
decreases with the growth of $\xi_0$, however  Eq. (\ref{prob})
does not hold anymore just indicating that these fluctuations may
grow. Certainly, correctly calculated probability of appearance of
a fluctuation with the size $\xi_0 (\tau =0)>\xi_0^{cr}$ should
decrease with increase of $\xi_0 (\tau =0)$. Being produced in
random processes, seeds evolve then according Eq. (\ref{ks}).

Two comments are in order. First, the profile (\ref{slab})
describes only one density distribution among various possible
configurations, which can be produced in fluctuations. We should
consider fluctuations involving many particles with the density
$\rho\simeq\rho_{liq}$ in the interior region  and
$\rho\simeq\rho_{gas}$ in the exterior (provided liquid is stable
phase). In difference with other configurations, the overcritical
droplet described by solution (\ref{slab}) gains in the Landau
free energy and conserves the form of the boundary layer,
$\mbox{tanh}(x)$, with time. Fluctuations with a different shape
of the boundary layer reach  $\mbox{tanh}(x)$- like shape  after
passage of a time, $\sim t_{\rm rec}$, necessary for a
reconstruction of the density profile. Second, various
fluctuations may have distinct velocities $\frac{d\xi_0}{d\tau}$
of the seed boundary, the latter being specified as the point,
where $\frac{\d^2 \rho}{d\xi^2} =0$. However fluctuations
containing a large number of particles, which we are interested
in, can be cooked with not a negligible probability  only provided
the velocity $\frac{d\xi_0}{d\tau}$ is zero or very small ($\ll
|\widetilde{\epsilon}|$). Otherwise the Landau free energy in Eq.
(\ref{prob}) (being $\propto |\widetilde{\epsilon}| \ll 1$, see
(\ref{de})) would acquire essential positive contribution that
would greatly diminish probability for the occurrence of such a
configuration. Fluctuations with initially zero or very small
values $\frac{d\xi_0}{d\tau}$, but with other density
distributions in the surface layer compared to (\ref{slab}), will
acquire a finite value of the velocity $\frac{d\xi_0}{d\tau}$ at
the seed  boundary after passage of a reconstruction time, $\sim
t_{\rm rec}$, when the density profile reaches the form of the
profile given by Eq. (\ref{slab}). For $t\gg t_{\rm rec}$ we may
simulate all these cases with the profile of  Eq. (\ref{slab}),
assuming different values of $\frac{d\xi_0}{d\tau}$ at $\tau\sim
\tau_{\rm rec}$.

\subsubsection{Dynamics of  slabs, $d_{\rm sol}=1$}

Consider dynamics of a slab of the stable phase (region
$0<|\xi|<\xi_0 (\tau)$) placed in a metastable surrounding
($|\xi|>\xi_0 (\tau)$). In this case one can find general solution
of Eq. (\ref{ks}). First, let us obtain solutions satisfying
initial conditions $\xi_0 (\tau =0)=\xi_0 (0)$,
$\frac{d\xi_0}{d\tau}|_{\tau =0}=0$, since appearance of such
seeds in fluctuations is more probable than, if
$\frac{d\xi_0}{d\tau}|_{\tau =0}$ were nonzero.  Moreover
$\frac{d\xi_0}{d\tau}|_{\tau =0}=0$, if the seed is formed  near
the boundary of the system (provided the boundary is flat).
Solution of Eq. (\ref{ks}) acquires the form
 \be\label{Slabsol}
 \xi_0 (\tau) =\xi_0
 (0)+\frac{3}{2}|\widetilde{\epsilon}|\tau -\frac{3}{4}|\widetilde{\epsilon}|\beta\left[1-\mbox{exp}
 \left(-2{\tau}/{\beta}\right)\right].
 \ee
 Thus {\em{slabs of an arbitrary
initial size (for $\xi_0 \gg 1$) grow with passage of time.}}

In dimensional units we obtain
 \be\label{Rt} R(t) =R_0 +u_{\rm asymp} t-\frac{3c^{1/2}|\epsilon|
 \beta}{2^{1/2}\lambda^{3/2}v^4}
 \left[1-\mbox{exp}
 \left(-\frac{\lambda v^2 m n_{cr}t}{(\widetilde{d}\eta_{\rm r}
 +\zeta_{\rm r})\beta}\right)\right],
 \ee
 where  $2R_0$ is an initial size of the slab and
 \be\label{uasi}
 u_{\rm asymp}&=&
{3|\epsilon|}{v^{-2}}\sqrt{{\beta}/{(2\lambda )}}, \\ u_{\rm
asymp}(\gamma\epsilon^{max})&=&\gamma
|\delta{\cal{T}}|^{1/2}\sqrt{3\beta {T_{cr}}/{(2m)}},\nonumber
 \ee
$u_{\rm asymp}$ is the velocity of the growth of the slab at large
values of time,
 $t\gg t_{\rm init}$. Second equality (\ref{uasi}) is valid for
the mVW EoS. Then
\be
   \frac{3\gamma\,c^{1/2}\epsilon^{max}
 \beta}{2^{1/2}\lambda^{3/2}v^4}&=&\frac{ \gamma\,\beta
 \sigma_0}{6\sqrt{3}\,n_{cr}T_{cr}|\delta{\cal{T}}|^{1/2}},\\
\frac{\lambda v^2 m n_{cr}}{(\widetilde{d}\eta_{\rm r}
 +\zeta_{\rm r})\beta}&=&\frac{72n_{cr}T_{cr}^2(\widetilde{d}\eta_{\rm r}
 +\zeta_{\rm r})|\delta{\cal{T}}|}{\sigma_0^2 m}.\nonumber
 \ee
 One can
distinguish two  stages of the evolution: an initial stage,
$t\lsim t_{\rm init}$  ($\tau \sim \tau_{\rm init} =\beta$ in
dimensionless units), and a subsequent stage, $t\gg t_{\rm init}$
(or $\tau \gg \tau_{\rm init}$ in dimensionless units). The
initial stage lasts
 \be\label{initt}
  t\sim t_{\rm
 init}&=&\frac{2(\widetilde{d}\eta_{\rm r}
 +\zeta_{\rm r})\beta}{\lambda v^2 m n_{cr}},\\
 t_{\rm
 init}&=&\frac{8\beta(\widetilde{d}\eta_{\rm r}
 +\zeta_{\rm r})}{9T_{cr}n_{cr}|\delta{\cal{T}}|}=
\frac{\sigma_0^2 m}{36 n_{cr}T_{cr}^2 (\widetilde{d}\eta_{\rm r}
 +\zeta_{\rm r})|\delta{\cal{T}}|}.\nonumber
  \ee
Second line (\ref{initt}) is  for the mVW EoS. For $t\ll t_{\rm
 init}$ from (\ref{Rt}) we obtain
  \be\label{slabw} R(t)\simeq R_0 +wt^2/2 ,\quad
  w=\frac{3|\epsilon|\lambda^{1/2}}{\sqrt{2c}},
 \ee
that corresponds to the growing of the slab with constant
acceleration.

Another typical time scale is $t_{\rho}=R/u_{\rm asymp}$, since
for $t\gg t_{\rm init}$ from  (\ref{Rt}) we obtain $R=R_0 +u_{\rm
asymp}t$ that corresponds to the growth of the size of the slab
with constant velocity.
 In the vicinity of the critical point for the mVW EoS the
velocity $u_{\rm asymp}(\gamma\epsilon^{max})\propto |\delta
{\cal{T}}|^{1/2}$ and
 \be\label{trslab}
 t_{\rho}=\frac{R}{u_{\rm asymp}(\gamma\epsilon^{max})}\propto R|\delta T|^{-1/2}.\ee
 Thus time scales  $t_{\rm
 init}\propto  |\delta{\cal{T}}|^{-1}$ and $t_{\rho}\propto R|\delta{\cal{T}}|^{-1/2}$ demonstrate that all
 processes freeze out at the critical point.

 For a large viscosity the time scale $t_{\rm
 init}\propto 1/(\widetilde{d}\eta_{\rm r}
 +\zeta_{\rm r})$ is rather short (excluding the vicinity of the critical point) and
the system rapidly reaches the asymptotic
 regime.
 Contrary, in case of an ideal liquid
 typical time scale  $t_{\rm
 init}$ is long and thus during a long time the size of the slab grows with acceleration.

For the initial condition corresponding to a nonzero velocity of
the seed boundary (at $\tau =\tau_{\rm rec}$) we  find
 \be\label{Slabsol1}
 \xi_0 (\tau) &=&\xi_0
 (\tau_{\rm rec})+\frac{3}{2}|\widetilde{\epsilon}|(\tau -\tau_{\rm rec}) -\frac{\beta}{2}
 \left(\frac{3}{2}|\widetilde{\epsilon}|-\frac{d\xi_0}{d\tau}|_{\tau =\tau_{\rm
 rec}}\right)\nonumber\\
 &\times&\left[1-\mbox{exp}
 \left(-2{(\tau -\tau_{\rm rec})}/{\beta}\right)\right]
 \ee
instead of (\ref{Slabsol}).

Another  comment is in order. From Eq. (\ref{divz}) using
(\ref{slab}) and the condition that the hydrodynamical velocity
(the velocity of the inflow of the surrounding matter) $u_{\rm
inflow}=0$ deeply inside the slab, we find
 \be\label{admis}
 u_{\rm inflow}(t)=-{l{t_0}^{-1}({d\xi_0}/{d\tau}})[\mbox{sgn}\xi \mbox{tanh}(\xi -\mbox{sgn}\xi\cdot\xi_0
 )+1].
  \ee
At the right boundary of the slab $u_{\rm inflow}(R =R_0
(t))=-{l}{t_0}^{-1}({d\xi_0}/{d\tau})=-dR(t)/dt$. For  $R\gg R_0
(t)$ at large $t$, $u_{\rm inflow}(R \gg R_0
(t))=-3\widetilde{\epsilon}{l}{t_0}^{-1}=-2u_{\rm asymp}$. In
case, if the initial shape of the seed differs from that given by
Eq. (\ref{slab}), the solution (\ref{admis}) is invalid for
$t<t_{\rm rec}$, where as before, $t_{\rm rec}$ is the time scale
of the reconstruction of the shape of the seed. From our numerical
solutions below we will see that for effectively large viscosity,
$\beta \ll 1$, the reconstruction period proves to be rather short
and in opposite limit, $\beta \gg 1$, of an effectively small
viscosity this time interval is very long.

\subsubsection{Initial stage of the seed evolution (for arbitrary $d_{\rm
sol})$}

Let us return to the consideration of the general case of an
arbitrary value $d_{\rm sol}$. As in previous subsection consider
dynamics of stable seeds in a metastable surrounding. Then we may
replace $\pm\widetilde{\epsilon}$ to $|\widetilde{\epsilon}|$ in
Eq. (\ref{ks}). We  search solution of Eq. (\ref{ks}) in the form
of the Taylor expansion
 \be\label{initxi}
 \xi_0 (\tau) =a_0 +a_1 \tau +a_2 \tau^2 +a_3 \tau^3
+...,
 \ee
  imposing initial conditions $\xi_0 (\tau =0) =\xi_0 (0)$,
$\frac{d\xi_0 (\tau) }{d\tau}|_{\tau =0}=0$. Then  we find
 \be\label{zero}
 \xi_0 (\tau) =\xi_0 (0)+\frac{1}{2}{w_{\tau}\tau^2} \left(1-\frac{2\tau}{3\beta}\right)+...,
 \,\,
\tau \ll \mbox{min}\left\{\tau_{\rm init}, \sqrt{{\xi_0
(0)}/{w_{\tau}}}\right\},
 \ee
 where
$ w_{\tau}=2\beta^{-1}\left[
 \frac{3}{2}|\widetilde{\epsilon}|-\frac{d_{\rm
sol}-1}{ \xi_0 (0)}\right].$
  For slabs this result follows from the
general solution (\ref{Slabsol}) at $\tau \ll \tau_{\rm init}$. In
dimensional units Eq. (\ref{zero}) is rewritten as
 \be\label{acs}
 R(t)=R_0 +\frac{w t^2}{2}\left(1-\frac{2t}{3t_0
 \beta}\right)+...,\quad t\ll  \mbox{min}\left\{
t_{\rm init},\left({R_0}/{w}\right)^{1/2}\right\}.
 \ee
The acceleration
 \be\label{axal}
 w={d^2 R}/{dt^2}=(d_{\rm sol}-1)\lambda
 v^2 \left({R_0}-R_{cr}\right)\left(R_{cr}{R_0}\right)
 \ee
changes sign at $R_0 =R_{cr}$. As we expected, seeds of
undercritical size shrink with passage of time, whereas
overcritical seeds grow. For  seeds of a nearcritical size the
process proceeds  slowly ($w\propto
|\delta{\cal{T}}|(R_0-R_{cr})/R_{cr}^{2}$). For undercritical
seeds of a small size, $w\propto -|\delta{\cal{T}}|/R_0$. Note
also that $w$ does not depend on the viscosity. For $d_{\rm
sol}\rightarrow 1$ from (\ref{acs}), (\ref{axal}) we recover
result (\ref{slabw}).

 One could think that for
slabs in case of perfect fluid  Eq. (\ref{acs}) is valid for all
times. Indeed, it follows from general solution (\ref{Rt}) for
$\beta\rightarrow
 \infty$ ($t_{\rm init}\rightarrow\infty$ in this case). But motion with constant
acceleration, see (\ref{axal}), becomes relativistic for large
times. Deriving Eq. (\ref{dimens}) from general system of
hydrodynamical equations we dropped quadratic terms in the
velocity. Thereby deriving (\ref{acs}) we additionally assumed
that  $t\ll \sqrt{R_0 /w}$.

 Notice, if we supposed that the velocity of the seed boundary
were finite at $\tau =0$, we would obtain
 \be\label{Slabsol12}
 \xi_0 (\tau) =\xi_0
 (0)+\left(\frac{d\xi_0}{d\tau}|_{\tau =0}\right)\tau +\frac{1}{\beta}\left[
 \frac{3}{2}|\widetilde{\epsilon}|-\frac{d_{\rm sol}-1}{ \xi_0
(0)}-  \frac{d\xi_0}{d\tau}|_{\tau =0} \right]\tau^2 +...
 \ee
instead of Eq. (\ref{zero}), compare Eqs. (\ref{Slabsol12}) and
(\ref{Slabsol1}) for $\tau_{\rm rec}=0$, $\tau \ll \tau_{\rm
init}$.

\subsubsection{Late stage of the  evolution ($t \gg t_{\rm
init}$) of large seeds ($R (t)\gg R_{cr}$)}

Let $\beta \ll \xi_0^2$. Then  we may drop the
 term $\frac{d^2 \xi_0 }{d
{\tau}^2}$ in  Eq. (\ref{ks}) and Eq. (\ref{ks}) simplifies as
 \be\label{xi0} {d\xi_0 (\tau)}/{d\tau}=
\frac{3}{2}|\widetilde{\epsilon}|-(d_{\rm sol}-1){\xi_0^{-1}
 (\tau)}.
  \ee
Note that solutions of Eq. (\ref{xi0}) do not satisfy necessary
condition $\frac{d \xi_0 }{d {\tau}}|_{\tau =0}=0$, except for the
case $\xi_0 (0)=\xi_0^{cr}$. Therefore initial stage $\tau \lsim
\tau_{\rm init}$ should be in any case  described by more general
Eq. (\ref{ks}), which is of the second order in time derivatives
(see also discussion above in subsection \ref{Reduction}). Thus
besides the condition $\beta \ll \xi_0^2$ we should still require
that $\tau \gg \tau_{\rm init}=\beta$.

For $\xi_0 (\tau)\gg \xi_0^{cr}$ (e.g. for initially overcritical
droplets/bubbles and rods of a very large size, $\xi_0 (0)\gg
\xi_0^{cr}$)  surface effects become unimportant. In this case one
can neglect the term $\propto 1/\xi_0$ in Eq. (\ref{xi0}). Then
for $\xi_0 (\tau)\gg \xi_0^{cr}$ we arrive at the solution
(\ref{Slabsol}). Thus large seeds grow with
 constant velocity,
 \be\label{long}
 \xi_0
(\tau)\simeq \xi_0 (0)+\frac{3}{2}|\widetilde{\epsilon}|\tau
,\quad R(t) \simeq R_0 +u_{\rm asymp}t, \quad t\gg t_{\rm init}.
 \ee
The time scale for the growth of the seed of the size $R$ is
 \be\label{trho1}
 \tau\sim \tau_{\rho}(\xi_0) =\frac{2\xi_0 }{3|\widetilde{\epsilon}|},\quad
 t\sim t_{\rho}(R)=\frac{R}{u_{\rm asymp}},
  \ee
for $t_{\rho}\gg t_{\rm init}$. Values $t_{\rm init}$ and
$t_{\rho}$ are the same as in (\ref{initt}) and (\ref{trslab}).
For the mVW EoS we estimate
 \be\label{trslab1}
 t_{\rho}(\gamma\epsilon^{max})=\frac{R}{u_{\rm asymp}(\gamma\epsilon^{max})}=\frac{4(\widetilde{d}\eta_{\rm r}
 +\zeta_{\rm r})}{3\gamma^2\,n_{cr}T_{cr}|\delta
{\cal{T}}|}\frac{R}{R_{cr}} \propto \gamma^{-2}R|\delta
T|^{-1}.\ee Both $t_{\rm init}$ and $t_{\rho}(R_{cr},\gamma
\epsilon^{max})$ are $\propto |\delta T|^{-1}$.

From (\ref{initt}) and (\ref{trslab1}) we see that for the system
near the critical point the  time scale
 $ t_{\rho}(R_{cr},\epsilon^{max}))$ is larger than $t_{\rm init}$ for $\beta \ll
1$. In case of effectively small viscosity, $\beta \gg 1$, the
 time scale $t_{\rm init}$  exceeds the value
 $t_{\rho}(R_{cr},\gamma\epsilon^{max})$. Then  the regime $t_{\rho}(R)\gg t_{\rm
 init}$ is reached only for $R\gg
 R_{cr}(\gamma\epsilon^{max})$.
Note that inequality $\beta \ll \xi_0^2$  is obviously fulfilled
for all $\xi_0 >\xi_0^{cr}$ for
$t_{\rho}(R_{cr},\gamma\epsilon^{max})\gg t_{\rm init}$, i.e.
$\tau_{\rho}(\xi_0^{cr})\gg \beta$, since the latter inequality
can be rewritten as $\beta \ll (\xi_0^{cr})^2$.

\subsubsection{Evolution
of seeds with nearcritical sizes (at $t \gg t_{\rm init}$)}

Let $d_{\rm sol}\neq 1$ and we continue to consider evolution  of
seeds of a stable phase in a metastable surrounding. Consider a
seed of initially nearcritical size,
 $ |\xi_0 (0)-\xi_0^{cr}|\ll \xi_0^{cr}$, at an intermediate stage
of its evolution, when the size of the seed still remains to be
close to the critical one,
 $|\xi_0 (\tau)-\xi_0 (0)|\ll\xi_0^{cr}$. We assume that conditions $\tau\gg \tau_{\rm init}=\beta$ and
$ \beta\ll \xi_0^2$ are again fulfilled.
 As we will see, evolution of the nearcritical seed proves to be very slow
(cf. Eq. (\ref{evolin}) below) and there exists a time interval,
where these conditions are fulfilled.
 Solution of (\ref{xi0}) is then as follows
 \be\label{ksil1}
&&\xi_0 (\tau) +\xi_0^{cr}
\mbox{ln}\frac{\xi_0 (\tau)-\xi_0^{cr}}{\xi_0
(0)-\xi_0^{cr}}=\xi_0 (0)+\frac{(d_{\rm
sol}-1)\tau}{\xi_0^{cr}},\\
&&{d\xi_0 (\tau)}/{d\tau}=({d\xi_0 (\tau)}/{d\tau})|_{\rm
n.cr}={(d_{\rm sol}-1)( \xi_0 (\tau)-\xi_0^{cr})}[{\xi_0^{cr}\xi_0
(\tau)}]^{-1}.\nonumber
 \ee

 Using condition that the size of the seed is close
to the critical size we find
 \be\label{xinearcr}
\xi_0 (\tau)=\xi_0 (0)+{(d_{\rm sol}-1)[{\xi_0^{cr}\xi_0
(0)}]^{-1}( \xi_0 (0)-\xi_0^{cr})\tau}.
 \ee
In dimensional units the latter equation renders
 \be\label{evolin}
 R(t)=R_0 +u_{\rm n.cr}t, \quad u_{\rm n.cr} = \frac{(d_{\rm sol}-1)c\rho_{\rm r} (R_0
-R_{cr})}{R_0 R_{cr} (\widetilde{d}\eta_{\rm r} +\zeta_{\rm r})}.
 \ee
As from Eq. (\ref{axal}), we see that seeds (droplets/bubbles and
rods) of overcritical size grow, whereas seeds of undercritical
size shrink. The velocity of the growth/shrinking of nearcritical
seeds (for $d_{\rm sol}\neq 1$, $|\xi_0 (0)- \xi_0^{cr}|/
\xi_0^{cr}\ll 1$) proves to be very low ($\propto |R_0 -R_{cr}|$).
Typical time, when initially nearcritical seed still continues to
be nearcritical, is
 \be
 t\sim t_{\rm n.cr}=\frac{(R_{cr})^3 (\widetilde{d}\eta_{\rm r}
 +\zeta_{\rm r})}{(d_{\rm sol}-1)c\rho_{\rm r} |R_0
-R_{cr}|}\gg t_{\rm init},
 \ee
 that justifies omitting  of the term $\frac{d^2 \xi_0
}{d {\tau}^2}$, as we have done it. In the vicinity of the
critical point for the mVW EoS, $t_{\rm
n.cr}(\gamma\epsilon^{max})\propto \gamma^{-3}|R_0
-R_{cr}|^{-1}|\delta{\cal{T}}|^{-3/2}$. Thereby dynamics of seeds
at this stage is very slow: $t_{\rm n.cr}\gg (t_\rho
(R_{cr},\gamma\epsilon^{max}),\, t_{\rm init}$).

 \subsubsection{Evolution of seeds of initially small size ($l\ll R_0 \ll R_{cr}$) at $t \gg t_{\rm init}$}
 For $\tau\gg \tau_{\rm init}$, $\xi_0^2\gg \beta$, describing seeds of a small size
($1\ll\xi_0 \ll\xi_0^{cr}$, $d_{\rm sol}\neq 1$) we can drop the
term $\propto \widetilde{\epsilon}$ in (\ref{xi0}). Then solution
satisfying initial condition $\xi_0 (\tau =0)=\xi_0 (0)$  acquires
the form
 \be\label{ksil1Us}\xi_0 (\tau)\simeq \sqrt{\xi_0^2
(0) - 2(d_{\rm sol}-1)\tau},\quad
\frac{d\xi_0 (\tau)}{d\tau}=-\frac{(d_{\rm
sol}-1)}{\sqrt{\xi_0^2 (0) - 2(d_{\rm sol}-1)\tau}}.
 \ee
In dimensional units
 \be\label{Rtdis} R(t)\simeq \sqrt{R_0^2  - 2(d_{\rm sol}-1){tl^2}/{t_0}}.
  \ee
 For $\tau\gg\tau_{\rm init}$,
$\xi_0^2 (0)\gg \beta$, i.e. for $t_{\rm init}\ll t\ll t_{\rm
dis}$, from (\ref{Rtdis}) we find
 \be\label{disLin} R(t)=R_0 +u_{\rm dis} t, \quad u_{\rm dis}= -\frac{(d_{\rm sol}-1)c\rho_0}{R_0
 (\widetilde{d}\eta_{\rm r}
+\zeta_{\rm r})} .
 \ee
The time scale of the dissolution  of the initial fluctuation of a
small size
 \be\label{disol}
t_{\rm dis} =  \frac{(\widetilde{d}\eta_{\rm r} +\zeta_{\rm r}
)R_0^2}{2(d_{\rm sol}-1)c\rho_{\rm r}}
=\frac{16n_{cr}T_{cr}(\widetilde{d}\eta_{\rm r} +\zeta_{\rm r}
)R_0^2}{(d_{\rm sol}-1)\sigma_0^2}
 \ee
(in dimensional units) proves to be  $\propto R^2_0$. Thereby,
fluctuations of sufficiently small sizes are easily produced and
then they are  rapidly dissolved. As we see from (\ref{Rtdis}),
(\ref{disLin}), at $t\gsim t_{\rm init}$ the speed of the
dissolution first reaches a constant value and then the process is
rapidly  accelerated.

Note that solving the problem  we assumed $\xi_0 \gg 1$ and
$\xi_0^2 (0)\gg \beta$. Both conditions  are  satisfied at least
in case of an effectively large viscosity, $\beta \ll 1$.

\subsubsection{Numerical solution of equation for the droplet/bubble boundary
}

\begin{figure}
\centerline{%
\includegraphics[height=6.0truecm] {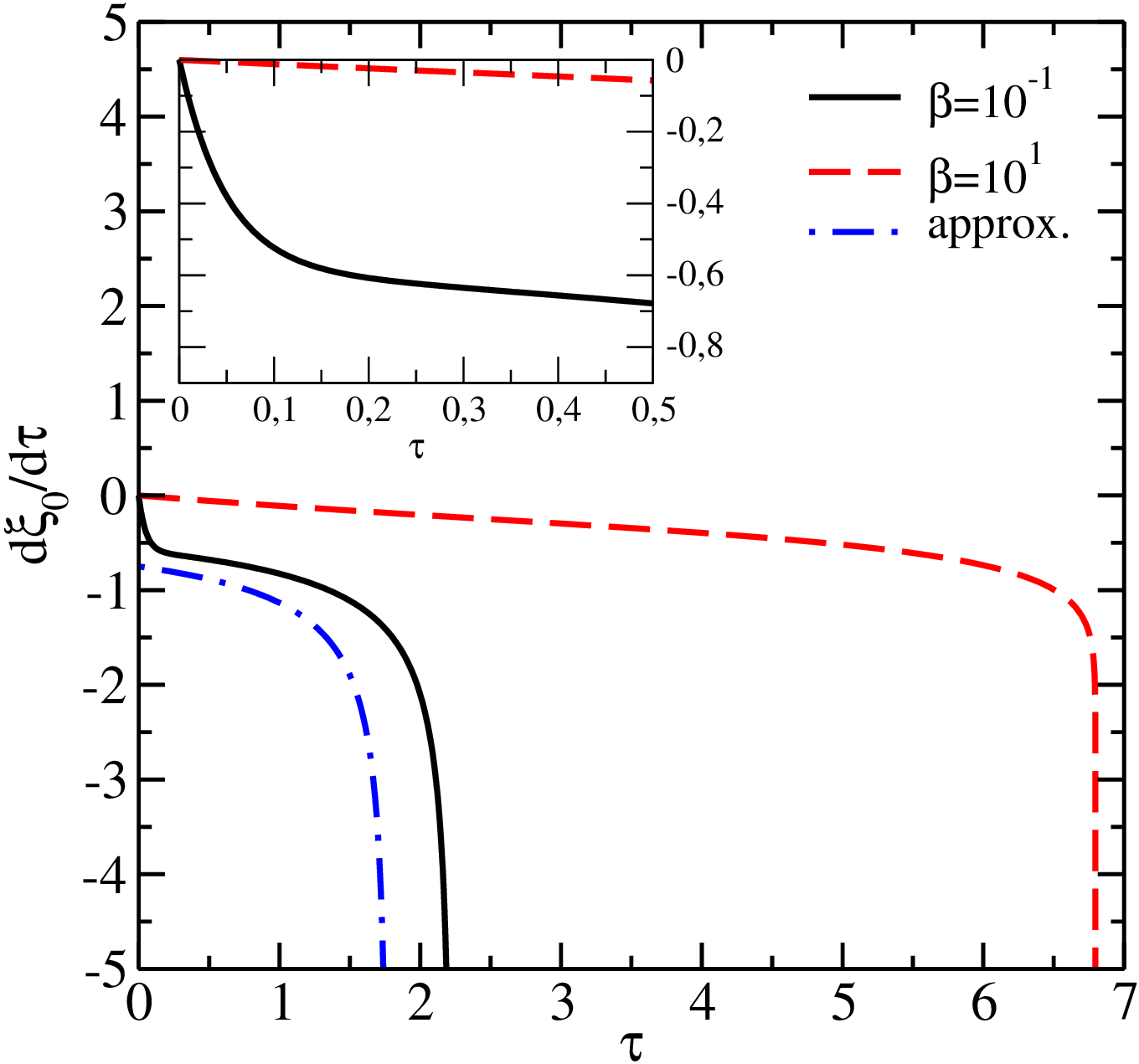} \hspace{0.3cm}
\includegraphics[height=6.0truecm]{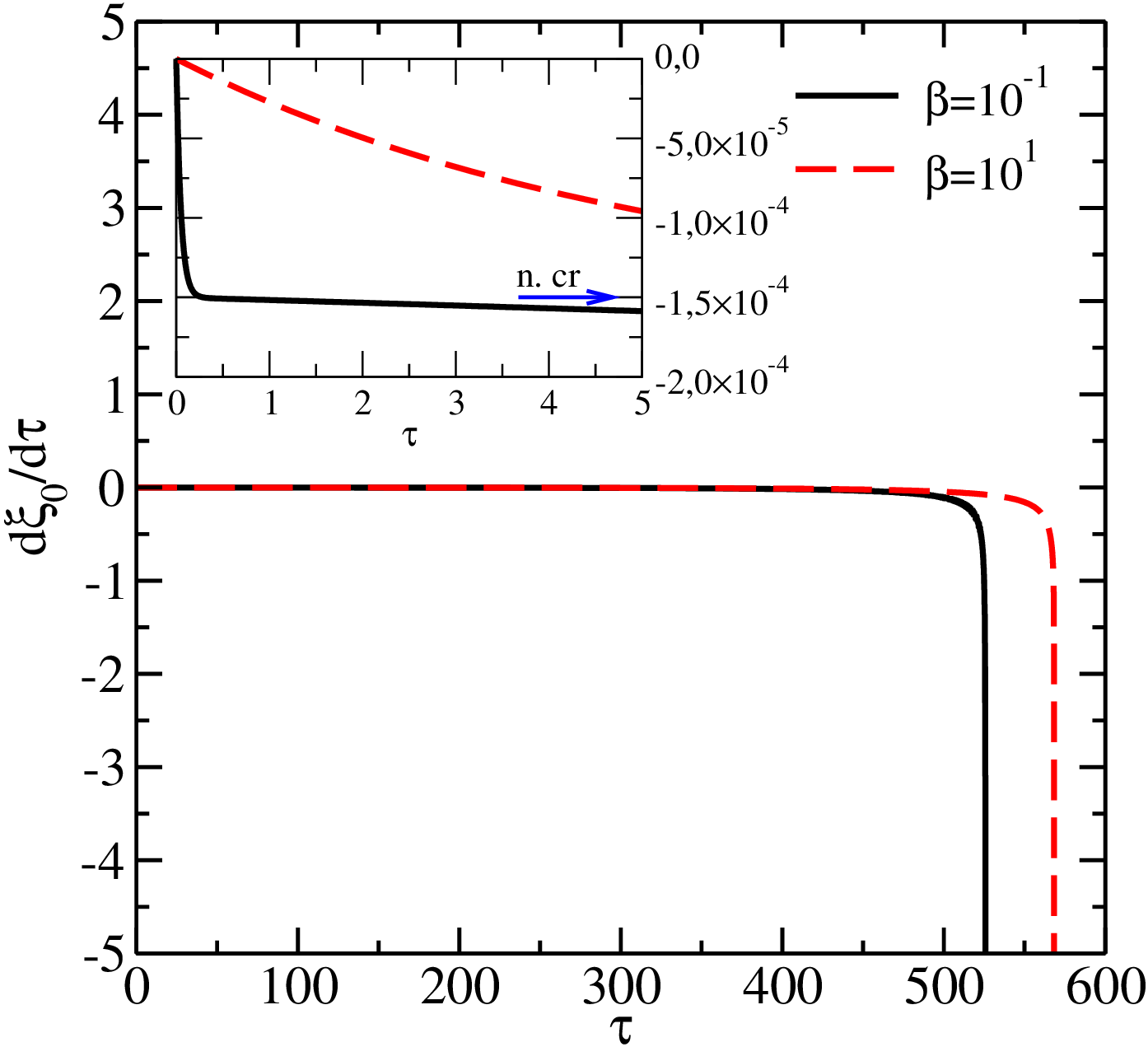}
}%
\caption{ Numerical solution of Eq. (\ref{ks}) for the
undercritical droplet of initial size $\xi_0 (0)=0.3 \xi_0^{cr}$
(left panel) and for the near-critical seed, $\xi_0 (0)=0.999
\xi_0^{cr}$, (right panel) for effectively large viscosity (solid
curves)  and for effectively small viscosity  (dash curves);
$\widetilde{\epsilon}=0.1$. Dash-dotted line is analytical
solution given by Eq. (\ref{ksil1Us}). }
\vspace{10mm}\label{dynamics}
\end{figure}

\begin{figure}
\centerline{%
\includegraphics[height=6.0truecm] {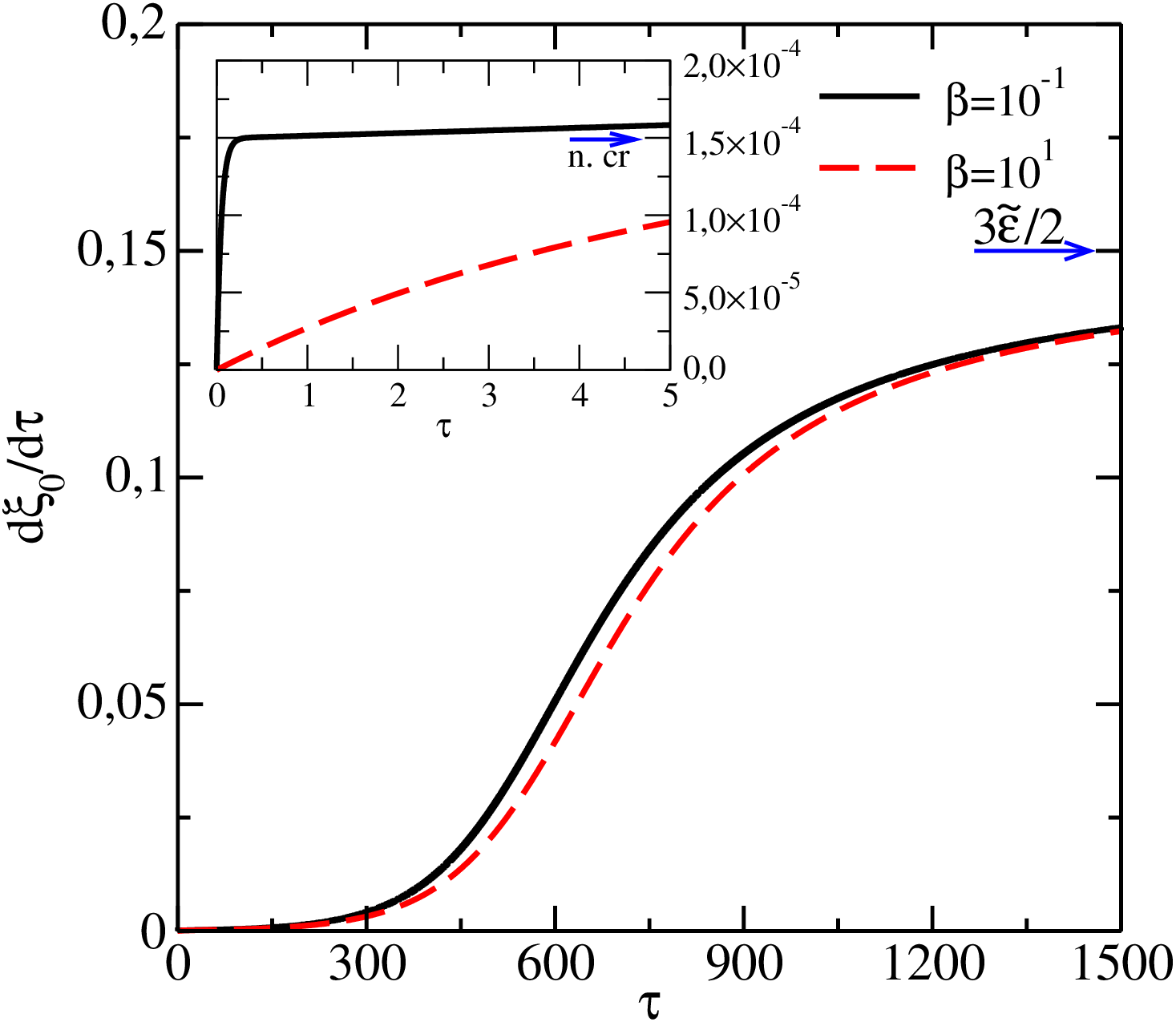}
\includegraphics[height=6.0truecm] {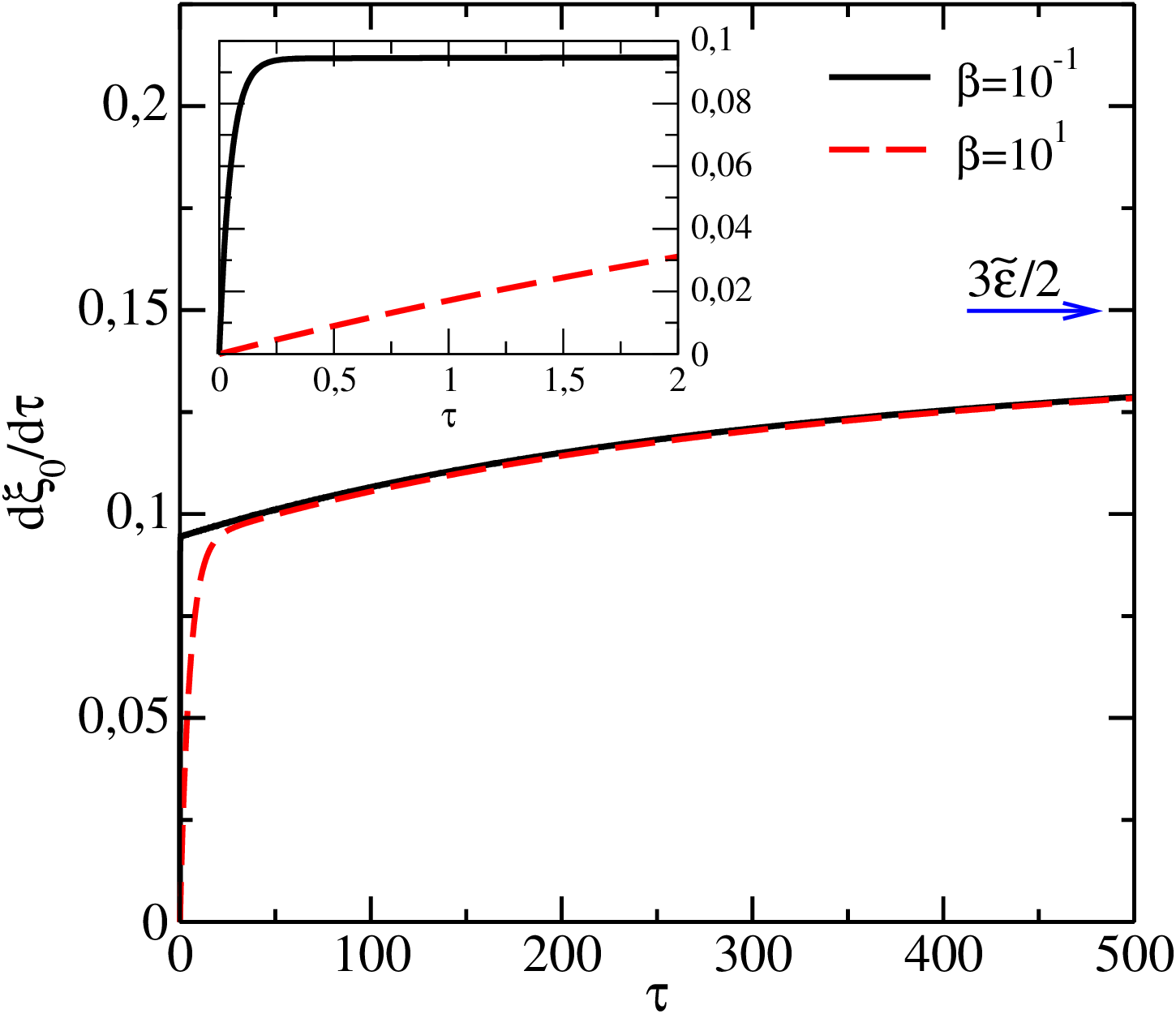}
}%
\caption{ Numerical solution of Eq. (\ref{ks}) for the
overcritical droplet of initial size $\xi_0 (0)=1.001 \xi_0^{cr}$
(left) and for $\xi_0 (0)=1.1 \xi_0^{cr}$ (right) for two values
of $\beta$. The arrow $3\widetilde{\epsilon}/2$ shows asymptotic
value of the velocity (in dimensionless units). }\label{dynamics1}
\end{figure}

Peculiarities of different regimes, which we have described
analytically  in limit cases, are demonstrated by numerical
solutions of Eq. (\ref{ks}) presented in Figs. \ref{dynamics} and
\ref{dynamics1} on example of the time dependence of dimensionless
velocity $\frac{d\xi_0 (\tau)}{d\tau}$. Insertions in Figs.
\ref{dynamics} and \ref{dynamics1} show evolution of seeds  at
 small  times.

 Fig. \ref{dynamics} shows numerical solution of Eq.
 (\ref{ks}) for undercritical seeds. In the left panel we consider
 evolution of
 the seed of a  small size ($\xi_0 (0)=0.3 \xi_{cr}$) and in the right panel, of the seed of a near-critical size
 ( $\xi_0 (0)=0.999
\xi_0^{cr}$).
 We see that undercritical
 seeds are shrinking with  time.
 For the
 seed of a sufficiently small size (left panel)  shrinking process is rather fast. The
 larger
 viscosity (smaller $\beta$), the faster is the process. One
 distinguishes two stages of the process: an initial stage ($\tau \lsim \tau_{\rm init}=\beta$), and a
 dissociation stage, $\tau_{\rm init}\ll
 \tau \lsim \tau_{\rm dis}$, see (\ref{disol}). The latter stage is subdivided by two stages:
  $\tau\ll \tau_{\rm dis}$ (characterizing by approximately constant velocity) and $\tau\sim \tau_{\rm dis}$
  (when the process is rapidly accelerated). The curve labeled as
 "approx" demonstrates   analytical solution  Eq. (\ref{ksil1Us}),
 being
 valid for $\tau\gg \tau_{\rm init}$, that requires   $\tau_{\rm dis}\gg \tau_{\rm init}$. The latter
 condition is not fulfilled for $\beta =10$ (effectively small viscosity), but it is
 fulfilled for $\beta =10^{-1}$ (effectively large viscosity). Deviation of the dash-dotted
curve from  the solid one for  $\beta =10^{-1}$  is due to the
fact that the ratio $\xi_0 (0)/\xi_0^{cr}=0.3$ is not yet much
less than unity, and $\xi_0 (0)\simeq 4$ is not yet much larger
than unity, whereas conditions $\xi_0 (0)/\xi_{cr}\ll 1$, $\xi_0
(0)\gg 1$  are required for validity of the analytical solution
(\ref{ksil1Us}). We have checked this with various choices of
parameters. Significant difference between the dash-dotted and the
dash curve in case of a small effective viscosity (large $\beta$)
is due to the fact that $\tau_{\rm init}>\tau_{\rm dis}$ in this
case. Thus {\em{for an effectively small viscosity dissolution of
the small size seed occurs at the time scale $\tau\sim \tau_{\rm
init}$.}}

 For the nearcritical seed (see right panel of Fig. \ref{dynamics}) one can distinguish three stages of the process:  initial stage
 ($\tau \lsim \tau_{init}=\beta$); an intermediate
 stage,
 when the seed still continues to have nearcritical size,
 $(\xi_0^{cr}-\xi_0 (\tau) )/\xi_0^{cr}\ll 1$ (it lasts very
 long, up to $\tau \lsim 500$); and a short  subsequent dissociation stage.
 Dependence on  the viscosity
 (the value $\beta$) does not manifest itself at the intermediate
 stage. The "n.cr." arrow in Figure insertion shows the value of the velocity
 which follows from
 our analytical solution (\ref{xinearcr}).

In Fig. \ref{dynamics1} we demonstrate numerical solution of Eq.
 (\ref{ks}) for overcritical seeds. We see that these
 seeds are growing with  time.
 For initially
 nearcritical seed (left panel),  there exist four stages of the process.
 Initial stage lasts a time
$\tau \lsim \tau_{\rm init}=\beta$, see Figure insertion. Next is
a prolonged stage,
 when $(\xi_0 (\tau)-\xi_0^{cr} )/\xi_0^{cr}\ll 1$ (velocity is almost constant for $\tau \lsim
 200$).
 Then there is a transition stage  (for $\tau\lsim 1500\div
 2000$), when   $(\xi_0 (\tau)-\xi_0^{cr} )/\xi_0^{cr}\sim 1$,
 and finally
 the asymptotic regime, when the velocity reaches new constant
 value $\frac{3}{2}\widetilde{\epsilon}$ (shown in Figure by arrow). On a so long time scale the
  memory about  initial stage of the evolution (the latter depends on $\beta$: $\tau \lsim \tau_{\rm init}=\beta$)
  is lost: dash and solid curves coincide.
 Note however that in dimensional units asymptotic regime is  achieved much faster
 in case of a more
 viscous fluid.

The right panel of Fig.  \ref{dynamics1} shows evolution of the
overcritical seed for $\xi_0 (0)=1.1\xi_0^{cr}$. The process
proceeds faster, than for  $\xi_0 (0)=1.001\xi_0^{cr}$, and in
three stages. From initial stage through a transition stage the
seed reaches  asymptotic regime. Transition regime lasts very
long.

As we see, presented numerical solutions of Eq.  (\ref{ks})
support  our simplified analytical considerations performed above.

\subsection{Evolution  of the shape of seeds}\label{Spherization}

To be specific consider solutions in $d=3$ space. Majority of
seeds, being produced in fluctuations, have near spherical form.
Then their shape evolves with passage of time. To describe this
evolution let us expand $\xi_0 (\theta , \phi, \tau )$ in
spherical functions:
 \be
 \xi_0 =\sum_{l,m}\xi_0 (l,m)Y_{l,m}(\theta , \phi ).
  \ee
It is convenient to
 use shorthand  notations $\xi_0 (l=0,m)=\xi_0 (l=0)$ and $\xi_0
(l\neq 0,m)=\xi_{0l}$. Further assume that the seed has almost
spherical form, i.e. $ \xi_{0l}\ll 1$. Using that $\Delta_{\xi}
=\frac{\partial^2}{\partial \xi^2}+\frac{2}{\xi}\frac{\partial
}{\partial \xi}+\frac{\hat{L}^2}{\xi^2}$, $\frac{\hat{L}^2
\xi_{0l} }{\xi^2}=\frac{l(l+1) \xi_{0l} }{\xi^2}$, in the linear
approximation in $\xi_{0l}$ from Eq. (\ref{ks}) we obtain equation
for the time evolution of the seed boundary (for $\xi_{0l}$). It
looks cumbersome.
Dropping  terms $\propto O(\frac{d\xi_{0}}{d \tau},
\frac{d^2\xi_{0}}{d \tau^2})$ we are able to  simplify  equation
as follows
 \be\label{linas}
\frac{1}{2}\frac{d^2\xi_{0l}}{d \tau^2}&=&
\frac{(2-l(l+1))\xi_{0l}}{\widetilde{\beta} \xi_0^2 (l=0, \tau
)}-\frac{1}{\widetilde{\beta}}\frac{d\xi_{0l}}{d \tau}, \quad
l\geq 1,\\
 \widetilde{\beta}&=&\beta\left[1+\frac{1}{2}{l(l+1)}{\xi_0^{-2}
(l=0,\tau)}\right]^{-1}.\nonumber
 \ee
For $\xi_0^2 (l=0,\tau )\gg l^2$, that we will further assume
(just to deal with simpler expressions) one has
$\widetilde{\beta}\simeq \beta$. The value $\xi_0 (l=0, \tau)$
obeys Eq. (\ref{ks}). Undamped mode with $l=1$ describes the
motion of the system as a
 whole.

\subsubsection{Initial stage of the evolution}

For
  $\tau \ll \mbox{min} \left(\tau_{\rm init}, \sqrt{\xi_0 (0)/w_{\tau}}\right)$, see solution
  (\ref{zero}),
  Eq. (\ref{linas}) is further simplified as
 \be
\frac{1}{2}\beta\frac{d^2\xi_{0l}}{d \tau^2}+\frac{d\xi_{0l}}{d
\tau}+\frac{(l(l+1)-2)\xi_{0l}}{ \xi_0^2 (l=0,\tau =0)}=0.
 \ee
Solution of this equation is as follows: $\xi_{0l}=C_1
e^{\lambda_1 \tau}+C_2 e^{\lambda_2 \tau}$. Using initial
condition $\frac{d\xi_{0l}}{d \tau}|_{\tau =0}=0$ one obtains $C_1
=-\frac{\lambda_2}{\lambda_1 -\lambda_2}\xi_{0l}(0)$, $C_2
=\frac{\lambda_1}{\lambda_1 -\lambda_2}\xi_{0l}(0)$,
 \be
 \lambda_{1,2}=-\frac{1}{\beta}
 \pm
 \sqrt{\frac{1}{\beta^2}  -\frac{2(l(l+1)-2)}{ \beta\xi_0^2 (l=0,\tau
 =0)}}.
 \ee
Thus we find two damping solutions. For
 \be\label{inequ}
 1\ll\xi_0 (l=0,\tau =0)< \sqrt{2\beta [l(l+1)-2]}, \quad l>1,\quad
 \ee
besides damping there occur oscillations. Inequalities
(\ref{inequ}) are fulfilled for  $\beta \gsim  \xi_0^2 (l=0,\tau
=0)$ (limit of effectively very small viscosity).

\subsubsection{Limit of an effectively   small viscosity}

The time scale of oscillations is
 $\tau \sim \tau^{\rm
id/osc}_l =|\mbox{min}(\lambda_{1,2})|^{-1}\sim \sqrt{\beta} \xi_0
(l=0, \tau =0)\lsim \tau_{\rm init}$ provided inequality
(\ref{inequ}) is fulfilled. Condition $\tau^{\rm id/osc}_l\ll
\tau_{\rm init}$ is satisfied only
 for very large  $\beta \gg |\widetilde{\epsilon}|^{-2}$ in case $\xi_0 (l=0, \tau =0)\sim \xi_0^{cr}$
 and  for $\beta \gg 1$ for seeds of a small size $\xi_0  (l=0, \tau =0)\gsim 1$.
In dimensional units the time scale
 \be
t^{\rm id/osc}_l \simeq \frac{2R_0}{3 \sqrt{
[l(l+1)-2]}}\left(\frac{m}{T_{cr}|\delta{\cal{T}}|}\right)^{1/2}\lsim
t_{\rm init}
 \ee
  does not depend on
the value of the viscosity.

Damping occurs at the time scale
$\tau^{\rm id/damp}_l \sim \beta , $
that corresponds to
\be
t\sim t^{\rm id/damp}_l \sim t_{\rm init}
 \ee
in dimensional units. Thus the time scale for the reconstruction
of the initial density profile of the seed  (if  initially it
deviates only slightly from the profile  Eq. (\ref{sol})), is
$t_{\rm rec}\sim t^{\rm id/damp}_l \sim t_{\rm init}$ in this
case.

\subsubsection{Limit of an effectively large viscosity}
 For $\beta \ll 1$ \ \ \  (effectively\ \    large\ \   viscosity)
   \ \ \ the \ \   \ time  \ \  scale \ \  $\mbox{min} \left(\tau_{\rm init}\right.$,  \ \ \   $\left.\sqrt{\xi_0 (l=0, \tau =0)/w_{\tau}}\right)$
  is very short and for
$\tau\gg  \tau_{\rm init}$  one can drop the term
$\frac{d^2\xi_{0}(l=0)}{d \tau^2}$ in l.h.s. of Eq. (\ref{linas}).
Then the latter equation is simplified as, cf. \cite{PS}:
 \be
 \frac{(2-l(l+1))\xi_{0l}}{ \xi_0^2
(l=0)}-\frac{d\xi_{0l}}{d \tau}=0.
 \ee
Its solution is as follows:
 \be\xi_0^l
 (\tau)=\xi_0^{l}(0)\mbox{exp}\left\{\int_{0}^{\tau}
 d\tau^{'}{[2-l(l+1)]}{\xi_0^{-2}
 (l=0,\tau^{'})}\right\}.
 \ee
All modes with $l>1$ prove to be damped. Thus an initially
deformed seed acquires spherical shape  with  time. Typical time
scale is
 \be\label{disoll}
\tau_{l}^{\eta}=\frac{\xi_0^2 (l=0,\tau =0)}{[l(l+1)-2]}, \quad
t_{l}^{\eta}=\frac{\rho_{cr}R_0^2}{\beta (\widetilde{d}\eta_{\rm
r} +\zeta_{\rm r})[l(l+1)-2]},\quad l>1.
 \ee
We find that $t_{l}^{\eta}\sim t_{\rm dis}$ for not too high $l$
and for $\xi_0 (l=0,\tau =0) <\xi_0^{cr}$, see Eq. (\ref{disol}).
The larger $l$, the more rapid is the process. For $\xi_0
(l=0,\tau =0) >\xi_0^{cr}$, $\xi_0 (l=0,\tau =0) -\xi_0^{cr}\sim
\xi_0^{cr}$ we obtain $t_l^{\eta}\sim t_{\rho}$.

\subsection{Dynamics of  density  fluctuations in spinodal
region}\label{spinodal}

Assume that  the system is driven to the spinodal region (region
between the minimum and the maximum of $\delta P
(\delta\rho)|_{T}$ in Fig. \ref{pres} (right). In the spinodal
region even fluctuations of an infinitesimal amplitude and size
may grow. To demonstrate this assume that the density is such that
the free energy $\delta F_L$ is close to  its maximum ($\delta F_L
\simeq 0$). Then we may linearize Eq. (\ref{dimens}) dropping
$\psi^3$ term. Setting
 \be\label{growing}
\psi=-\frac{\widetilde{\epsilon}}{2}+\mbox{Re}\{\psi_0
e^{\gamma_{\psi} \tau +i\vec{k}\vec{\xi}}\},
 \ee
where  $\psi_0$ is an arbitrary but small real constant, we find
two solutions of linearized Eq. (\ref{dimens}),
 \be\label{geng}
  \gamma_{\psi}(k)=({2\beta})^{-1}(-k^2
  \pm \sqrt{{k^4}(1-4\beta)
+8{\beta}{k^2}}\,).
 \ee
 Growing modes correspond to the choice of "$+$"-sign and $k^2 <2$. Most rapidly growing mode
 is
described by   the maximum value  of $\gamma_{\psi}(k)$ and
$k=k_{m}$:
 \be\label{geng1}
 k_m^2 =\frac{2\sqrt{\beta}}{2\sqrt{\beta}+1}, \quad \gamma_{\psi} (k_m)=\frac{2}{2\sqrt{\beta}+1}.
  \ee
 Thus the time scale for the growth of seeds in aerosol (spinodal
region) is
 \be\label{aerozole}
 t_{\rm aer}= t_0/\gamma_{\psi}(k_m).
  \ee
Increase of the wave amplitude is stopped, when the matter is
exhausted in the regions, where $\psi_0 e^{\gamma_{\psi} \tau}
\mbox{cos}(\vec{k} \vec{\xi}) <0$. It occurs for $\psi_0
e^{\gamma_{\psi} \tau}\sim 1$. After that,  seeds (regions $\psi_0
e^{\gamma_{\psi} \tau} \mbox{cos}(\vec{k} \vec{\xi}) >0$ with
typical size $l/\gamma_{\psi}$), for $\psi_0 e^{\gamma_{\psi}
\tau}\sim 1$, begin  to expand and flow together, until the stable
phase is cooked. With logarithmic accuracy the time scale of the
formation of the aerosol-like state is estimated as $t_{\rm aer}$,
following Eq. (\ref{aerozole}). The well known phenomenon
illustrating exponential growth of fluctuations in the spinodal
region is the boiling of the champaign occurring, when one opens
the bottle.

\subsubsection{Limit of an effectively large viscosity}

For  effectively large viscosity ($\beta \ll 1$) we get two
solutions:
 \be\label{large-gmet}
\gamma_{\psi}^{(1)}(k)\simeq 2\left[1-{k^2}/{2}-\beta k^2
\left({2}/{k^2}-1\right)^2\right],
\quad\gamma_{\psi}^{(2)}(k)\simeq -{k^2}/{\beta},
 \ee
for $k^2 \gg 4\beta$. For $k<\sqrt{2}$, $\gamma_{\psi}^{(1)}(k)$
describes growing mode. The mode $\gamma_{\psi}^{(2)}(k)$ is
damped. The most rapidly increasing mode corresponds to the
momentum $k_m =\sqrt{2}\beta^{1/4}$ (see (\ref{geng1})), and
$\gamma_{\psi}^{(1)}(k_m )\simeq 2$. The time scale characterizing
 growth of this mode is
 \be\label{aeroeta}
 t_{\rm aer}^{\eta}\sim t_0 /2 = {(\widetilde{d}\eta_{\rm r}
+\zeta_{\rm r})}[{\lambda v^2\rho_{\rm r}
}]^{-1}={4(\widetilde{d}\eta_{\rm r} +\zeta_{\rm r}
)}[{9n_{cr}T_{cr}|\delta {\cal{T}}|}]^{-1}.
 \ee
Second equality in (\ref{aeroeta}) is valid for the mVW EoS. The
larger the viscosity and the smaller $-\delta {\cal{T}}$, the
slower is the time evolution. Since $k_m =\sqrt{2}\beta^{1/4}$,
the size scale  of seeds is
 \be
 R_{\rm aer}^{\eta}\simeq l/(\sqrt{2}\beta^{1/4}).
 \ee
  Thus {\em{the size of
seeds in aerosol increases with increase of the viscosity.}}
For  $k^2
>2$ both  modes are damped.

\subsubsection{Limit of an effectively small viscosity}
 In case of  effectively small viscosity ($\beta \gg 1$) we get
 \be\label{gamsm}
\gamma_{\psi}(k)\simeq \pm k\sqrt{{2}/{\beta}} \sqrt{1-
{k^2}/{2}},
 \ee
 and the maximum value  ($\gamma_{\psi}^{max}$)
corresponding to $k_m =1$ is determined by
 \be
\gamma_{\psi}^{max}(1)={1}/{\sqrt{ \beta}}.
 \ee
 As we see, the time scale characterizing the growing mode,
  \be\label{aeroid}
  t_{\rm aer}^{\rm id}\sim t_0
/\gamma_{\psi} ={2c^{1/2}}/({\lambda v^2})\gg t_0 ,
 \ee
  does not depend
on the viscosity in this limit.  For the mVW EoS one has $v^2
\propto |\delta T|$ and $t_{\rm aer}^{\rm id} \propto 1/ |\delta
T|$.  Evolution then is slow near the critical point. Conditions
for the validity of the non-relativistic approximation, which we
have used, are fulfilled. For  $t >t_{\rm aer}^{\rm id}$, growth
of modes becomes exponentially fast, see (\ref{growing}). Since
$k_m =1$, the typical size of  seeds in aerosol is
\be
R_{\rm aer}^{\rm id}\simeq l, \quad R_{\rm aer}^{\rm id}<R_{\rm
aer}^{\eta}.
 \ee
Modes with $k^2
>2$ correspond to oscillations. Such fluctuations do not grow to
the stable phase.

\subsection{Evolution of the density fluctuations in the system close to
equilibrium}

Assuming that $\psi$ is
close to
 its new equilibrium value, $\psi_{eq}\simeq \pm 1 +\widetilde{\epsilon}/4$, we put $\psi
 =\psi_{eq}
 +\delta\psi$ in Eq. (\ref{dimens}) and linearize the latter equation in
 $\delta\psi$:
 \be\label{newst}
  -\beta\frac{\partial^2 \delta\psi}{\partial{\tau}^2} =\Delta_{\xi} \left(
\Delta_{\xi} \delta\psi -4 \delta \psi -\frac{\partial
\delta\psi}{\partial{\tau}} \right).
 \ee
Setting
 \be \delta\psi= \mbox{Re}\{\psi_0  e^{\gamma_{\phi} \tau
+i\vec{k}\vec{\xi}}\},
 \ee
where $\psi_0$ is an arbitrary but small real constant, we find
 \be\label{gamst} \gamma_{\psi}(k)=({2\beta})^{-1}\left(-{k^2}\pm \sqrt{k^4 (1-4\beta)
 -{16\beta k^2}}\right).
  \ee

Dependence on the quantity $\widetilde{\epsilon}$ disappears from
(\ref{newst}) and (\ref{gamst}).

\subsubsection{Limit of an effectively large viscosity}

In case of effectively  large viscosity ($\beta \ll 1$) and for
$k^2
> 16\beta$ there are only  damped solutions. For $k^2 \gg 8\beta$:
 \be\label{large-gst}
\gamma_{\psi}^{(1)}(k)\simeq -(k^2 +4),\quad
\gamma_{\psi}^{(2)}(k)\simeq -{k^2}/{\beta} .
 \ee
Fluctuations with large $k$ rapidly dissolve with  time.
{\em{Existence of long living short-wave excitations  is unlikely
in the viscous medium.}} However there remain long-wave damped
oscillations, for  $k^2 < 16\beta$.

\subsubsection{Limit of an  effectively small viscosity}
 In  case of effectively small viscosity ($\beta \gg 1$) we get
 \be\label{smeq}
\gamma_{\psi}(k)= -({2\beta})^{-1}\left({k^2}\pm {4i \sqrt{
\beta}\,k} \sqrt{1+{k^2}/{4}}\right),
 \ee
 that corresponds to oscillating and slowly damped modes
near the final equilibrium state,  rather than to unstable modes.
Since $t_0 \sqrt{\beta}$ does not depend on the viscosity and $t_0
\beta \rightarrow \infty$ for $\eta , \zeta\rightarrow 0$, in case
of the ideal fluid rapid oscillations  continue till the energy is
transported to the surface of the system (the process is governed
by the heat transport) or till the energy is radiated away in the
course of direct reactions. Thus {\em{in case of effectively small
viscosity  the stable phase is covered by fine ripples  during
some rather long period of time.}}

\subsection{The Reynolds number and turbulence}

The transition between laminar and turbulent flows occurs, when
the Reynolds number
 \be\label{Re}
 {\rm Re} =u \Lambda \rho_{fl}/\eta
  \ee
exceeds  the critical Reynolds number ${\rm Re}_{cr}$, which is
very large ($\gsim 1000$). Here $u$ is the mean fluid velocity,
$\Lambda$ is the characteristic diameter of the body embedded in
the fluid and $\rho_{fl}$ is the density of the fluid. In
particular problem of the growth of the droplet/bubble of the
stable phase developing in the metastable surrounding, $u$ means
the velocity of the growth of the seed, $\rho_{fl}$ is the density
of the metastable phase ($\simeq \rho_{cr}$ in case of small
overcriticality) and $\Lambda =2R$. Supposing $R=\alpha R_{cr}$,
$\alpha =const$, for the typical radius of the seed,  taking $u$
from Eq. (\ref{uasi}) and replacing these values in (\ref{Re}) we
find
 \be\label{Re1}
  {\rm Re} =\frac{16}{3}\alpha\beta .
  \ee
We see that  for relevant values of $\alpha \lsim 10$  factor $
{\rm Re}$ may reach  critical value only for unrealistically large
values of the parameter $\beta$ (for a tiny effective viscosity).
For  values of $\beta$ and $\delta{\cal{T}}$ with
which we are concerned, one has ${\rm Re} \ll {\rm Re}_{cr}$.

\section{Numerical integration of the system of hydrodynamical
equations in one and two spatial dimensions} \label{Numerical}
\subsection{Setup}

 To diminish computing time we
consider solutions in $d=2$ space.
 As is seen from
analytical solutions presented above, qualitative description of
the dynamics of the first-order phase transition remains similar
for $d=3$ and $d=2$. E.g., Eq. (\ref{Fsol}) continues to hold  for
$d=2$, provided one replaces $\delta F_L^{(d=3)}[d_{\rm
sol}=2;1]/(2L)$ to $\delta F_L^{(d=2)}[d_{\rm sol}=2;1]$. Further,
we assume validity of the isothermal approximation neglecting the
heat transport effects. In this case  evolution is governed by
Eqs. (\ref{Navier}) and (\ref{cont}).  This means that in the
present paper we will not simulate the late stage of the evolution
of fluctuations, when a fog-like state is formed.

 To solve the
problem we need to know EoS and transport coefficients. Since  we
will focus on demonstration of a qualitative behavior of the
system  undergoing the first-order phase transition, to avoid
extra complications we will consider simplest case assuming that
the viscosity does not depend on the density and the temperature.
As we have mentioned in Introduction, for nuclear systems   shear
and bulk viscosities, as well as the heat conductivity, are poorly
known and most probably the bulk viscosity is smaller than the
shear one.  Note that in analytical expressions, which we have
derived, viscosities enter  in  combination $
(\widetilde{d}\eta_{\rm r} +\zeta_{\rm r})$. Thereby, and in order
to diminish uncertainties we further put zero the bulk viscosity
and vary the shear viscosity in broad limits.

 The Landau free
energy and the pressure of the uniform matter  behave as shown in
Fig. \ref{pres}. Note that in general case the Helmloltz  free
energy density of the uniform matter does not produce two minima.
Rather it fulfills the double-tangent construction (at values of
densities corresponding to the Maxwell construction on the
isotherm $P(1/\rho)$). E.g., it is so for the original VW EoS and
for the EoS of the relativistic mean field Walecka model. We use
the mVW EoS, $P=f(T)P_{VW}$, see Appendix A. Extra function $f(T)$
is introduced  for generality since temperature dependence of the
purely VW EoS is too simple to describe behavior of nuclear
matter. At constant $T$ for all
$f(T)$ hydrodynamical descriptions are self-similar. E.g., the description 
of the dynamics in the framework of 
the original VW EoS can be obtained with the help of  the
scaling $t\rightarrow t \sqrt{f(T)}$ and $(\widetilde{d}\eta_{\rm
r} +\zeta_{\rm r})\rightarrow (\widetilde{d}\eta_{\rm r}
+\zeta_{\rm r})/\sqrt{f(T)}$, as it  follows from Eq. (\ref{v-t}).
In our numerical calculations we specified $f(T)$ in order
the Helmholtz free energy had two minima, as well as the Landau
free energy. Such a modification is quite not necessary since
hydrodynamical equations enters only gradient of the pressure,
which in both cases has the same  form. 
Moreover we use $f(T)\rightarrow 1$ for
$T\rightarrow T_{cr}$. In this case all the results valid in the
vicinity of the critical point do not depend on $f$ and are the
same as for the original VW EoS. If we wanted to recover 
$f$-dependence for $f(T\to T_{cr})=f_0\ne1$ in the vicinity 
of the critical point, we could do it with help of the 
replacement $t_0 \to t_0/f_0$, $\beta\to\beta f_0$ in
Eq.~(\ref{dimens}).

In the present paper we consider dynamics of the phase transition
at fixed pressure  at the system boundary. In order to describe
possible first-order phase transitions in heavy-ion collisions one
still should take into account expansion of the fireball, see
\cite{Toneev}.  Moreover, one needs to use a more realistic EoS,
e.g. that constructed in \cite{KTV}. These problems will be
considered in forthcoming publications.

The surface contribution to the pressure (the so called Laplace
pressure) is taken into account in the low gradient approximation:
\begin{equation}
P = P_{\rm mVW}  - c \nabla^2 \rho,
\end{equation}
with $\rho=m n$. The coefficient "$c$" can be expressed in terms
of the surface tension, see (\ref{c}). We examine both one- and
two-space dimensional solutions. In general case  no assumptions
about cylindrical symmetry are used. To illustrate the dynamics of
the overcritical and undercritical seeds we consider a fluctuation
in infinite matter in both spatial directions $x$ and $y$. Initial
conditions  correspond to a stable spot placed in the homogeneous
metastable medium.

\subsection{Evolution of disks ($d_{\rm sol}=2$) in $d=2$ spatial dimensions}

First consider dynamics of an initially static axial-symmetric
seed with the conserved number density  given by
\begin{equation}\label{densprof}
\rho (x,y;t=0) = \rho_{out} + (\rho_{in}-\rho_{out})
\Theta({R_0}-r), \quad r=\sqrt{x^2+y^2}.
\end{equation}
Such a seed would correspond to  the rod in case $d=3$. Densities
$\rho_{in}$ and $\rho_{out}$ are taken to be those in homogeneous
phases.
We  call these configurations liquid or gas disks  in dependence,
if $\rho_{in}> \rho_{out}$ or vise versa.

\begin{figure}
\centerline{%
{\includegraphics[height=6.0truecm] {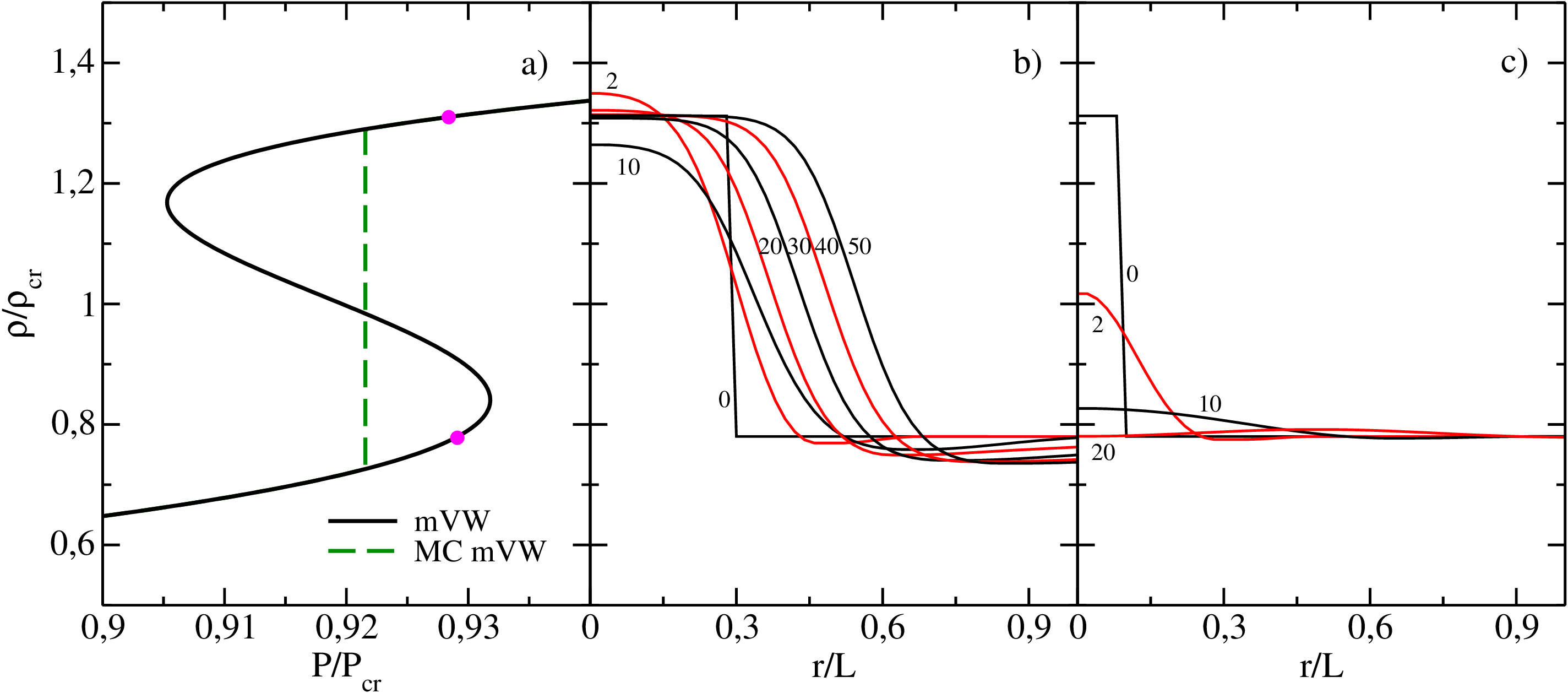}   } } \vspace{11mm}
\centerline{{\includegraphics[height=6.0truecm] {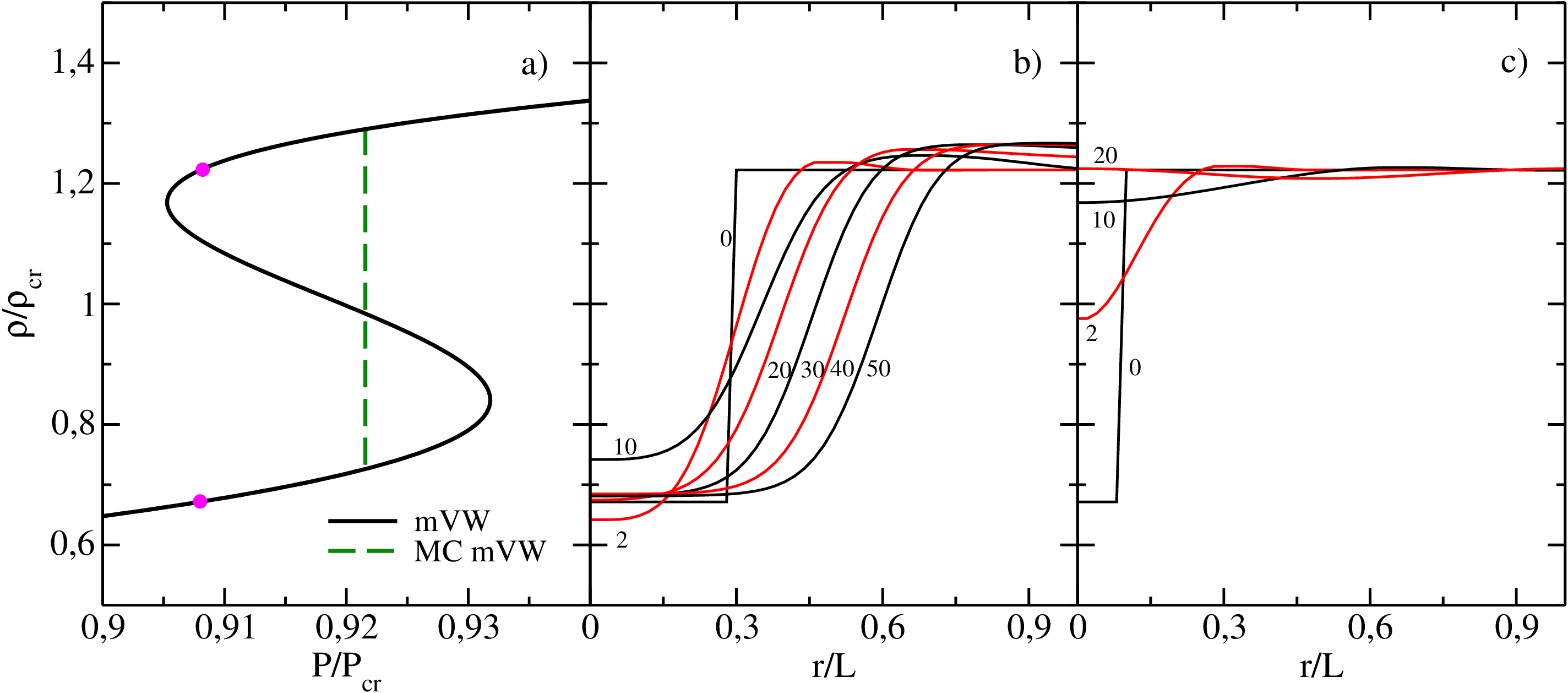} } }
\caption{ Isotherm for the pressure as function of the density for
mVW EoS, see Appendix A, with initial and final configurations
shown by dots  (left column).
 Dash vertical  line corresponds to the
Maxwell construction (MC) on the curve $P(1/\rho)$.
In upper panel initial state relates to stable liquid phase disk
in metastable super-cooled gas and on lower panel it corresponds
to the stable gas phase disk in metastable super-heated liquid.
Middle column demonstrates time evolution of the density profiles
for the
 overcritical liquid disk (upper panel) and gas disk (lower panel).
Numbers near curves (in $L$) are time snapshots; $r=\sqrt{x^2
+y^2}$, $|\delta {\cal{T}}|=0.02$, $L= 30$ fm. Right column, the
same for initially undercritical liquid or gas disks.}
\label{EoS_dynamics}
\end{figure}

\begin{figure}
\centerline{%
{\includegraphics[height=6.0truecm] {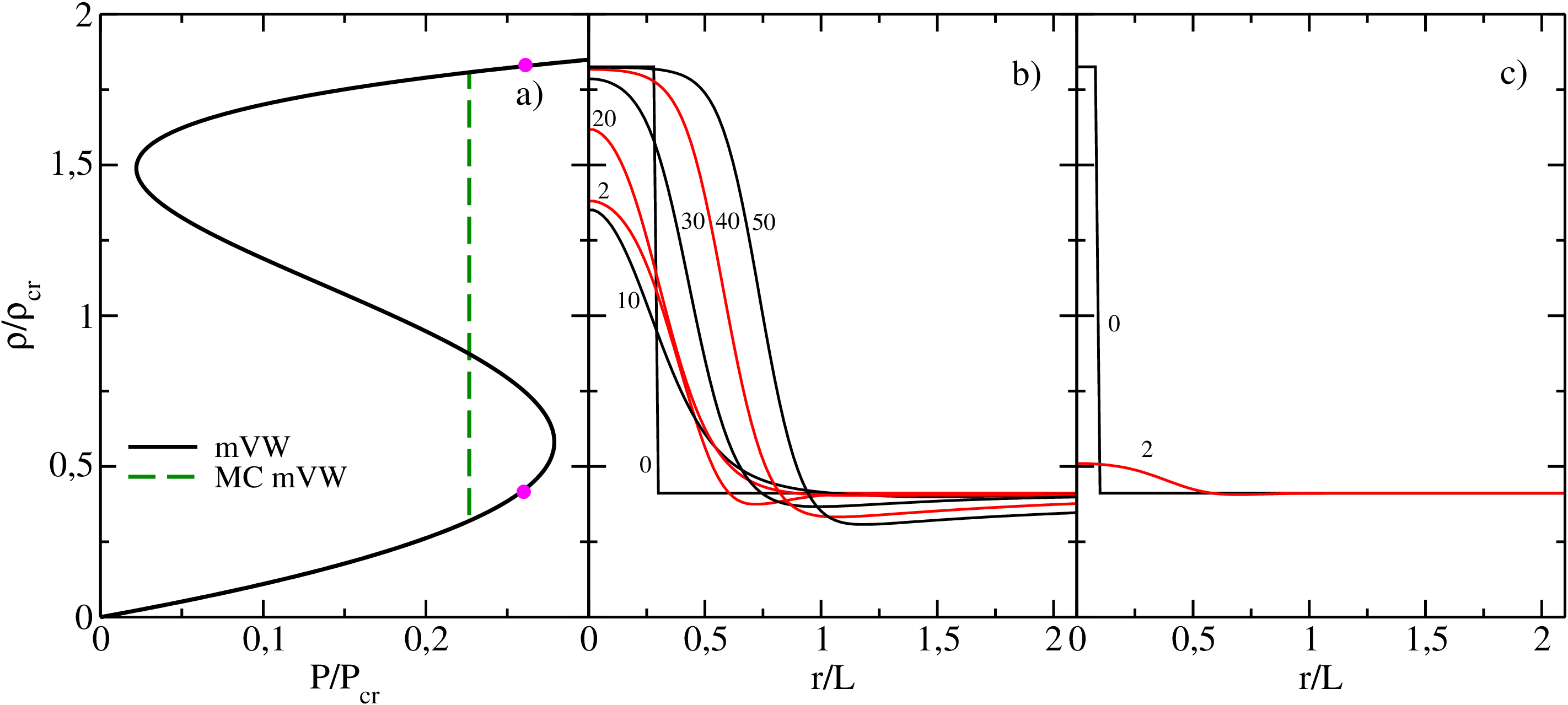}   } } \vspace{12mm}
\centerline{{\includegraphics[height=6.0truecm] {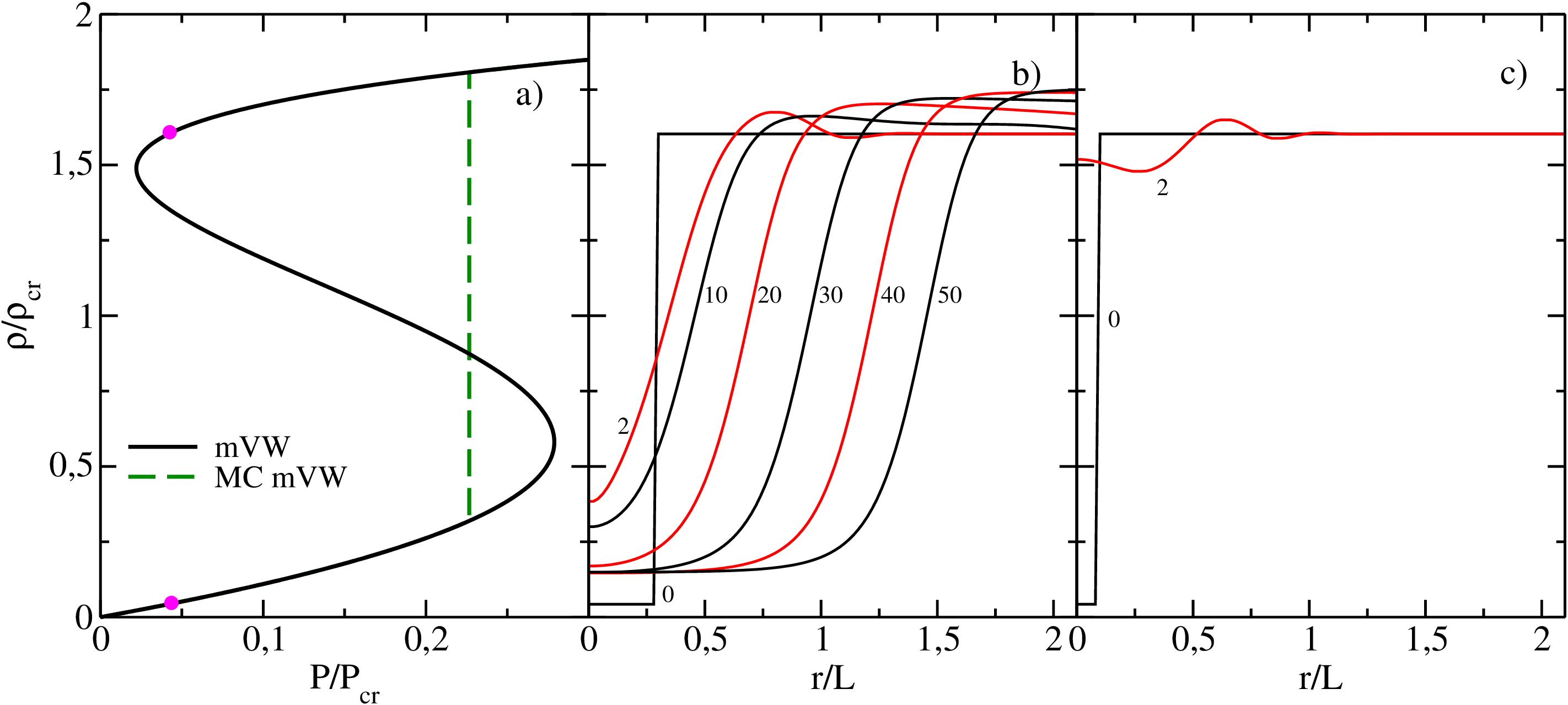} }
} \caption{ The same as in Fig. \ref{EoS_dynamics} but for
$|\delta {\cal{T}}|=0.15$, $L=5$~fm. } \label{EoS_dynamics1}
\end{figure}

 For $d=2$, expansion of the mVW
EoS near the critical point is valid only for $|\delta
{\cal{T}}|\ll 1/7$, see Appendix A, and analytical solution
(\ref{sol}) is applicable for $|\delta {\cal{T}}|\ll 1/16 \,(\xi_0
\gg 1)$. Therefore we first take $T$ very close to the critical
temperature,
 in order one could quantitatively compare results of
computing with analytical expressions.  Results are presented in
Fig. \ref{EoS_dynamics} for $T/T_{cr}=0.98$.    Parameters of EoS
are chosen, as for the NGL phase transition: $T_{cr}=18.6$~MeV,
$n_{cr}/n_{sat}=0.42$, $n_{sat}=0.16$ fm$^{3}$ is the nuclear
saturation density. The configuration is computed for  values of
kinetic parameters chosen as $\eta \simeq 3.2$~MeV$/$fm$^{2}$ and
$\beta \simeq 12.6$ (effectively small viscosity). For a large
viscosity general behavior of  solutions remains the same but
velocity of the seed evolution proves to be significantly smaller.
Some peculiarities of the $\beta$ dependence will be illustrated
in next figures. In the upper panel of Fig. \ref{EoS_dynamics} we
demonstrate the time evolution of initially liquid disk  and in
the  lower panel, of a gas disk. Left column demonstrates initial
and final configurations on the curve $P(\rho)$. In the middle
column we show dynamics of the initially overcritical seed $R_0
=0.3~L>R_{cr}\simeq 0.2~L$ and in the right column, of the
undercritical seed $R_0 =0.1~L$, in units of a relevant length
scale $L=30$~fm. The time snapshots are shown by numbers near
curves in units $L$ (i.e. $2$ means $\Delta t = 2L$, etc.).  We
see from the middle column that in case $R_0
> R_{cr}$ (in this example $R_0 \simeq 1.5~R_{cr}$) disks slowly grow with  time.
The initially selected distribution (\ref{densprof}) acquires
$\mbox{tanh}$-like shape (see (\ref{sol})) for $t\gsim 20L\simeq
600$ fm.  This is very large time (although in dimensionless units
it corresponds to a rather short time scale  $\tau \gsim 6$, of
the order of the duration of the initial stage, see  the right
panel of Fig. \ref{dynamics1}).
 As we see from the right column of Fig.  \ref{EoS_dynamics},
 seeds of an initially small size $R_0 <R_{cr}$
(in this example $R_0 \simeq 0.5~R_{cr}$) dissolve  for $t\gsim
20~L\simeq 600$~fm. This value agrees with the total time of the
shrinking process described  by the dash curve in the left panel
of Fig. \ref{dynamics}. The typical time scale characterizing the
dynamics is $t\sim t_{\rm init}\gg t_{\rm dis}$.

We have checked  that the time evolution occurs  in a line with
above analytical consideration. The only difference is that in
analytical treatment of the problem with initial distribution of
the $\mbox{tanh}$- form, we do not get values $\rho <\rho_{out}$
for droplets and $\rho >\rho_{out}$ for bubbles.  As follows from
our numerical solution, due to infall of the surrounding matter to
 the disk surface during the shape reconstruction, the density
decreases in the liquid disk neighborhood  below  the value of the
density in the homogeneous metastable matter and it increases in
the gas disk surrounding above the value of the density in the
homogeneous metastable matter (see the middle column of Fig.
\ref{EoS_dynamics}).

Note that  it is unlikely to find the system  in heavy-ion
collisions in so narrow vicinity of the critical point as we
considered, $|\delta{\cal{T}}|=0.02$, since the typical time of
the fireball expansion is much shorter than typical time of the
evolution of fluctuations that we found. Therefore in Fig.
\ref{EoS_dynamics1} we also demonstrate the time evolution of
disks for $T/T_{cr} =0.85$. For easier comparison with Fig.
\ref{EoS_dynamics} initial and final configurations are selected
at approximately the same deviations $(P-P_{\rm min})/(P_{\rm
max}-P_{\rm min})$. In this Figure we take $L=5$~fm, as the length
unit. The system described by the VW EoS is already rather far
from the critical point at $T/T_{cr} =0.85$. Indeed, for
$T/T_{cr}>T_{comp}/T_{cr}=\frac{27}{32}\simeq 0.844$ there appears
a region of negative pressures, that may cause some extra
peculiarities in the processes under consideration. Different
physical situations occurring for $T<T_{comp}$ were discussed in
\cite{SVB}. In this work we will avoid further discussion   of the
regime $T<T_{comp}$. Although for  $T/T_{cr} =0.85$
 the dynamics looks qualitatively the same, as in case demonstrated by
Fig. \ref{EoS_dynamics}, the critical radius proves to be
significantly smaller ($R_{cr}\simeq 1$~fm instead of 15~fm in
previous case) and the time scale characterizing the process is
reduced.
 Therefore for $T/T_{cr} =0.85$ seeds
evolve much faster compared to the case $T/T_{cr} =0.98$.
For overcritical discs the initially selected distribution
(\ref{densprof}) acquires the $\mbox{tanh}$-like shape (see
(\ref{sol})) for $t\gsim (20\div 40)L = 100\div 200$~fm. The
typical time scale is $t\sim t_{\rm init}$, which in case
$|\delta{\cal{T}}|=0.15$ is much shorter than in case
$|\delta{\cal{T}}|=0.02$ considered above. Initial disks of a
small size ($R_0 \simeq 0.5~R_{cr}$) almost disappear for $t\gsim
10L = 50$~fm. In this case $t_{\rm init}\sim t_{\rm dis}$. All
these values of   time scales are larger or of the order of the
time scale characterizing the fireball expansion in low energy
heavy-ion collisions.

\begin{figure}
\centerline{%
\includegraphics[height=6.0truecm] {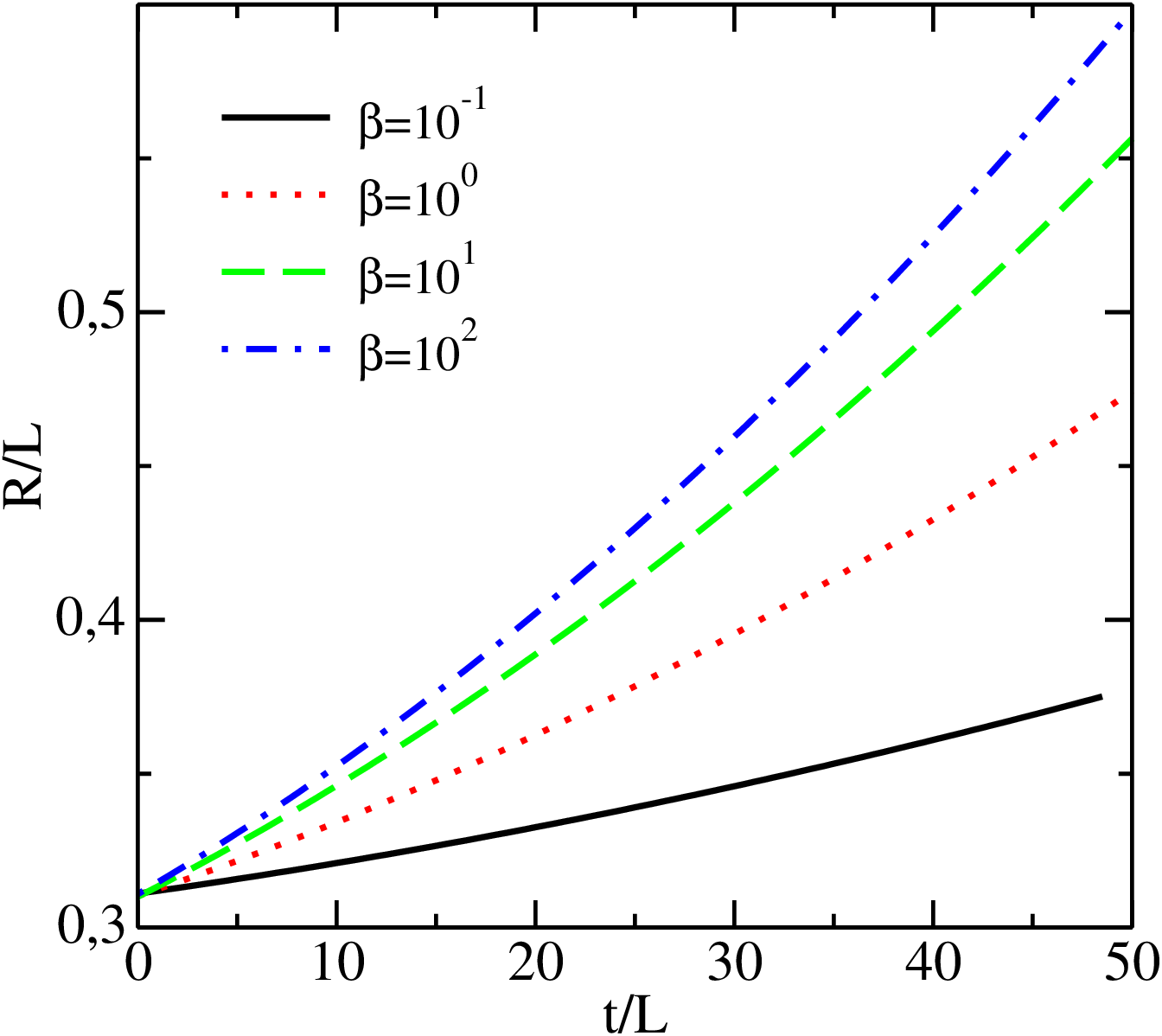}
\hspace{5mm}
\includegraphics[height=6.0truecm] {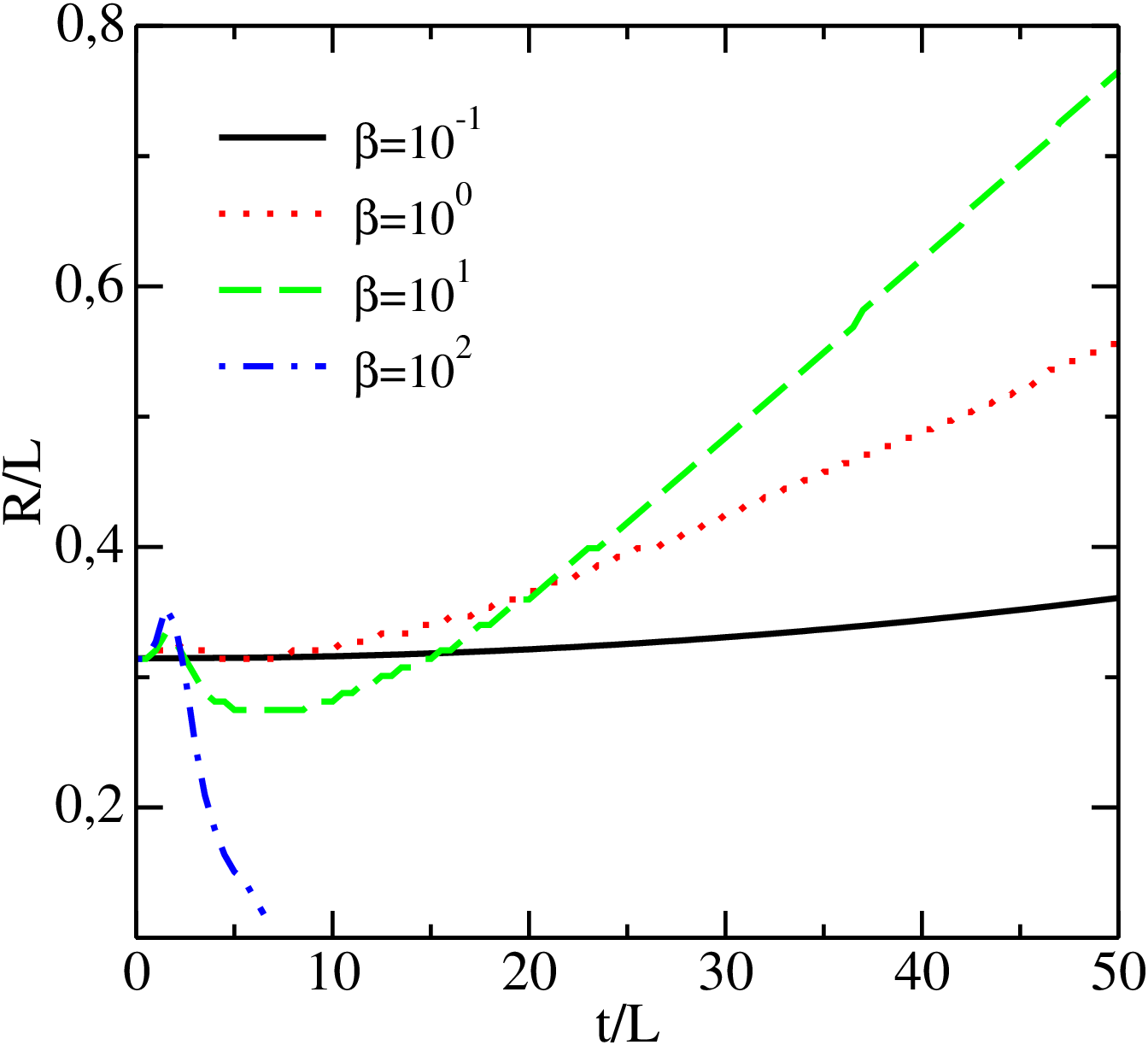}
} \caption{ Solution $R(t)$ for the liquid disk boundary,
$\rho(R(t);t)=\rho_{cr}$, for several values of the viscosity.
In left panel all  parameters except viscosity are taken the same
as in Fig. \ref{EoS_dynamics} ($\delta{\cal{T}}=0.02, L=30$) and
in right panel, as in Fig. \ref{EoS_dynamics1}
($\delta{\cal{T}}=0.15, L=5$).
 }
\label{interface}
\end{figure}

We further varied parameters $T_{cr}$, $n_{cr}$, $\eta$, $\beta$
in broad limits (in the range relevant for the NGL phase
transition) and checked that it does not change the qualitative
picture presented in Figs. \ref{EoS_dynamics},
\ref{EoS_dynamics1}. We demonstrate it in Fig. \ref{interface}
showing the evolution of the disk surface with time for $R_0
>R_{cr}$. The disk surface is specified as the boundary, where
density achieves the critical value ($\rho =\rho_{cr}$). In the
left panel results are presented for $T/T_{cr}=0.98$. The velocity
of the growth of the seed (slop of the curve) increases with
increase of $\beta$. Large values of the dimensional time
presented in Fig. \ref{interface}, $t\sim 50 L$, correspond to the
value of the dimensionless time $\tau \sim 5\sqrt{\beta}$. The
latter value corresponds to the transition regime in the right
panel of Fig. \ref{dynamics1} (for all values of $\beta$ presented
in Fig. \ref{interface}). In this regime the velocity of the seed
surface follows linear law. The asymptotic regime is achieved at
much larger values of time. In the right panel of  Fig.
\ref{interface} results are presented for $T/T_{cr}=0.85$. For
$t\gg  t_{\rm rec}$ the behavior is similar to that  in the left
panel. However
  at  smaller values of time ($t\lsim  t_{\rm rec}$)
  for large values of $\beta$ (effectively small
viscosity) there arise peculiarities. These peculiarities are
associated with reconstruction of the initial density profile
(\ref{densprof}), occurring at $t\lsim  t_{\rm rec}$. During this
reconstruction period the typical radius of the seed may decrease.
 The dash curve  demonstrates the same behavior, as is seen from
the density profiles for time snapshots 2$L$, 10$L$ and 20$L$ in
the middle column of the upper panel of Fig. \ref{EoS_dynamics1}.
If  during the reconstruction process the size of the initially
overcritical seed becomes smaller than the critical size, it
causes subsequent shrinking of the seed. Namely this case  is
shown by the dash-dotted curve in the right panel of Fig.
\ref{interface}. To check this statement we increased initial size
of the seed at given $\beta$ and the behavior became similar to
that shown by the dash curve. The larger  $\beta$ and
$|\delta{\cal{T}}|$, the higher is the seed velocity, cf. Eq.
(\ref{uasi}). For fixed $\beta$, evolution of the seed shown in
the left panel ($T/T_{cr}=0.98$, $L=30$~fm) is slower than that
demonstrated in the right panel ($T/T_{cr}=0.85$, $L=5$~fm). Thus
peculiarities of the case of effectively low viscosity ($\beta \gg
1$), which are seen in the right panel of the figure, are due to a
larger inertia than in cases presented in the left panel.

\begin{figure}
\centerline{%
{\includegraphics[height=6.0truecm] {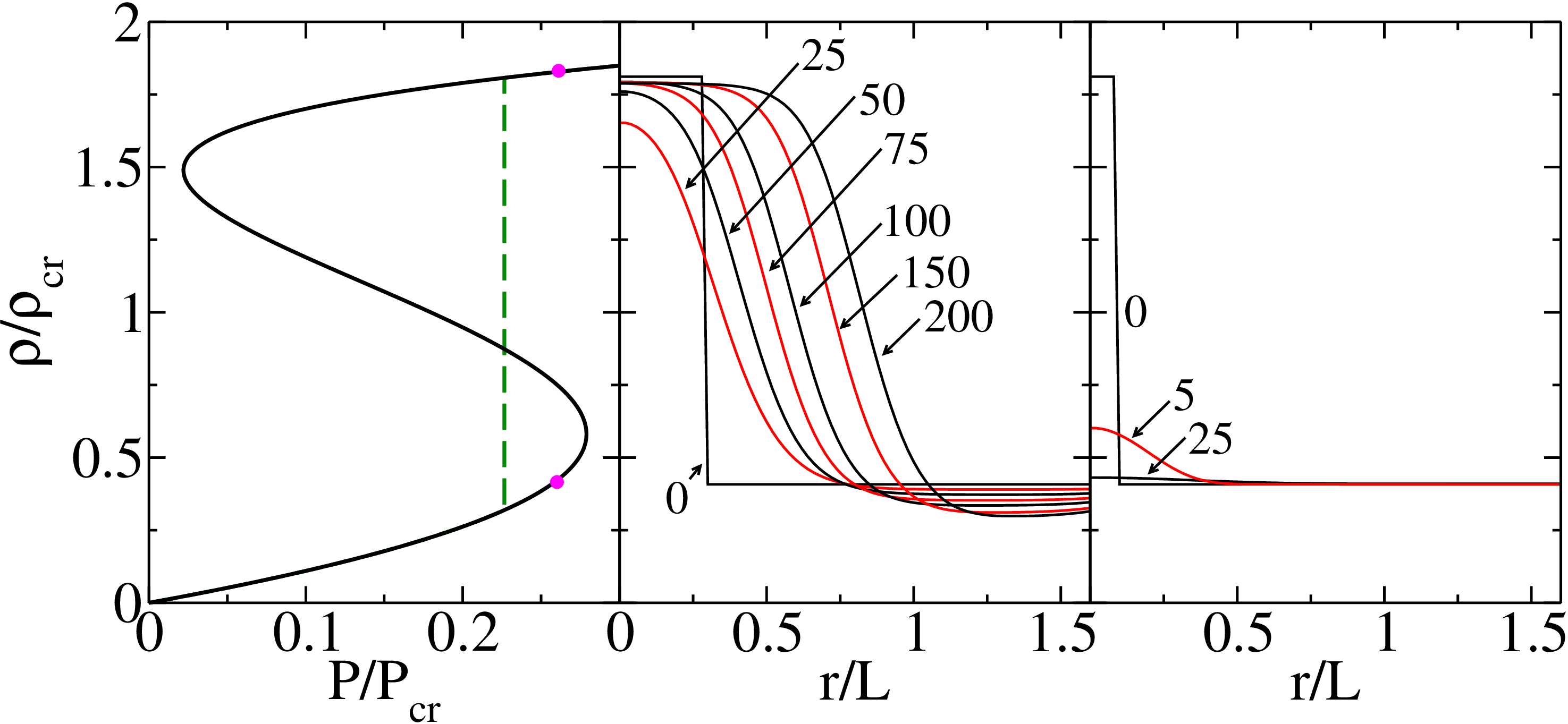}   } }
\vspace{9.0mm} \centerline{{\includegraphics[height=6.0truecm]
{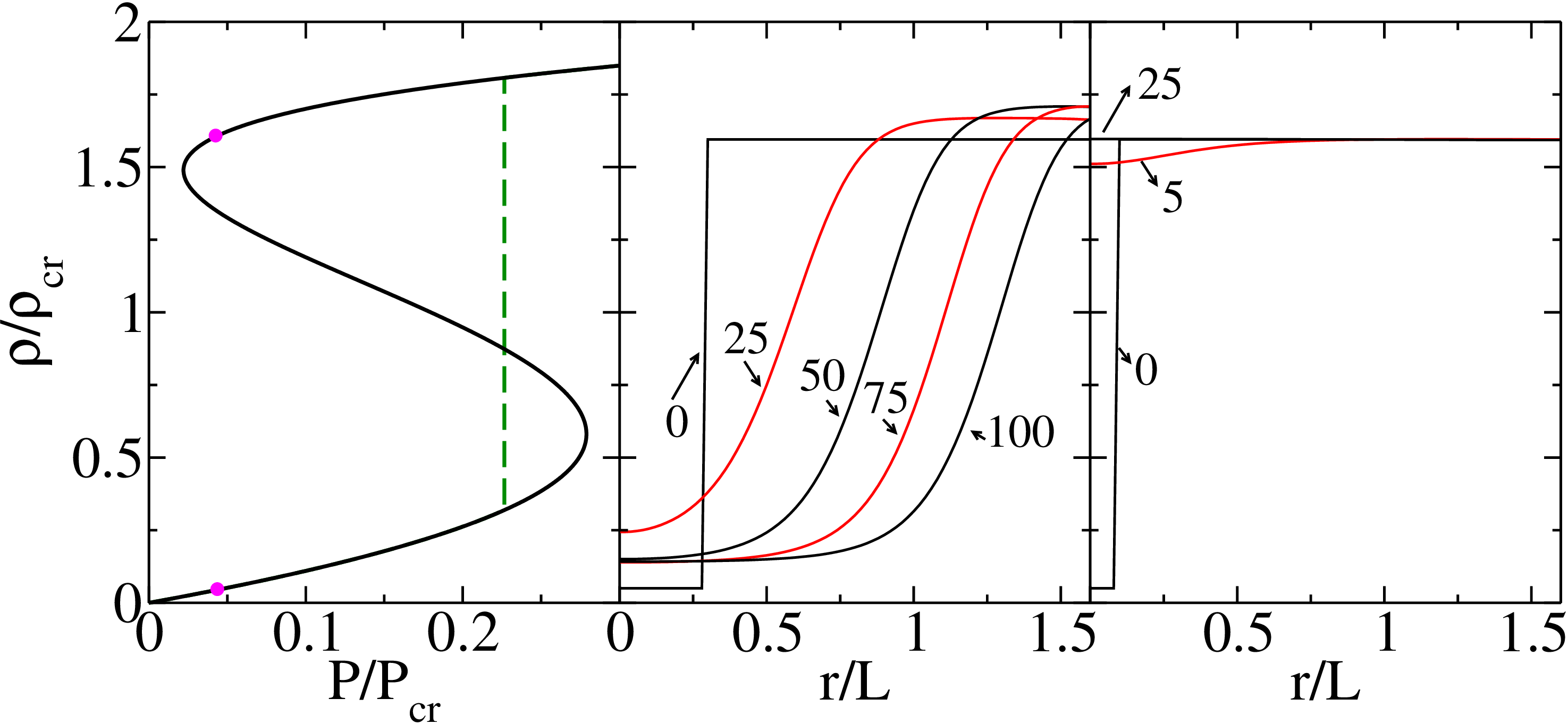} } } \caption{The same as in Fig.
\ref{EoS_dynamics1} but for  $T_{cr}=162$~MeV, $n/n_{sat}=1.3$,
$\eta \simeq 45$MeV$/\mbox{fm}^2$ and for $\beta =0.2$.
}\vspace{0.0mm} \label{EoS_dynamicsQ}
\end{figure}
\begin{figure}
\centerline{%
\rotatebox{0}{\includegraphics[height=5.0truecm] {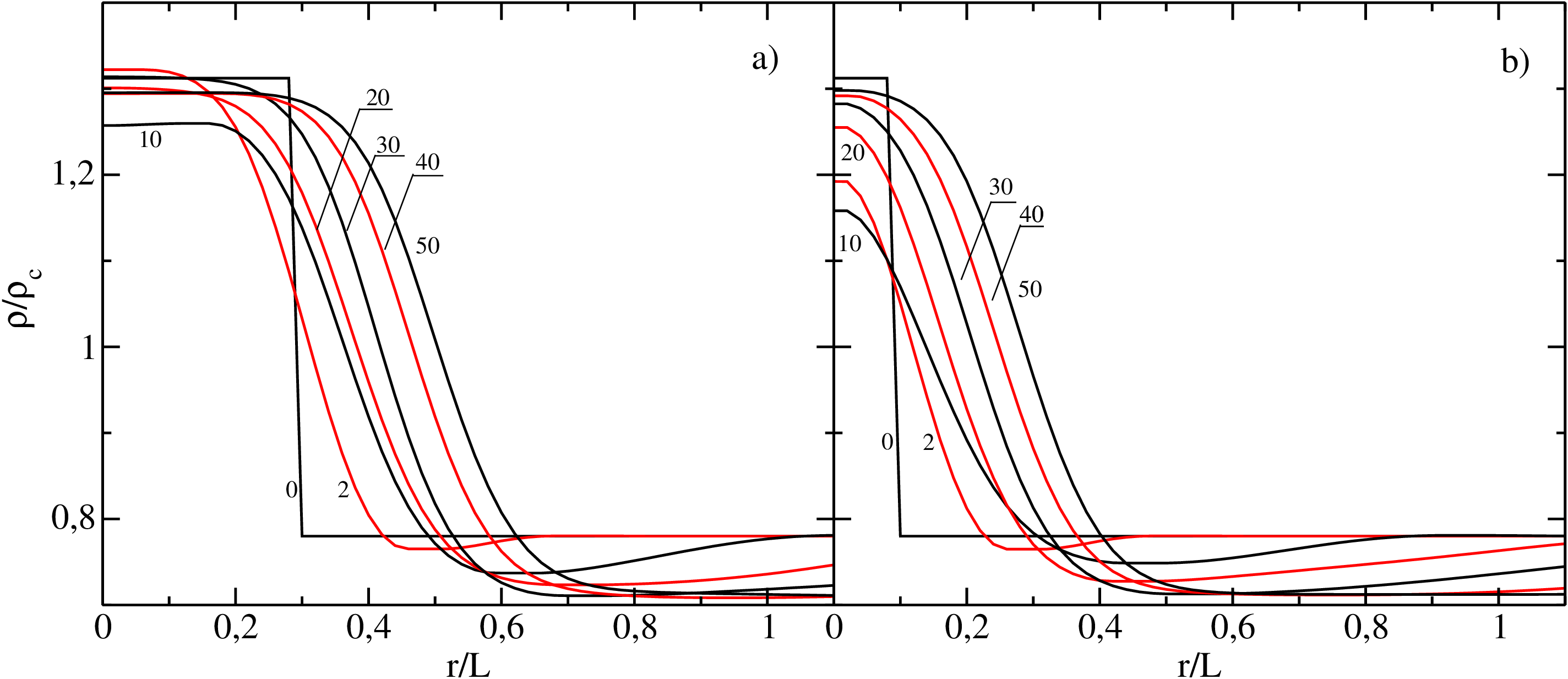}   }  }
\caption{Evolution of  bands ($d_{\rm sol}=1$) of initially large
(left) and  small (right) sizes, $r =|x|$. Parameters are taken
the same as in Fig.
\ref{EoS_dynamics1}.
}
\label{cracks}
\end{figure}

In order to show how much the seed dynamics is sensitive to the
choice of parameters of the EoS we take an another parameter
choice $T_{cr}=162$~MeV, $n/n_{sat}=1.3$, relevant for the
hadron-sQGP phase transition. Results are presented in Fig.
\ref{EoS_dynamicsQ} for configurations with approximately the same
$(P-P_{\rm min})/(P_{\rm max}-P_{\rm min})$ as shown in previous
figures. As in Fig. \ref{EoS_dynamics1} we take $T/T_{cr}=0.85$
and compute the configuration for $\eta \simeq
45$MeV$/\mbox{fm}^2$ and for $\beta =0.2$ (effectively large
viscosity). As we see, typical time scales and the shapes of
configurations  look similar to the case presented in Fig.
\ref{EoS_dynamics1}, in spite of the parameter sets are completely
different.

\subsection{Evolution of bands ($d_{\rm sol}=1$) in $d=2$ spatial dimensions}
 In Fig.
\ref{cracks} we show dynamics of liquid bands in metastable gas
phase. Values $|\delta {\cal{T}}|=0.02$, $L=30$~fm, are taken the
same as in Fig. \ref{EoS_dynamics1} and we again choose
$T_{cr}=18.6$~MeV, $n_{cr}/n_{sat}=0.42$, $\eta \simeq
3.2$~MeV$/$fm$^{2}$ and $\beta \simeq 12.6$ (effectively small
viscosity). These solutions are similar to  slabs in $d=3$.
 Left panel
shows time evolution of a  band of a large initial size ($R_0
=0.3L$), whereas right panel demonstrates evolution of a band
having initially rather small size ($R_0 =0.1L$). In difference
with disks (solutions with $d_{\rm sol}\neq 2$) in both cases
(for large and small initial sizes of bands) dynamics looks
similar: bands of the stable phase, being prepared in the
metastable phase, undergo growth to the new phase.
Nevertheless, we also see that during the shape reconstruction the
slab first begins to dissolve and then  grows. This peculiarity
appeared since initial form of the density distribution  that we
exploit in numerical calculations deviates from the form given by
analytical solution (\ref{slab}). Thus even for slabs {\em{there
might exist a small critical size, that depends on peculiarities
of the initial density profile. }} Slabs having sizes smaller than
this critical size could then completely dissolve.

\begin{figure}
\centerline{%
\rotatebox{0}{\includegraphics[height=5.0truecm] {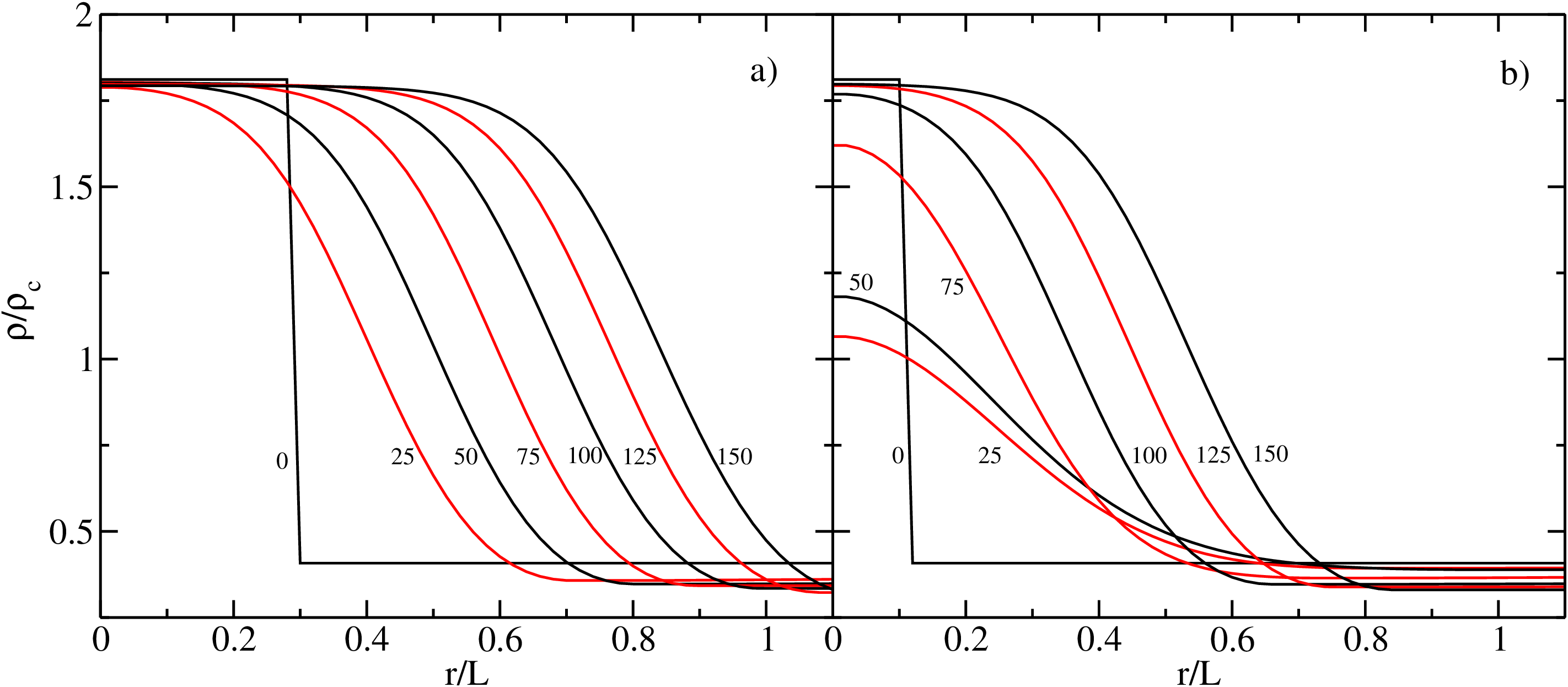}   }
} \caption{The same as in Fig. \ref{cracks} but for $|\delta
{\cal{T}}|=0.15$, $L=5$~fm,  $\eta \simeq 23$~MeV$/$fm$^{2}$ and
$\beta \simeq 0.042$.
}\label{cracks0.15}
\end{figure}

Fig.  \ref{cracks0.15} shows the same as Fig. \ref{cracks}, but
for $|\delta {\cal{T}}|=0.15$, $L=5$~fm,  $\eta \simeq
23$~MeV$/$fm$^{2}$ and $\beta \simeq 0.042$ (effectively large
viscosity). We see that the qualitative picture of the time
evolution remains the same.

 The
probability to prepare a band in a fluctuation is tiny. However,
 bands of the stable phase could be
formed near the system boundary, provided the latter is flat.

\begin{figure}
\centerline{%
\includegraphics[height=5.0truecm] {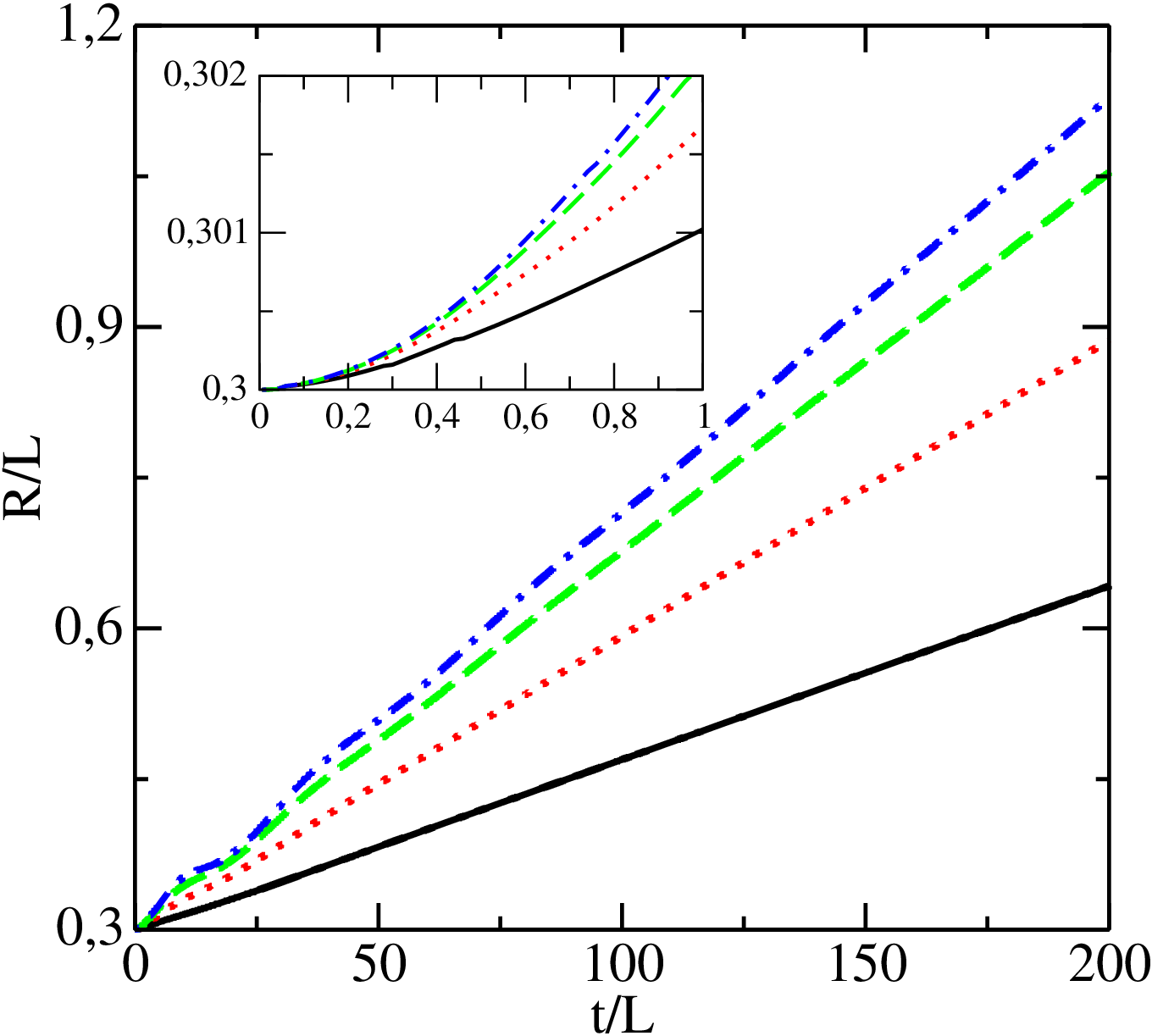}
\includegraphics[height=5.0truecm] {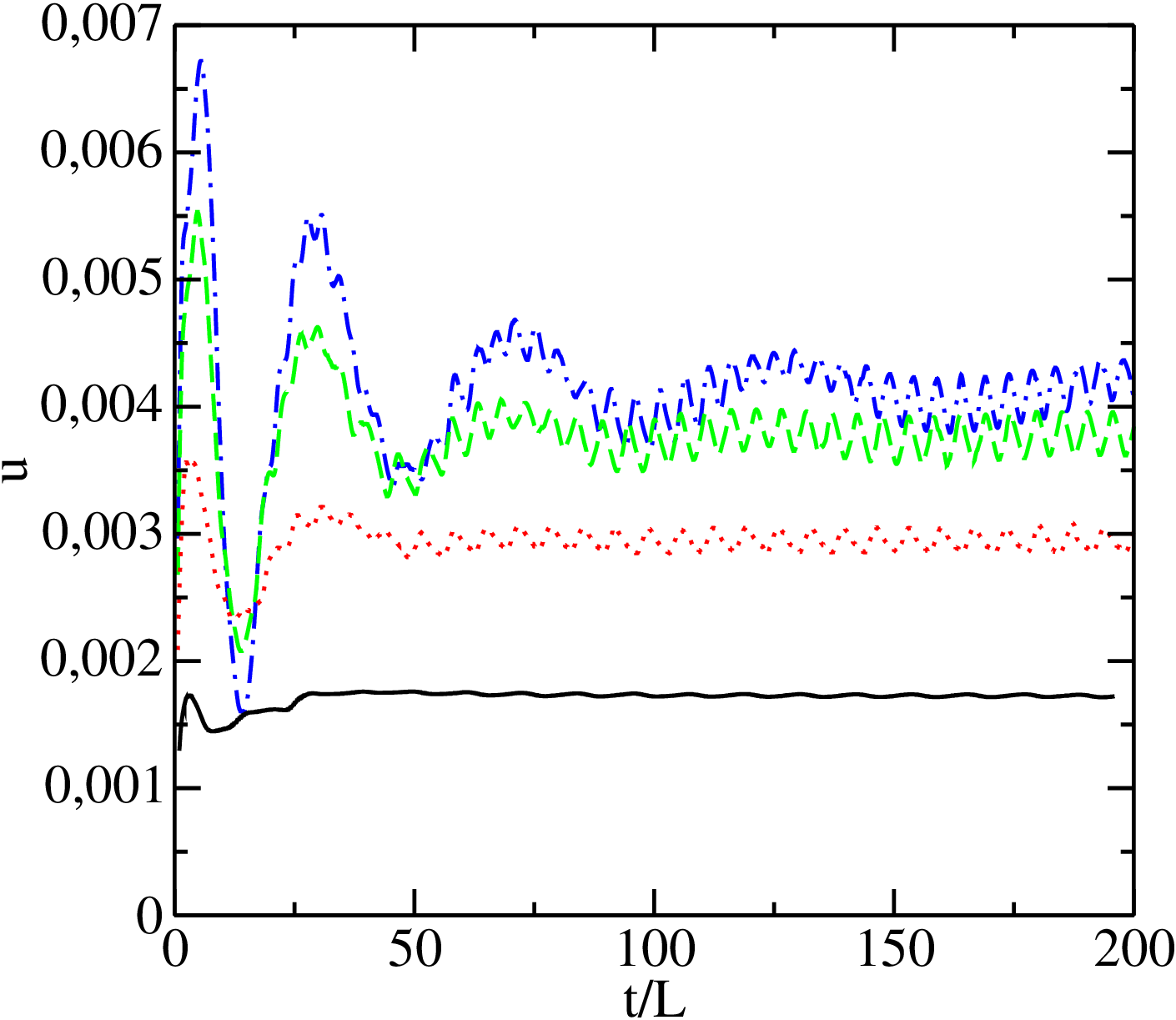}
 } \caption{Solution $R(t)$ (left panel) and $u=dR/dt$ (right
panel) for the band boundary for several values of the viscosity.
Values of $\beta$ and other parameters are taken the same as in
left panel of Fig. \ref{interface}.}\label{band}
\end{figure}

In Fig. \ref{band}    we demonstrate the law for the growing with
time of the band boundary $R(t)$ (in the left panel) and the
velocity of the boundary $u=dR/dt$ (in the right panel) for
different values of the viscosity. As for discs, the band boundary
is specified as the point, where the density achieves the critical
value ($\rho =\rho_{cr}$). Results are presented for
$T/T_{cr}=0.98$, $L=30$~fm. For small values of time (see Figure
insertion)  $R(t)$ obeys the quadratic law, as it follows from Eq.
(\ref{Slabsol}). Solid curve (effectively large viscosity, $\beta
=0.1$) follows the law (\ref{Slabsol}) for $t>t_{\rm rec}\sim
100$~fm. The higher $\beta$, the larger is deviation from this law
since the reconstruction time increases then  as $t_{\rm
rec}\propto \sqrt{\beta}$. It is clearly demonstrated in the right
panel, where we present the time dependence of  the velocity of
the seed growth. As follows from the Figure, even for large times
the velocity $u$ does not obey the scaling law, $u\propto
\sqrt{\beta}$ (as it would follow from Eq. (\ref{Rt})). This is
so, because values of time $t\sim 200L$ still correspond to the
transition regime in the right panel of Fig. \ref{dynamics1} (for
all values of $\beta$ presented in Fig. \ref{band}). In this
regime the velocity of the seed surface still slowly increases
with time. The asymptotic regime is achieved at  larger values of
time (or for smaller  $\beta$ at values of time shown in Figure).

Another important issue is presence of the damped  long-wave
oscillations which are clearly seen for all values of the
effective viscosity. They occur at $t\lsim t_{\rm rec}$. Besides,
in case of effectively small viscosity short-wave oscillations are
clearly seen. As follows from Eq. (\ref{smeq}), stable phase is,
indeed, covered by fine ripples.

\subsection{Evolution of fluctuations in spinodal region}

As the initial density profile,  we take now a wave developing  on
the background of a density $\bar{\rho}$ ( the latter value is
chosen somewhere in the spinodal region), cf. sect. \ref{Small}.
 The time dependent solution has the form
\be\label{waves}
 \rho (t)=\bar{\rho}+A_0 f(t)\mbox{sin} (\vec{k}\vec{r}),
 \ee
 where $A_0$ is a small constant and $f(0)=1$.

We checked that   in case of an effectively low viscosity  ($\beta
\gg 1$) for $k>k_{cr}=\sqrt{2}$ there appear oscillating modes and
at an effectively  high viscosity  ($\beta \ll 1$) there are only
damped modes. For  $k<k_{cr}=\sqrt{2}$ there are  growing modes.
 Various
regimes, as corresponding to growing, oscillating and damping
initial disturbances, are indicated in Table 1.
\begin{table}
\begin{center}
\begin{tabular}{ | c | c | c |}
  \hline
 Viscosity/Wave number &  Small & Large  \\
  \hline
  $k>k_{cr}$& oscillation  & damping  \\ \hline
  $k<k_{cr}$& growth  & growth  \\
\hline
\end{tabular}
\end{center}\vspace{3mm}
\label{Tgrowing}\caption{Different scenarios of  the evolution of
initial disturbances in spinodal region. } \vspace{1mm}
\end{table}

\begin{figure}
\centerline{%
\includegraphics[height=6.0truecm]{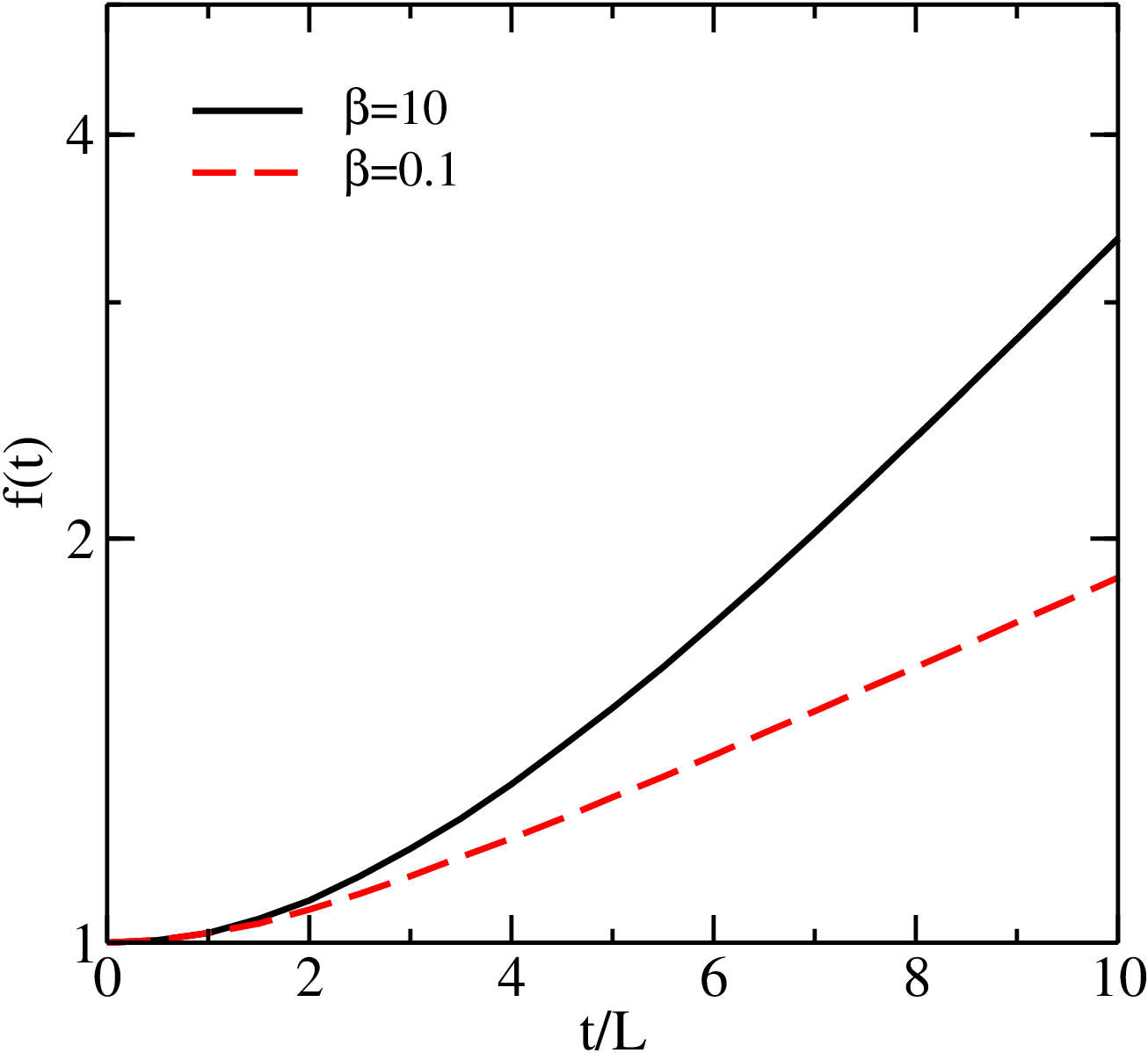}
\hspace{5mm}
\includegraphics[height=6.0truecm] {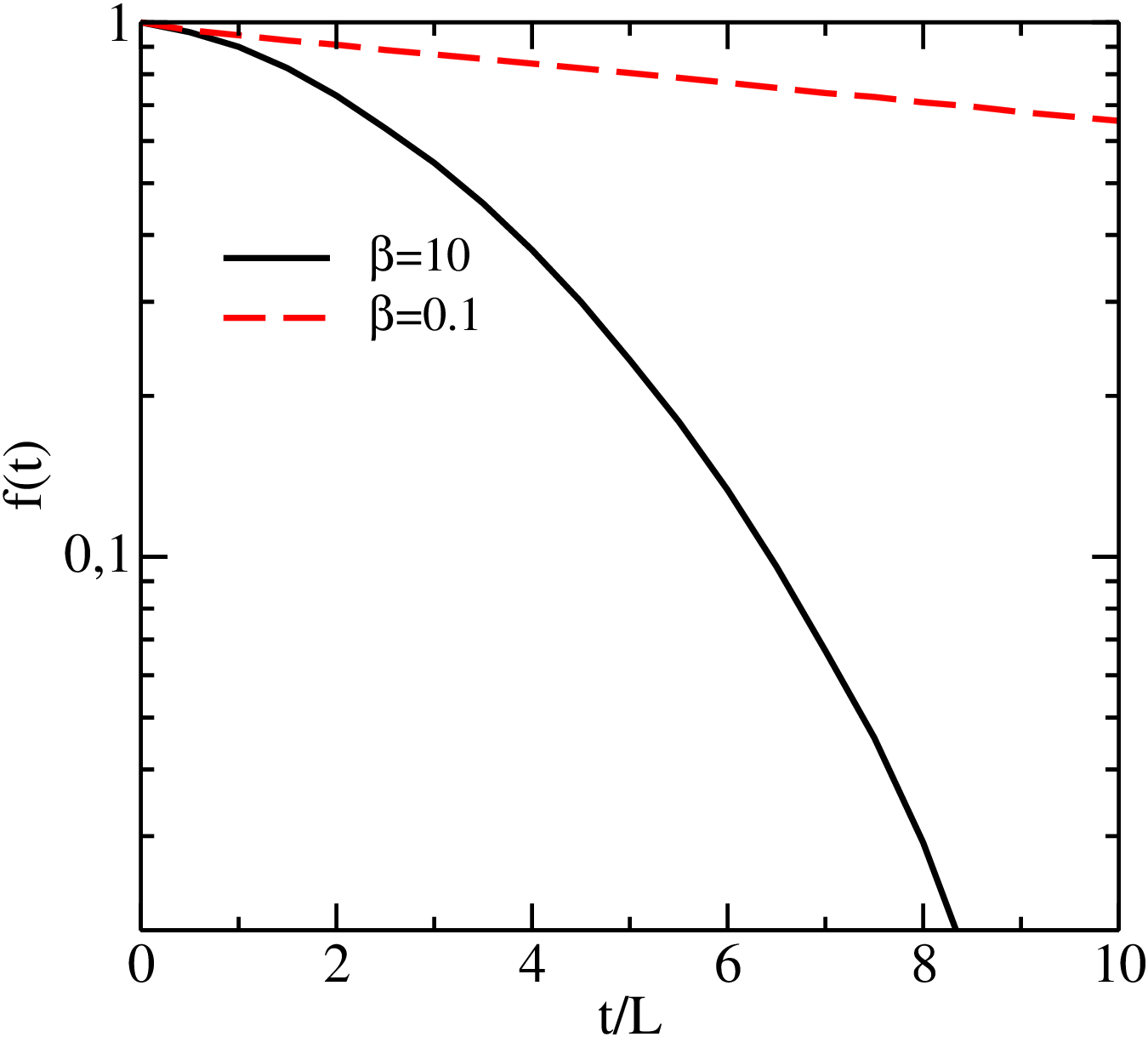} }
\caption{ Time evolution of the wave amplitudes $f(t)$, see Eq.
(\ref{waves}), normalized to the amplitude of the  initial
disturbance. Solid line  is for effectively small viscosity
($\beta =10$) and dash line, for the large viscosity ($\beta
=0.1$). Left panel: the undercritical wave number $k=2l/L$
(growing modes).
Right panel: the overcritical value $k=8l/L$ (oscillation modes
for large $\beta$ and damped modes for small $\beta$).  Other
parameters are taken the same, as in Fig \ref{EoS_dynamics1}.
}\label{waveampl}\vspace{3.0mm}
\end{figure}

In Fig. \ref{waveampl}  we show time evolution of  wave amplitudes
given by  Eq. (\ref{waves}) for an undercritical value of the wave
number $k$ (left panel) and for an overcritical value (right
panel). We  take $T_{cr}=18.6$~MeV, $n_{cr}/n_{sat}=0.42$, $c
\simeq 5.56\cdot10^{-3}$ fm$^{2}$.
 In case of the overcritical value $k$ and an effectively small viscosity
($\beta =10$) we demonstrate the change of the amplitude in the
half-period of the oscillation. Such a behavior
 fully
agrees with that follows from our analytical treatment of the
problem, see (\ref{large-gmet}), (\ref{large-gst}),  (\ref{smeq}).
 We have checked that in case of a
 small overcriticality (for $|\delta {\cal{T}}| =0.02$) slopes of the curves coincide
 up
to the third digit with values of $\gamma$ calculated in subsect.
\ref{spinodal}. Even in case of  sufficiently large deviation from
the critical point (for  $|\delta {\cal{T}}| =0.15$, as presented
in Fig. \ref{waveampl}), the difference of the curves $f(t)$
obtained numerically  from those  calculated analytically is less
than 30\%.

\begin{figure}\vspace{6mm}
\centerline{%
\includegraphics[height=6.0truecm]{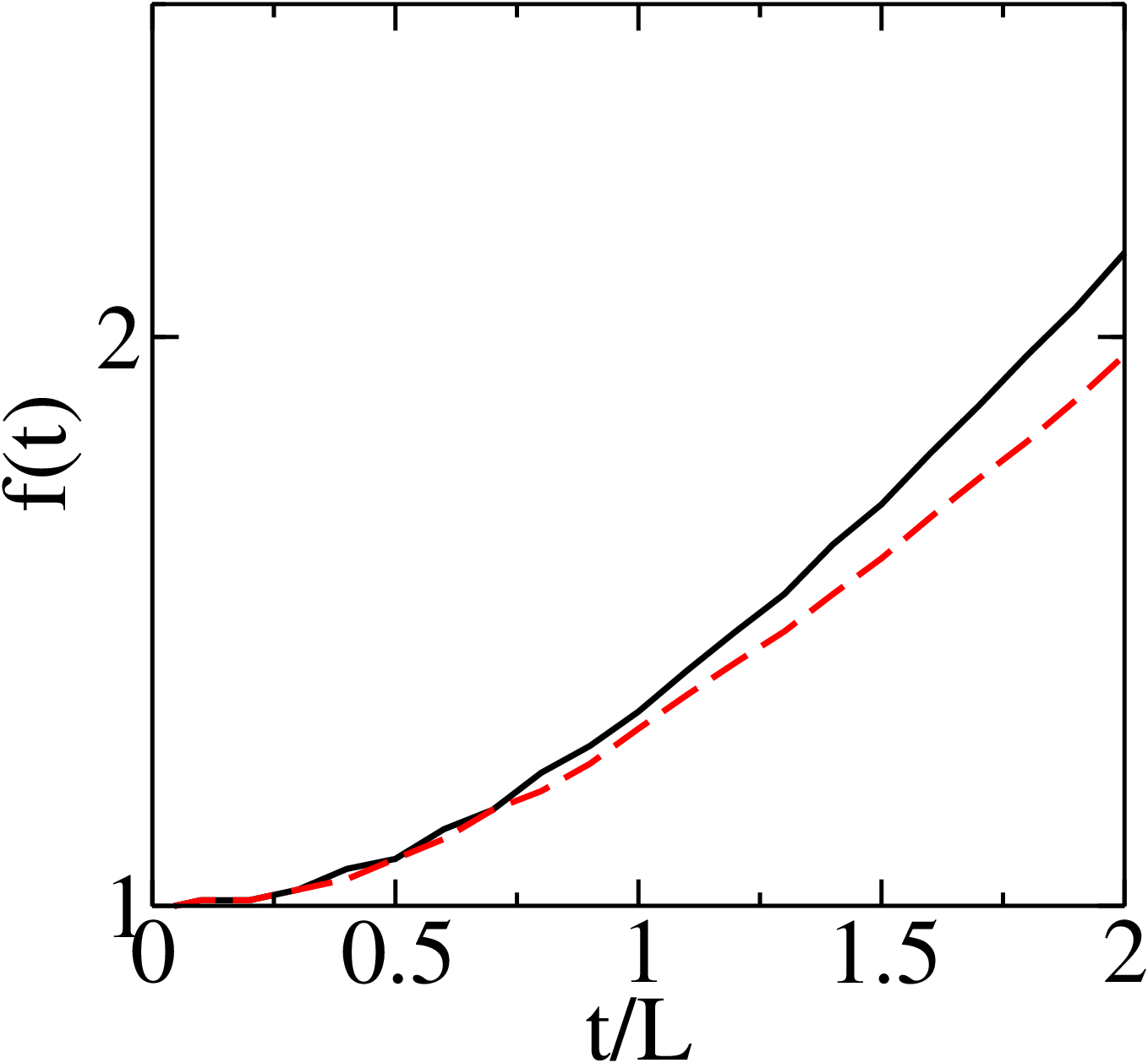}
\hspace{-1.0mm}
\includegraphics[height=6.0truecm] {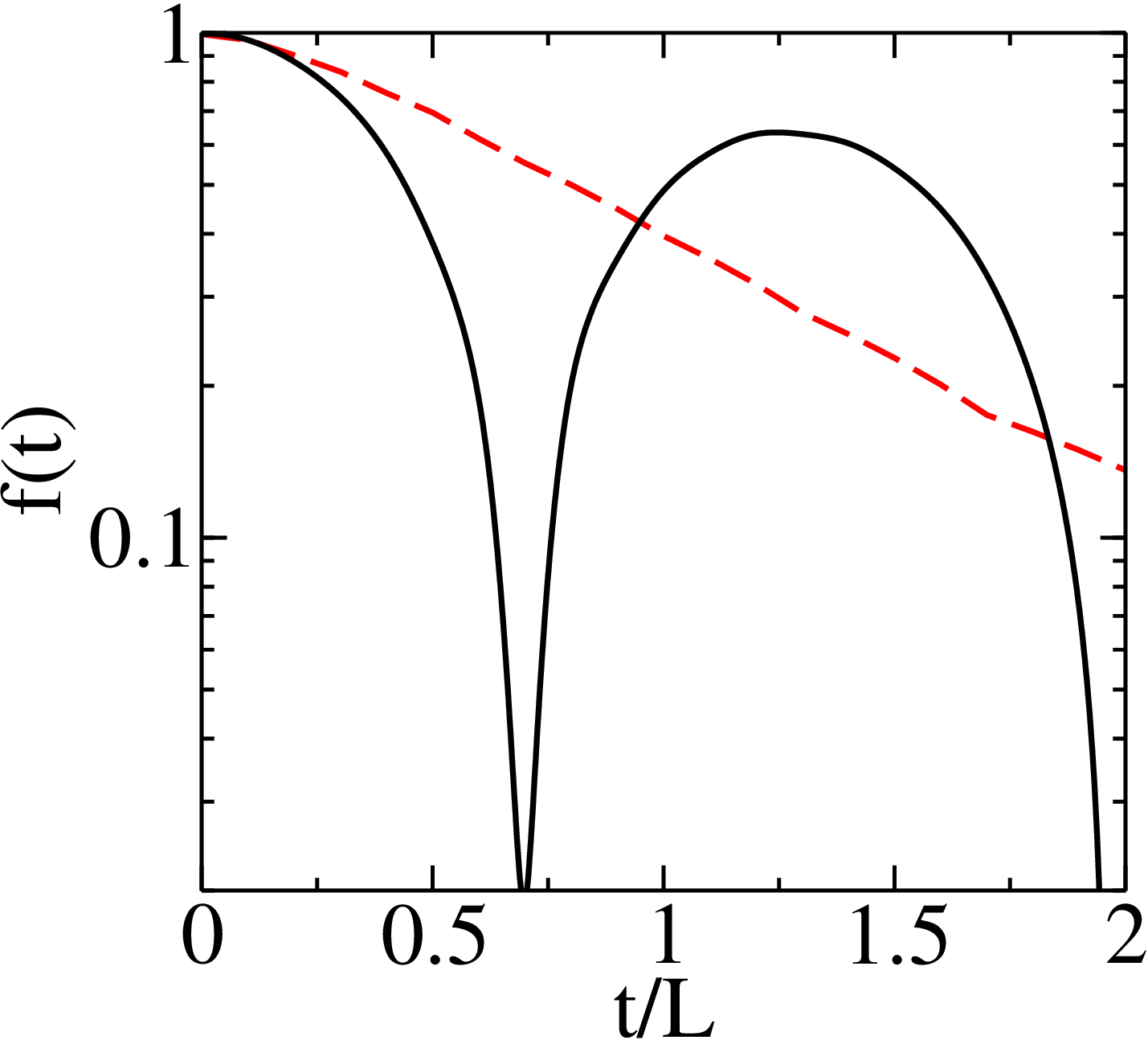} }
\caption{ The same as in Fig. \ref{waveampl} but for
$T_{cr}=162$~MeV, $n/n_{sat}=1.3$. }\label{waveamplQ}\vspace{20mm}
\end{figure}

In Fig. \ref{waveamplQ}  we show time evolution of the wave
amplitudes given by (\ref{waves}), for an undercritical value of
the wave number $k$ (left panel) and for an overcritical value
(right panel) for the parameter choice $T_{cr}=162$~MeV,
$n/n_{sat}=1.3$,
as for the
hadron-sQGP phase transition.

We see that in case relevant for the hadron-quark phase transition
the evolution is more rapid compared to the example of the
configuration presented in Fig. \ref{waveampl}  relevant for the
NGL phase transition. Nevertheless even in the former case the
time evolution remains sufficiently slow, especially at an
effectively large viscosity. Then the typical time scale $t\sim
 10$~fm is comparable with the total time of the fireball expansion.
 Thus {\em{in heavy-ion collisions during expansion of the
fireball the system may linger in the old phase for a while even
at $T< T_{cr}$. This means that the equilibrium value of the
critical temperature of the phase transition might be
significantly higher than the value which may manifest in the
growth of fluctuations in experiments.}}

\subsection{Evolution of asymmetric spots ($d_{\rm sol}=2$) in
$d=2$ space}

\begin{figure}
\includegraphics[width=14.0truecm]{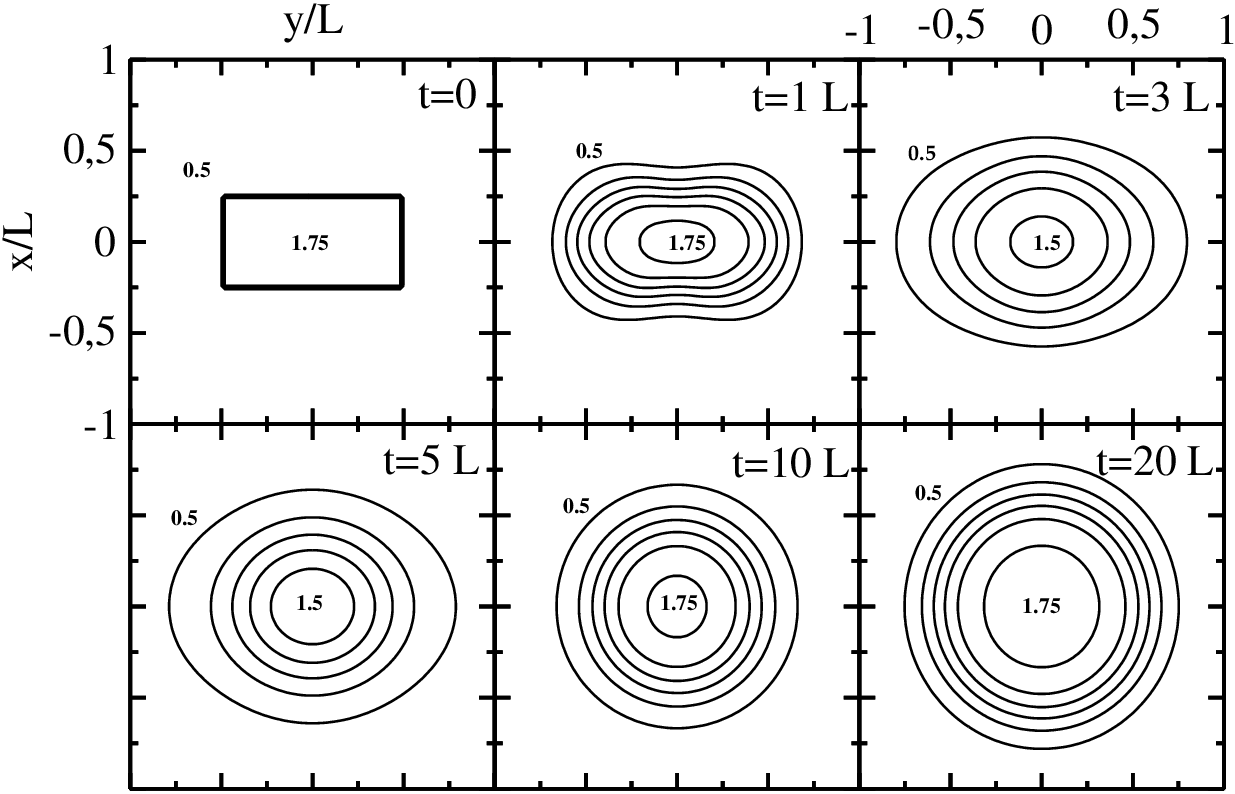}
\caption{ Isolines of the density $n/n_{cr}$ demonstrating the
time evolution of initially asymmetric disk of overcritical size
(in both directions) for effectively large viscosity ($\beta =
0.1$). Density isolines are drawn with an increment equal 0.25.
Time snapshots are indicated in the sub-figures, $|\delta
{\cal{T}}|=0.15$, $L=5$ fm. } \label{LB_assym}
\end{figure}

\begin{figure}
\includegraphics[width=14.0truecm]{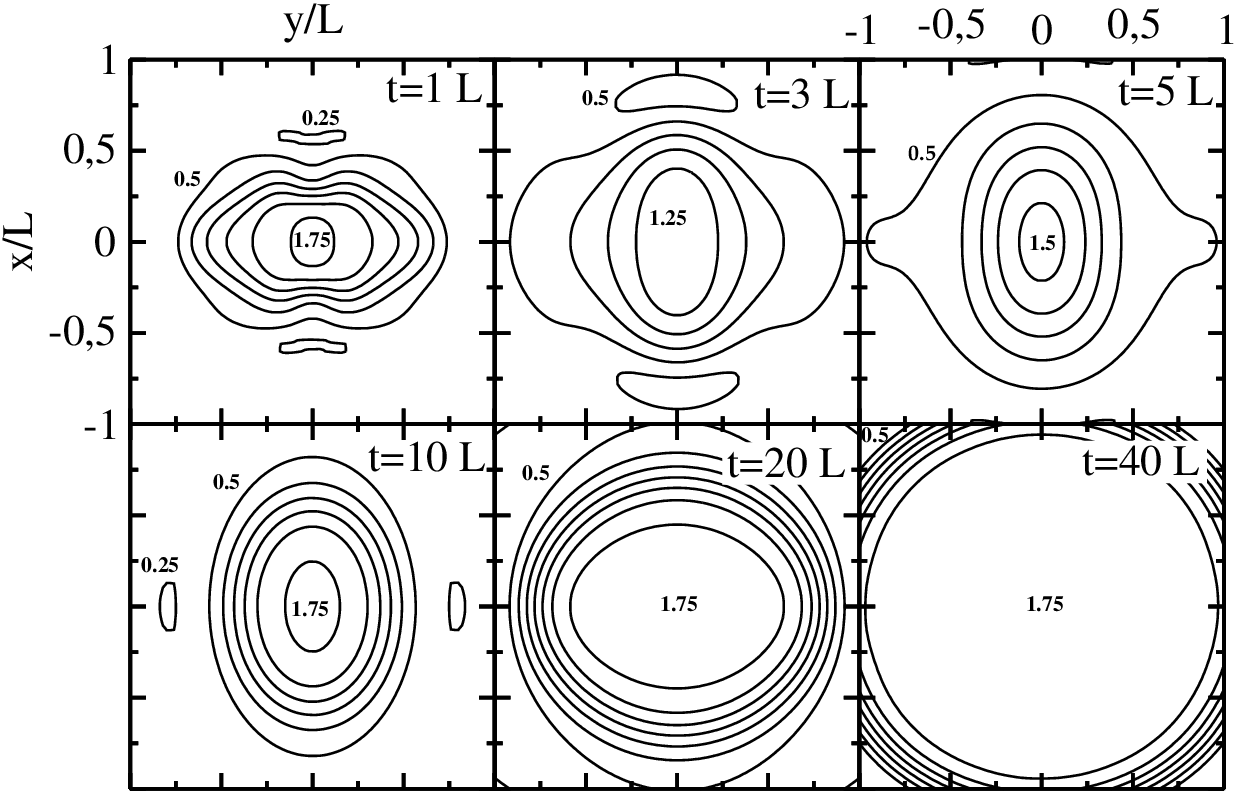}
\caption{The same as in Fig. \ref{LB_assym}, but for effectively
very small viscosity $\beta=10^3$.} \label{HB_assym}
\end{figure}

Now let us demonstrate how  seeds having shapes significantly
different from  discs  become the discs with passage of time. In
Figs. \ref{LB_assym} and \ref{HB_assym} we show time evolution of
the initially rectangular seed for effectively large ($\beta
=0.1$) and very small ($\beta =10^{3}$) values of the viscosity,
respectively. In case of an effectively large viscosity the shape
of the seed monotonously transforms to the spherical one for
typical time $t_{\rm rec}\sim 30$~fm. For an effectively small
viscosity the dynamics is a more peculiar. In a line with our
findings, see Eq. (\ref{inequ}), one can recognize oscillations of
the form. In the process of oscillations some pieces of matter
first decouple with the growing seed and then fly away. Besides,
the process lasts longer than in case of a large viscosity,
$t_{\rm rec}\sim 150$~fm.

\section{The NGL and hadron -- sQGP first-order phase
transitions}\label{Quark}

\subsection{The  NGL phase transition}

For  the description of the NGL first-order phase transition we
take values $T_{cr} \simeq 18.6$~ MeV, $n_{cr}/n_{sat}\simeq
0.42$, $n_{sat}\simeq 0.16$~fm$^{-3}$, similar to those one uses
in the RMF models describing this  transition, e.g. see
\cite{SVB}. We use the mVW model for the EoS, see Appendix A.
Parameters of the EoS are then as follows: $a\simeq 3.14\cdot
10^{2}\mbox{MeV}\cdot \mbox{fm}^{3}$, $b\simeq 5$~fm$^{3}$,
$\lambda \simeq 2.85\cdot
10^{-6}{\mbox{fm}^{6}}/{\mbox{MeV}^{2}}$, $v^2 \simeq 1.56\cdot
10^{4}|\delta {\cal{T}}| {\mbox{MeV}^{2}}/{\mbox{fm}^{6}}$,
$\epsilon^{max} \simeq 2.16 (\delta
{\cal{T}})^{3/2}{\mbox{MeV}}/{\mbox{fm}^{3}}$.

In order to estimate the gradient coefficient "$c$" we suppose
that $l(T=0)=d_{\rm dif}\simeq 0.5$~fm,  see Eq. (\ref{dimens}),
where $d_{\rm dif}$ is the diffusion length, as it follows from
the Woods-Saxon parameterization of the density profile of the
nucleus. Then $c\simeq 5.56\cdot 10^{-3}$~fm$^{2}$, and for an
appropriate value $|\delta {\cal{T}}|\simeq 0.15$ we find $l\simeq
1.3$~fm. In order to obtain $l(T=0)\simeq 0.5$~fm we should take
$\sigma_0 \simeq 4$~MeV$/\mbox{fm}^2$. Then for $T=0.85~T_{cr}$ we
get $\sigma \simeq \frac{1}{4}$~MeV$/\mbox{fm}^2$. Using these
values and  taking $\eta\simeq 23$~MeV$/$fm$^{2}$ following Ref.
\cite{Danielewicz}, we obtain $\beta_{\rm NGL} \simeq 0.023$ {\em{
that corresponds to the limit of effectively very large
viscosity.}}

Then we evaluate the time scale $t_0 \simeq 20
|\delta{\cal{T}}|^{-1}$~fm, see Eq. (\ref{dimens}). Also we are
able to estimate values $t_{\rho}$, $t_{\rm dis}$ as they follow
from Eqs. (\ref{trho1}),
 (\ref{disol}). We find
$t_{\rho}(\gamma\epsilon^{max})\simeq 40R\gamma^{-1}
|\delta{\cal{T}}|^{-1/2}$,
$t_{\rho}(R_{cr},\gamma\epsilon^{max})\simeq 30\gamma^{-2}
|\delta{\cal{T}}|^{-1}$, $t_{\rm dis}\simeq 20 R (R/\mbox{fm})$.
The initial stage of the process occurring for $t\lsim t_{\rm
init}\simeq 0.06 |\delta{\cal{T}}|^{-1}$, see (\ref{initt}),
proves to be very short ($t_{\rm init}\ll t_{\rho}$).

Further following \cite{Danielewicz} we evaluate thermal
conductivity $\kappa \simeq 0.08 $fm$^{-2}$
 and estimate typical time for the thermal transport
$t_T$.  From (\ref{kT}) we find $t_T \simeq 3 R (R/\mbox{fm})$.
Here we used that for symmetric nuclear matter
 $c_V \simeq
3n$ for $T\gg \epsilon_{\rm F}$ and $c_V \simeq 2(\pi/3)^{2/3}
mn^{1/3}T$ for $T\ll\epsilon_{\rm F}$. Thus the heat transport
might become operative for $R>R_{\rm fog}(\gamma \epsilon^{ma})
\simeq 13\gamma^{-1}|\delta{\cal{T}}|^{-1/2} $~fm, where $R_{\rm
fog}$ is a typical radius of  seeds in the "nuclear fog". Using
that $R_{cr}(\gamma\epsilon^{max})\simeq 0.9\gamma^{-1}|\delta
{\cal{T}}|^{-1/2}$~fm, see Eq. (\ref{crR}), we obtain $R_{\rm fog}
\gg R_{cr}$. Thereby there is a  long time interval, where the
heat transport is not yet efficient and solutions presented in
this paper are valid.

 Assuming that the fireball reaches spinodal region,  following
 (\ref{aeroeta})  we may estimate  the time scale
 for the formation of the aerosol,
$t_{\rm aer}^{\eta}\sim 10 |\delta{\cal{T}}|^{-1}
$~fm,  and the size scale for seeds in aerosol, $R^{\eta}_{\rm
aer} \simeq  2^{-1}\beta^{-1/4}l\simeq 0.6
|\delta{\cal{T}}|^{-1/2}$~fm.
 With the
help of Eq. (\ref{Re1}) for $\beta \simeq 0.023$, we  evaluate
typical Reynolds numbers ${\rm Re}\sim (0.1 \div 1)\ll {\rm
Re}_{cr}$), for $\alpha \sim 1\div 10$.

We see that all relevant time scales for the formation of  seeds
are  long ($\gsim 10$fm)  increasing, when the system comes closer
to the critical point. Thus it is likely that only an initial and
might be an intermediate stage of the NGL first-order phase
transition may manifest itself in the course of heavy-ion
collisions. Also {\em{ it seems unlikely to observe effects of a
narrow  vicinity of the critical point of the phase transition,
since the fireball is located in this region a short time compared
with the time scale characterizing the growth of seeds.}} This
also means that  the critical temperature of the first-order phase
transition calculated at assumption of the thermal equilibrium
might be significantly higher than the value, which may manifest
in the growth of fluctuations in experiments. Note that  values of
the critical temperature calculated in different models are in the
range $T_{cr}\sim 15\div 20$~MeV, whereas experimental value, as
it follows from the analysis of the multi-fragmentation is
$T_{cr}\sim 5\div 8$~MeV, see \cite{Randrup}. One usually
associates this difference with the finite size effects. We point
out that at least partially the difference can be explained by the
dynamical effects.

\subsection{The hadron--sQGP phase transition}

In case of the hadron-sQGP first-order phase transition critical
values $T_{cr}$ and $n_{cr}$ are rather unknown and can be varied
in a broad range. Values $T_{cr}=(80\div 170)$ MeV and
$n_{cr}=(1\div 6)n_{sat}$ are used in different models. For rough
estimates we take values $T_{cr} \simeq 162$~ MeV,
$n_{cr}/n_{sat}\simeq 1.3$, as they are obtained following lattice
calculations, cf. \cite{FK}. We again use the mVW model for the
EoS, see Appendix A. Parameters of the EoS are then as follows:
$a\simeq 8.76\cdot 10^{2}\mbox{MeV}\cdot \mbox{fm}^{3}$, $b\simeq
1.60$~fm$^{3}$, $\lambda \simeq 7.80\cdot 10^{-5}
q^{-3}{\mbox{fm}^{6}}/{\mbox{MeV}^{2}}$, $v^2 \simeq 1.56\cdot
10^{4} q^2 |\delta {\cal{T}}| {\mbox{MeV}^{2}}/{\mbox{fm}^{6}}$,
$\epsilon^{max} \simeq 58.4\cdot  |\delta {\cal{T}}|^{3/2}
{\mbox{MeV}}/{\mbox{fm}^{3}}$, $m_q$ is the effective quark mass,
$q=m_q /(300\mbox{MeV})$. Further, we obtain $l(T=0)\simeq 0.2$~fm
(radius of confinement) for $\sigma_0 \simeq
40$~MeV$/\mbox{fm}^2$. If one used $\sigma_0 \simeq
100$~MeV$/\mbox{fm}^2$, one would estimate  $l(T=0)\simeq 0.5$~fm.

Next we estimate $s(T)\simeq 7T^3 (T/T_{cr})$, $c_V \simeq 28 T^3
(T/T_{cr})$ at $T$ near $T_{cr}$, as it follows from the lattice
data \cite{Aoki:2005vt}. Assuming minimal value of the viscosity
$\eta_{\rm min} =s/(4\pi)=60$MeV/fm$^{2}$, $\zeta_{\rm min}=0$ we
evaluate maximum value of $\beta$: $\beta_{\rm sQGP}^{\rm max}
\simeq 0.015 q$
for $\sigma_0 \simeq 40$~MeV$/\mbox{fm}^2$~\cite{Berger:1986ps}, {\em{ that corresponds
to the limit of effectively very large viscosity.}} Even for
$\sigma_0 \simeq 100$~MeV$/\mbox{fm}^2$, $m_q =600$~MeV we would
get $\beta_{\rm sQGP}^{\rm max} \simeq 0.2 \ll 1$.  Note that
following \cite{Kharzeev} the bulk viscosity diverges in the
critical point. If were so ($\beta \rightarrow 0$), the
quark-hadron system would behave as absolutely viscous fluid, like
glass, in near critical region. Contrary, Refs.
\cite{SR,Shuryak:2008eq} argue for a smooth behavior of the bulk
viscosity.

With $\beta =0.015$,
 we further estimate $t_0 \simeq 2 |\delta{\cal{T}}|^{-1}$~fm, $t_{\rho}(\gamma\epsilon^{max})$ $\simeq 9.1
 R\gamma^{-1}
q^{1/2}|\delta{\cal{T}}|^{-1/2}$ and $t_{\rm dis}\simeq 14q ({R}_0
/{\mbox{fm}})R_0$. The time scale
 for the formation of the aerosol is
$t_{\rm aer}^{\eta}\simeq  |\delta{\cal{T}}|^{-1}
$~fm,  and the size scale for  seeds in aerosol is $R^{\eta}_{\rm
aer} \simeq 0.24|\delta{\cal{T}}|^{-1/2}
$~fm. Only $t_{\rm init}^{\eta}\simeq  0.03
q|\delta{\cal{T}}|^{-1}$ fm proves to be small (excluding quite
small $\delta{\cal{T}}$).

For the thermal conductivity  we  use an estimation $\kappa
\simeq\alpha_0 \eta/m$, see \cite{GIK}. Factor $\alpha_0$ depends
on the EoS used. We will take $\alpha_0 =3$. Then one recovers
appropriate relation between values of $\kappa$ and $\eta$ for NGL
transition, which we have used above, see \cite{Danielewicz}. For
the hadron--sQGP transition this estimation renders $\kappa_{\rm
sQGP} \simeq 3\eta/m_q$. Then we are able to evaluate the scale of
the heat transport time, $t_T \simeq 26
q\left({R}/{\mbox{fm}}\right)^2$ fm.
 The
heat transport becomes operative for $R>R_{\rm fog}(\gamma
\epsilon^{max}) \simeq 0.3\gamma^{-1}
q^{-1/2}|\delta{\cal{T}}|^{-1/2}$~fm. Here $R_{\rm fog}$ is the
scale of size of the seed in the quark (or hadron) fog-like state.
Using that $R_{cr}(\gamma\epsilon^{max})\simeq
0.3\gamma^{-1}|\delta {\cal{T}}|^{-1/2}(\sigma_0/(40
\mbox{MeV}/\mbox{fm}^2))$~fm, for $\sigma_0 =(40\div 100)
$\,~MeV$/\mbox{fm}^2$ we obtain $R_{\rm fog} \lsim R_{cr}$.
Thereby, {\em{the heat transport might be always operative for the
description of the evolution of overcritical seeds in hadron--sQGP
phase transition.}} The value $R_{\rm fog}$ proved to be very
small ($\lsim 0.1\div 1$~fm). However  typical time $t_T$ is
rather long.  Therefore,
 the system most probably would have no time to fully  develop  a fog-like state in a
 hadron-quark phase transition in heavy-ion collisions.

For the system located in the vicinity of the critical point all
estimated time scales (except $t_{\rm init}$) are very large. If
the system trajectory paths   rather far from
 the critical point,
all time scales, except $t_T$, become of the order or less than
the typical life-time of the fireball
 ($\sim 10$~fm at RHIC conditions).
Reynolds numbers are $\lsim 1$, being  much smaller than ${\rm
Re}_{cr}\gsim 1000$. Thereby, turbulence regime is not reached.

\section{Conclusion}\label{Conclusion}

In this paper we studied the  dynamics of systems undergoing
first-order phase transitions. We formulated analytical description
of the problem  (in general case in $d+1$-dimensional space-time)
valid for systems  in the vicinity of the critical
point.
The analytical solutions were derived for configurations of certain symmetries
(droplets/bubbles, rods and slabs for $d=3$).

Then, the  general system of equations of non-ideal
non-relativistic hydrodynamics was numerically solved for a
modified Van der Waals equation of state. Results for the original
Van der Waals equation of state can be obtained by simple
re-scaling of the time and the viscosity.
 Since  there exist many different regimes for the relevant
 processes,  in this paper   we partially restricted our analysis. We did not incorporate the heat transport,
which governs evolution of seeds of a large size. Also we focused
on the case, when  concentration of seeds of the new
phase in the old phase is still rather small and one can ignore
their coalescence. For simplicity we did not
 generate fluctuations in  random processes  considering evolution
 of the given seed  after it has been produced in a fluctuation. Generalizations will
 be presented elsewhere.
 With mentioned reservations we revealed and studied general features of the dynamics of  first-order phase
 transitions. These main features are as follows:

\begin{itemize}

\item[(i) ]  Essential role in the
dynamics of the first-order phase transitions is played by the
viscosity effects. This might be very important, since existing
three-dimensional hydrodynamical schemes pretending to study phase
transitions exploit  equations of ideal hydrodynamics, whereas
viscosity effects are simulated only implicitly, e.g., with the
help of phenomenological coefficients responsible for a friction
of fluids.

\item[(ii)]  Because of  the surface
 tension (except for one-dimensional slabs)
  there exists a critical size for the seeds of the stable phase. Seeds of overcritical sizes
 grow with  time, while undercritical fluctuations dissolve.  The
 closer to the critical point, the slower are processes.
Even far from the critical point  overcritical seeds grow slowly.

\item[(iii)]  We have shown that seeds with an asymmetric shape  become spherical  with
time. This process is slow. For systems with effectively large
viscosity the shape of the seed changes steadily to the spherical
one.
  For systems with effectively small viscosity the seed
 undergoes long-wave and short-wave damped oscillations  in the process of acquiring spherical quasi-equilibrium shape.
 Short-wave oscillations damp very slowly.

\item[(iv)]  In the spinodal region the system is unstable against
 generation of  waves (with not too high wave-number).
 However instability develops  slowly,  if the system is
 close to the critical point
 (the time scale tends to infinity at the critical point). {\em{Far from the critical
 point  processes become more rapid. The latter observation can be very
 important for studying   possible signatures of the gas-liquid
 and
 hadron-quark first-order phase
 transitions in heavy-ion collisions.}}

 \end{itemize}

  As a signal of the phase transition, some
 models suggest to search anomalies in the behavior of the derivatives of thermodynamic quantities,
 e.g. specific heat. These
 anomalies appear due to fluctuations, whereas we have shown that those
 {\em{anomalies in fluctuations may not have
 sufficient time to develop.}} Moreover we conclude that in   heavy-ion collisions {\em{the system may linger in
the old phase (e.g. in the QGP state, or in the gas state) longer
during the fireball expansion, even when  $T(t)$ has already decreased
below the corresponding value of the equilibrium critical
temperature of the phase transition.}}
 In another words, this means that the  value of
the critical temperature calculated within equilibrium
thermodynamical models might be significantly higher than the
value, which may manifest in the growth of fluctuations in
experiments.

Typical dimensionless parameter $\beta$ that separates effectively
viscous and perfect fluid regimes proves to be   $\propto
\sigma_0^2 /(\frac{4}{3}\eta +\zeta)^2 $, where $\sigma_0$ is the
surface tension at $T=0$ and $\eta$ and $\zeta$ are the shear and
bulk viscosities. According to  our estimates {\em{the system
undergoing nuclear gas-liquid phase transition in the course of
heavy-ion collisions at low energies represents effectively very
viscous fluid}} ($\beta \ll 1$). There exist arguments that
strongly coupled quark-gluon plasma state, which is, as commonly
expected, formed in heavy-ion conditions at RHIC, represents
almost perfect fluid in the cross-over region (see e.g.
\cite{Romatschke:2007mq}). Estimating the ratio of the viscosity
to the entropy density as $\eta /s <0.2$ it was concluded
\cite{Shuryak:2008eq} that  strongly coupled quark-gluon plasma is
the most perfect liquid known. It is usually believed that this
property of the plasma will survive at finite baryon density for
systems in the vicinity of the critical point of the first-order
phase transition (critical end point).
 In contrast our estimates show that  {\em{the system undergoing
the hadron-quark first-order phase transition in the course of
violent heavy-ion collisions represents effectively very viscous
fluid}} ($\beta \ll 1$).

We found  that {\em{ the heat transport effects may play
important role in description of the hadron-sQGP  phase transition
dynamics, whereas these effects are much less pronounced in the
case of the nuclear gas-liquid transition.}}

In  future we plan to use a realistic  equation of state to
study the heavy-ion collision dynamics.

After our paper has been submitted to the journal there appeared
interesting paper \cite{RandrupGQP} devoted to the description of
fluctuations in the spinodal region at the hadron-sQGP first-order
phase transition which well complements our study.

\vspace*{5mm} {\bf Acknowledgements} \vspace*{5mm}

We are grateful to  B. Friman,  Y.B. Ivanov, E.E. Kolomeitsev, J.
Randrup, and V.D. Toneev  for numerous discussions and valuable
remarks. Especially we are grateful to D. Blaschke and L. Grigorenko for the reading
of the manuscript and making numerous useful comments. This work
was supported by the
Russian Foundation for Basic Research RFBR grant
08-02-01003-a and  the BMBF/WTZ project  RUS 08/038.

\section{Appendix A. Modified Van der Waals EoS}
The best known example to illustrate principal features of the
first-order phase transition is the Van der Waals fluid. The
pressure is given by
 \be
\label{prVan} P_{\rm VW}[V,T]=\frac{NT}{V-Nb}-\frac{N^2
a}{V^2}=\frac{nT}{1-bn}-n^2 a ,
 \ee
  where parameter "$a$" governs the strength of the mean field attraction and
  "$b$"
 controls a short-range repulsion. Obviously realistic EoS of nuclear matter
 has much more complicated temperature dependence. Due to this
 we will
exploit a mVW EoS with
 \be \label{prVanm} P[V,T]=f(T)P_{\rm VW}[V,T],
\ee
 where $f(T)$ is a function of the temperature. In the given
paper we do not consider the heat transport, assuming $T=const$.
In this case our solutions are self-similar. Doing replacement
$t\rightarrow t\sqrt{f(T)}$ and  $(\widetilde{d}\eta_{\rm
r}+\zeta_{\rm r})\rightarrow(\widetilde{d}\eta_{\rm r}+\zeta_{\rm
r})/\sqrt{f(T)}$ we recover results for the original VW EoS.

We choose $\rho_{\rm r} =\rho_{cr}$, $T_{\rm r} =T_{cr}$ as the
reference point (see notations in subsect. \ref{Reduction}) and
perform expansion of the pressure in the vicinity of this point.
The critical liquid-gas point (evaporation point)  is determined
from the conditions $\partial P/\partial V =\partial^2 P/\partial
V^2 =0$, $\partial^3 P/\partial V^3 <0$. Critical parameters are
 \be\label{point} T_{cr}=\frac{8a}{27b}, \quad
V_{cr}=3Nb, \quad P_{cr}=\frac{a}{27b^2}, \quad
n_{cr}=\frac{1}{3b}.
 \ee
 For the given EoS they coincide with those for purely VW EoS.

 For $T=T_{comp}=\frac{27}{32}T_{cr}\simeq 0.844
T_{cr}$, $n_{comp}=\frac{3}{2}n_{cr}$  pressure (\ref{prVan})
touches zero. For $T<T_{comp}$ there arises density interval (from
$(\frac{3}{2}-\sqrt{\frac{9}{4}-\frac{8}{3}\frac{T}{T_{cr}}})n_{cr}$
to
$(\frac{3}{2}+\sqrt{\frac{9}{4}-\frac{8}{3}\frac{T}{T_{cr}}})n_{cr}$),
where pressure becomes negative. {\em{In this paper we will
restrict our consideration by taking $T_{cr}>T>T_{\rm comp}$.}}


For a more
convenient analytical treatment of the problem let us, selecting
corresponding  function $f(T)$ additionally fulfill condition
$\frac{\partial P}{\partial T}|_{(n_{cr},T_{cr})}=0$. It allows to
parameterize the Helmholtz free energy, as it has been done in the
paper body for the Landau free energy, with two minima deviating
only little from each other.
 Then we may specify function $f$, e.g., as
 \be f(\delta{\cal{ T}})\simeq C(\delta{\cal{ T}})\left[1- 2 \delta
\cal{T}\right]^2
. \label{f92}
 \ee
Although our analytical consideration is valid only
  in the vicinity of the critical point $0<-\delta{\cal{ T}} \ll 1$ and we perform numerical calculations for
  $0<-\delta{\cal{ T}}$,   let us specify the
pre-factor $C(\delta {\cal{T}})=[1+4(\delta {\cal{T}})^2]^{-1}$ to
reproduce the ideal gas EoS  for sufficiently low $n$ and high
$T$.

  Expanding the pressure in  $|\delta
{\cal{T}}|\ll 1$, $\delta n /n_{cr}\ll 1$ we get
 \be\label{modif}
 \delta P=\frac{9T_{cr}\delta{\cal{ T}}\delta n
}{4}+\frac{9T_{cr}}{16n_{cr}^2} (\delta n)^3  
- 6 T_{cr} n_{cr} (\delta{\cal{ T}})^2 +...
 \ee
 Last term in (\ref{modif}) is actually unimportant since addition
 of any constant (at $T=const$ last term is constant) does not
 change equations of motion. Let us count $P$ from its
 value in the final equilibrium state (reaching at $t\rightarrow \infty$).
 Namely this difference $  P - P_{f}$ has the meaning
 of the thermodynamical force driving the system to the final
 equilibrium state (see (\ref{rs})). Then
we may construct the Landau free energy (\ref{fren}) such that
$\delta (\delta F_L)/\delta(\delta n) = P - P_{f}$. Comparing
(\ref{modif}) with (\ref{pre}) we find relations between
coefficients:
 \be\label{parame}
 a&=&{9}{T_{cr}}/({8}{n_{cr}}), \,\,
 b={1}/({3n_{cr}}),\,\,
 v^2 =-{3m^2 T_{cr}\delta{\cal{ T}}}/({2ab})=- 4 \delta{\cal{T}} n_{cr}^2 m^2
 ,\nonumber\\
 \lambda &=&\frac{3ab}{2m^3}=\frac{9}{16}\frac{T_{cr}}{n_{cr}^2
 m^3},\quad \epsilon =n_{cr}(\mu_{in}-\mu_{f}),\nonumber\\
 t_0
 &=&{8(\widetilde{d}\eta_{\rm r} +\zeta_{\rm r})}({9n_{cr}T_{cr}|\delta{\cal{T}}|})^{-1}.
  \ee
Since we used $f(T\to T_{cr})=1$ in Eq.~(\ref{f92}), these relations 
 are the same as for the original VW EoS.

\section{Appendix B. Mean field and fluctuation region}
In the paper body we considered dynamics of fluctuations assuming
that thermodynamical characteristics like pressure, free energy
etc., are given (mean field approximation) and not modified by
fluctuations  (mean field approximation). In order to estimate a
possible influence of fluctuations on thermodynamical
characteristics of the equilibrium uniform system at $T\neq 0$ let
us compare mean field and fluctuation contributions to the density
of the specific heat $\delta c_V =-T[\partial^2 \delta
{\cal{F}}/(\partial T)^2]_V$. Using (\ref{de}), (\ref{parame}) for
$T$ near $T_{cr}$ we find
 \be\label{cMF}
 c_V^{\rm MF} \simeq
 9n_{cr}/2
  \ee
  for configurations corresponding $|\epsilon|\ll \epsilon^{max}$.

The fluctuation contribution to the specific heat density
$c_V^{'}$ can be found with the help of the functional integration
 \be \mbox{exp}[\delta F^{'}/T]=\int D\delta n^{'} \mbox{exp}(\delta F^{'}[\delta
 n^{'}]/T),
 \ee
where following (\ref{fren}) we have
 \be
\delta F^{'}=\frac{T}{2}\int \frac{d^3
k}{(2\pi)^3}\mbox{ln}[\vec{k}^{\,2} +\alpha ],\quad \alpha
=2\lambda v^2 /c.
 \ee
From here using (\ref{c}), (\ref{parame}) we obtain
 \be\label{cfl}
 c_V^{'}=\frac{\alpha^{3/2}}{16\pi|\delta{\cal{T}}|^{1/2}}=\frac{108}{\pi}\frac{n_{cr}^3
 T_{cr}^3}{\sigma_0^3|\delta{\cal{T}}|^{1/2}}.
  \ee
Equating (\ref{cMF}) and (\ref{cfl}) (Ginzburg -- Levanyuk
criterion) we estimate the Ginzburg number
 \be\label{Gi}
 \mbox{Gi} =\frac{T_{cr}-T_{\rm
 fl}}{T_{cr}}=\left(\frac{24}{\pi}\frac{n_{cr}^2 T_{cr}^3}{\sigma_0^3}\right)^{2}.
  \ee
  Fluctuation region is narrow provided $ \mbox{Gi}\ll 1$ and it
  is broad for  $ \mbox{Gi}\gsim 1$. Similar estimate  to  $ \mbox{Gi}\ll 1$
  follows from the so-called Ginzburg criterion
$\frac{4\pi}{3}l^3|\delta {\cal{F}}^{\rm MF}|\gg T_{cr}$. In the
fluctuation
  region ($T_{cr}>T>T_{\rm fl}$)
  fluctuation effects   may modify $\delta{\cal{T}}$ dependence of coefficients in Eq.  (\ref{modif}).

  Substituting in (\ref{Gi})
  typical values of parameters for the hadron-quark phase
  transition we estimate $ \mbox{Gi}\gsim 1.4 \left(100\mbox{MeV}\cdot\mbox{fm}^{-2}/\sigma_0\right)^6$,
  i.e. fluctuation region is  broad.
  In case of the NGL phase transition we estimate  $ \mbox{Gi}\sim 10\left(T_{cr}/18.6 \,\mbox{MeV}\right)^6$
  and fluctuation region is also broad.

One can construct  description of the  fluctuation region
incorporating effect of long-range fluctuations directly in the
phenomenological expressions for the free energy $\delta F$ and
the pressure $\delta P$. One can do it with the help of the
replacements $v^2 \rightarrow v_{\rm ren}^2 =v^2
(T=0)|\delta{\cal{T}}|^{2/3}$, and $\lambda \rightarrow
\lambda_{\rm ren}=\lambda (T=0)|\delta{\cal{T}}|^{2/3}$ (similarly
one incorporates fluctuations in description of the superfluid
He$^4$). Other coefficients in expressions of sub-section \ref{ES}
remain unchanged. With these replacements dependence on
$|\delta{\cal{T}}|$ disappears from the Ginzburg criterion, and
both values $c^{\rm MF}_V$ and $c^{'}_V$ do not vanish for
$\delta{\cal{T}}\rightarrow 0$. The Ginzburg -- Levanyuk criterion
then reads as $\left(\frac{128}{3\pi}\frac{n_{cr}^2
T_{cr}^3}{\sigma_0^3}\right)^2 \ll 1$. Effect of not included
fluctuations is small provided this inequality is fulfilled.

Finally it is worthwhile to mention that it takes a long time
(typically $\sim t_0$) to develop critical fluctuations. Therefore
if the system  passes the fluctuation region during a time
$t_{evol}$, as it occurs in a course of heavy ion collisions, and
$t_{evol}<t_0$, critical fluctuations will not have enough time to
develop. Thus for $t<t_0$ it is legitimate to use mean field EoS
to describe evolution of the system.


\begin{thebibliography}{99}
\bibitem{LP}
E.M. Lifshiz, and L.P. Pitaevsky, "Physical Kinetics", Pergamon,
1981.
\bibitem{PS}
A.Z. Patashinsky, and B.I. Shumilo, JETP {\bf 50}, 712 (1979).
\bibitem{MSTV90} A.B. Migdal, E.E. Saperstein, M.A. Troitsky, and
D.N. Voskresensky, Phys. Rep. {\bf 192}, 179 (1990).
\bibitem{V93}
D.N. Voskresensky, Phys. Scripta {\bf 47}, 333 (1993).
\bibitem{Onuki}
A. Onuki, Phys. Rev. {\bf E7}, 036304 (2007).
\bibitem{SG} H. Stoecker, and W. Greiner, Phys. Rep., {\bf 137},
277 (1986).
\bibitem{IR}
Yu.B. Ivanov, V.N. Russkikh, and V.D. Toneev, Phys. Rev. {\bf
C73}, 044904 (2006).


\bibitem{IdealRHIC}
D. Teaney, J. Lauret, and E.V. Shuryak, Phys. Rev. Lett., {\bf
86}, 4783 (2001); P.F. Kolb, U.W. Heinz, P. Huovinen, K.J. Eskola,
and K. Tuominen, Nucl. Phys. {\bf A696}, 197 (2001); T. Hirano,
and K. Tsuda, Phys. Rev. {\bf C66}, 054905 (2002); P. Kolb, and R.
Rapp, Phys. Rev. {\bf C67}, 044903 (2003).
\bibitem{Teaney} D. Teaney, Phys. Rev. {\bf C68}, 034913 (2003).
\bibitem{Romatschke}P.~Romatschke, Eur. Phys. J. {\bf C52}, 203
(2007)
\bibitem{Romatschke:2007mq}
  P.~Romatschke, and U.~Romatschke,
  Phys.\ Rev.\ Lett.\  {\bf 99}, 172301  (2007).

\bibitem{Shuryak:2008eq}
  E.~Shuryak,
  Prog. Part. Nucl. Phys. {\bf 62}, 48 (2009).
\bibitem{Muronga} A. Muronga, and D. H. Rischke, arXiv:
nucl-th/0407114.
\bibitem{Lallouet}Y. Lallouet, D. Davesne, and C. Pujol, Phys. Rev. {\bf C67}, 057901 (2003).
\bibitem{NA} C. Nonaka, and M. Asakawa,  Phys. Rev. {\bf
C71}, 044904 (2005); Nucl. Phys. {\bf A774}, 753 (2006).

\bibitem{Dumitru}
K. Paech, H. Stoecker, and A. Dumitru, Phys. Rev. {\bf C68},
044907 (2003); K. Paech, and A. Dumitru, Phys. Lett. {\bf B623},
200 (2005);  C.E. Aguiar, E.S. Fraga, and T. Kodama, J. Phys. {\bf
G32}, 179 (2006).

\bibitem{GIK}
V.M. Galitsky, Yu.B. Ivanov, and V.A. Khangulian, Sov. J. Nucl.
Phys. {\bf 30}, 401 (1979).
\bibitem{Danielewicz}
L. Shi, and P. Danielewicz, Phys. Rev., {\bf C68}, 064604 (2003).



\bibitem{CKM} L.P. Csernai, J.I. Kapusta, and L.D. McLerran, Phys.
Rev. Lett., {\bf 97}, 152303 (2006).
\bibitem{Kharzeev} D. Kharzeev, and K. Tuchin, JHEP 0809, 093 (2008);
F. Karsh, D. Kharzeev, and K. Tuchin, Phys. Lett., {\bf B663},
217 (2008).
\bibitem{SR} C. Sasaki, and K. Redlich, arXiv:
0806.4745 [hep-ph]; 0811.4708 [hep-ph].


\bibitem{Gubster} S.S. Gubster, S.S. Pufu,  and F.D. Rocha,
arXiv:0806.0407 [hep-th].


\bibitem{Sakai}
S. Sakai, and A. Nakamura, PoS LAT2007:221 (2007).

\bibitem{Berges}J. Berges, and K. Rajagopal, Nucl. Phys. {\bf B538}, 215
(1999).
\bibitem{Stephanov} M.A. Stephanov, K. Rajagopal, and E.V.
Shuryak, Phys. Rev. Lett., {\bf 81}, 4816 (1998).
\bibitem{Alton} C.R. Allton et al., Phys. Rev. {\bf D71}, 054508
(2005); R.V. Gavai, and S. Gupta, Phys. Rev. {\bf D71}, 114014
(2005)
\bibitem{FK} Z. Fodor, and S.D. Katz, JHEP {\bf 0404}, 050
(2004);
  F.~Csikor, G.~I.~Egri, Z.~Fodor, S.~D.~Katz, K.~K.~Szabo and A.~I.~Toth,
  JHEP {\bf 0405}, 046 (2004)
\bibitem{KMR} V. Koch, A. Majumder, and J. Randrup, Phys.
Rev. {\bf C72}, 064903 (2005).

\bibitem{Berdnikov} B. Berdnikov, and  K. Rajagopal, Phys. Rev. {\bf D61}, 105017 (2000).

 \bibitem{FrimanRedlich} C. Sasaki,
B. Friman, and K. Redlich, Phys. Rev. Lett. {\bf 99}, 232301
(2007); Phys. Rev. {\bf D77}, 034024 ( 2008); Mod. Phys. Lett.,
{\bf A23}, 2469 (2008).

\bibitem{Scavenius} .
O. Scavenius, A. Dumitru, and A.D. Jackson, Phys. Rev. Lett. {\bf
87}, 182302 (2001); G. Torrieri, B. Tomasik, and I. Mishustin,
Phys. Rev. {\bf C77}, 034903 (2008).

\bibitem{Bondorf}G. R\"opke, L. M\"unchow, and H. Schulz, Phys. Lett.
{\bf B110}, 21 (1982).
\bibitem{SVB} H. Schulz, D.N. Voskresensky, and
J. Bondorf, Phys. Lett. {\bf B133}, 141 (1983).
\bibitem{Siemens}
A.D. Panagiotou, M.W. Curtin, H. Toki, D.K. Scott, and P.J.
Siemens, Phys. Rev. Lett. {\bf 52}, 496 (1984).

\bibitem{Agostino}
M. D'Agostino et all., Phys. Lett. {\bf B473}, 219 (2000); M.
Schmidt et all., Phys. Rev. Lett. {\bf 86}, 1191 (2001).
\bibitem{Randrup}
P. Chomaz, M. Colonna, and J. Randrup, Phys. Rep. {\bf 389}, 263
(2004).

\bibitem{Glendenning}
N.K. Glendenning, Phys. Rev. {\bf D46}, 1274 (1992);
Phys. Rep. {\bf 342}, 393 (2001).

\bibitem{VYT}
H. Heiselberg, C.J. Pethick, and E.F. Staubo, Phys. Rev. Lett.
{\bf  70} 1355 (1993); D.N. Voskresensky, M. Yasuhira, and T.
Tatsumi, Nucl.Phys. {\bf A723}, 291 (2003); Toshiki Maruyama, T.
Tatsumi, D.N. Voskresensky, T. Tanigawa, and S. Chiba, Phys. Rev.
{\bf C72}, 015802 (2005); Toshiki Maruyama, T. Tatsumi, D.N.
Voskresensky, T. Tanigawa, T. Endo, and S. Chiba, Phys. Rev. {\bf
C73}, 035802 (2006).
\bibitem{Watanabe}
G. Watanabe, Phys. Rev. {\bf A73}, 013616 (2006).
\bibitem{SV08} V.V. Skokov, and D.N. Voskresensky, arXiv:
0811.3868 [nucl-th].
\bibitem{CH}
J. W. Cahn, and J. E. Hilliard, J. Chem. Phys {\bf 28}, 258
(1958).
\bibitem{Gavin}D. Bower, and S. Gavin, Phys. Rev. {\bf C64}, 051902
(2001);
\bibitem{Kapusta}L.P. Csernai, and J. Kapusta, Phys. Rev. {\bf D46}, 1379  (1992).
\bibitem{Koide}
T. Koide, G. Krein, and R. O. Ramos, Phys. Lett. {\bf B636}, 96
(2006).
\bibitem{astro}
S. Coleman, Phys. Rev. {\bf D15}, 2929 (1977); E.J. Copeland, M.
Gleiser, and H.R. M\"uller, Phys. Rev. {\bf D52}, 1920  (1995).
\bibitem{Beysens}
D.A. Beysens, and Y. Garrabos, Physica {\bf A281}, 361 (2000).
\bibitem{Toneev}
K. Morawetz, M. Ploszajczak, and V.D. Toneev, Phys. Rev. {\bf
C62}, 064602 (2000).
\bibitem{KTV} A.S. Khvorostukhin, V.D. Toneev, and D.N. Voskresensky, Nucl. Phys. {\bf
A791}, 180 (2007); Nucl. Phys. {\bf A813}, 313 (2008).
\bibitem{Aoki:2005vt}
  Y.~Aoki, Z.~Fodor, S.~D.~Katz, and K.~K.~Szabo,
  JHEP {\bf 0601}, 089  (2006).






\bibitem{Berger:1986ps}
  M.~S.~Berger, and R.~L.~Jaffe,
  Phys.\ Rev.\  C {\bf 35}, 213  (1987) ; C{\bf 44}, R566 (1991).






\bibitem{RandrupGQP}
J. Randrup, arXiv:0903.4736; Phys. Rev. {\bf C79}, 054911 (2009).
\end{thebibliography}
\end{document}